\documentclass[12pt]{elsart}
\usepackage{epsfig}
\usepackage{colordvi}
\begin{document}

\begin{frontmatter}

\title{
Systematics of pion emission in heavy ion collisions in the $1A$ GeV regime.}

\author[GSI]{W.~Reisdorf, \thanksref{info}},
\author[HEID]{M.~Stockmeier},
\author[GSI]{A.~Andronic},
\author[HEID]{M.L.~Benabderrahmane},
\author[GSI]{O.N.~Hartmann},
\author[HEID]{N.~Herrmann},
\author[GSI]{K.D.~Hildenbrand},
\author[GSI]{Y.J.~Kim},
\author[GSI,ZAG]{M.~Ki\v{s}},
\author[GSI]{P.~Koczo\'{n}},
\author[GSI]{T.~Kress},
\author[GSI]{Y.~Leifels},
\author[GSI]{X.~Lopez},
\author[HEID]{M.~Merschmeyer},
\author[GSI]{A.~Sch\"{u}ttauf},
\author[CLER]{V.~Barret},
\author[ZAG]{Z.~Basrak},
\author[CLER]{N.~Bastid},
\author[ZAG]{R.~\v{C}aplar},
\author[CLER]{P.~Crochet},
\author[CLER]{P.~Dupieux},
\author[ZAG]{M.~D\v{z}elalija},
\author[BUD]{Z.~Fodor},
\author[ITEP]{Y.~Grishkin},
\author[KOR]{B.~Hong},
\author[KOR]{T.I.~Kang},
\author[BUD]{J.~Kecskemeti},
\author[WAR]{M.~Kirejczyk},
\author[ZAG]{M.~Korolija},
\author[ROSS]{R.~Kotte},
\author[ITEP]{A.~Lebedev},
\author[WAR]{T.~Matulewicz},
\author[ROSS]{W.~Neubert},
\author[BUC]{M.~Petrovici},
\author[STRAS]{F.~Rami},
\author[KOR]{M.S.~Ryu},
\author[BUD]{Z.~Seres},
\author[WAR]{B.~Sikora},
\author[KOR]{K.S.~Sim},
\author[BUC]{V.~Simion},
\author[WAR]{K.~Siwek-Wilczy\'nska},
\author[ITEP]{V.~Smolyankin},
\author[BUC]{G.~Stoicea},
\author[WAR]{Z.~Tymi\'{n}ski},
\author[WAR]{K.~Wi\'{s}niewski},
\author[ROSS]{D.~Wohlfarth},
\author[GSI,IMP]{Z.G.~Xiao},
\author[IMP]{H.S.~Xu},
\author[KUR]{I.~Yushmanov},
\author[ITEP]{A.~Zhilin}

(FOPI Collaboration)
\address[GSI]{Gesellschaft f\"ur Schwerionenforschung, Darmstadt, Germany}
\address[BUC]{National Institute for Nuclear Physics and Engineering,
Bucharest,Romania}
\address[BUD]{Central Research Institute for Physics, Budapest, Hungary}
\address[CLER]{Laboratoire de Physique Corpusculaire, IN2P3/CNRS, and
Universit\'{e} Blaise Pascal, Clermont-Ferrand, France}
\address[ROSS]{ Institut f\"ur Strahlenphysik, Forschungszentrum Rossendorf,
Dresden, Germany}
\address[HEID]{Physikalisches Institut der Universit\"at Heidelberg,Heidelberg,
Germany}
\address[ITEP]{Institute for Theoretical and Experimental Physics,
Moscow,Russia}
\address[KUR]{Kurchatov Institute, Moscow, Russia}
\address[IMP]{Institute of Modern Physics, Chinese Academy of Sciences,
Lanzhou, China}
\address[KOR]{Korea University, Seoul, South Korea}
\address[STRAS] {Institut Pluridisciplinaire Hubert Curien, IN2P3-CNRS,
Universit\'e Louis Pasteur, Strasbourg, France}
\address[WAR]{Institute of Experimental Physics, Warsaw University, Poland}
\address[ZAG]{Rudjer Boskovic Institute, Zagreb, Croatia}

\thanks[info]{Email:~W.Reisdorf@gsi.de}

\begin{abstract}
Using the large acceptance apparatus FOPI, we study pion emission in
the reactions (energies in $A$ GeV are given in parentheses):
$^{40}$Ca+$^{40}$Ca (0.4, 0.6, 0.8, 1.0, 1.5, 1.93),
$^{96}$Ru+$^{96}$Ru (0.4, 1.0, 1.5),
$^{96}$Zr+$^{96}$Zr (0.4, 1.0, 1.5),
$^{197}$Au+$^{197}$Au (0.4, 0.6, 0.8, 1.0, 1.2, 1.5).
The observables include 
longitudinal and transverse rapidity distributions and stopping,
polar anisotropies,                                              
pion multiplicities,
transverse momentum spectra,
ratios $(\pi^+ /\pi^-)$ of average transverse momenta and of yields,
directed flow,
elliptic flow.
The data are compared to earlier data where possible and to transport model
simulations.

\end{abstract}

\begin{keyword}
 heavy ions, pion production, rapidity, stopping, flow, isospin

 \PACS 25.75.-q, 25.75.Dw, 25.75.Ld
\end{keyword}
\end{frontmatter}

\section{Introduction}

The quest to use energetic heavy ion collisions to infer properties of infinite
nuclear matter, such as the equation of state (EOS)  under conditions
of density (or pressure) and temperature
significantly different from the ground state conditions has proven to be a
difficult task, although an impressive amount of data has been obtained using
experimental setups of increasing sophistication.
At incident energies per nucleon on the order or smaller than the rest masses,
systems consisting of two originally separated nuclei are small in the sense
that surface effects cannot be neglected even if some kind of transient
equilibrium situation were achieved.

In particular,
the mean free paths of pions~\cite{lee02}, \cite{mayer93}, the most
abundantly created particles,
with momenta below 1 GeV/c are neither large nor
small compared to typical nuclear sizes.
Therefore methods of analysis resting on the validity of either of
these two extremes in the
hope to simplify the theoretical description are bound to lead only to
qualitative success at best.
As a result,  event simulation codes based on microscopic transport
theory were soon developed~\cite{bertsch88} 
that allowed for multiple elementary collisions as the
heavy ion reaction proceeds, without however requiring a priori full
equilibration.

An important lesson that was learned in the last two decades was that
definite  conclusions on nuclear matter properties based on a
single observable had proven to be premature and/or of limited
accuracy, aside from not being sufficiently convincing.
As an example,
original hopes~\cite{harris87},\cite{stock86} to use deficits in pion
production relative to 
expectations based on compression-free scenarios were not supported by 
transport theoretical simulations~\cite{bertsch84}, \cite{kruse85},
\cite{kitazoe86}.

On the other hand transport calculations~\cite{bass95a} showed that pion 
azimuthal correlations, 'flow', 
qualify as an observable that could contribute significant constraints on the
EOS.
However, as pion production in the 1A GeV regime (at SIS) is not as copious
as it is in higher energy regimes and as pion azimuthal correlations 
turn out to be rather small (an effect of a few percent), 
it was concluded~\cite{bass95a}
that
{\em 'very high statistics and high-precision impact parameter classification
are necessary'} to exploit the sensitivity of pion flow to the EOS.

This requires the need for large acceptance detection systems
capable of registering under exclusive conditions
a much larger number of events than had been possible with the Berkeley Streamer
Chamber that was operated  in the 80's to obtain pion data at the BEVALAC
accelerator~\cite{fung78}, \cite{sandoval80}, \cite{harris85},
\cite{harris87}.
The first electronic $4\pi$-detector capable of measuring pions was
DIOGENE~\cite{alard87} installed at the Saturne synchrotron in Saclay.
The pion emission data presented in this work were obtained with a large
acceptance, high granularity device, FOPI~\cite{gobbi93,ritman95} installed
at the SIS accelerator in Darmstadt.
Particle identification is
based on time-of-flight,
energy loss and magnetic rigidity measurements with use of large volume drift
chambers and scintillator arrays.
A second large acceptance device~\cite{wieman91},
based on the use of a time projection
chamber, was also operated in the nineties at the BEVALAC accelerator in
Berkeley.
In addition, more specialized devices were build and used at the SIS
accelerator in Darmstadt: KaoS~\cite{senger93},
a high resolution magnetic spectrometer and
TAPS~\cite{novotny91}, based on
arrays of BaF$_2$ detectors allowing to identify neutral pions
and $\eta$ particles via their two-photon decay branches.

The experimental situation concerning the pion observable before the advent of
these newer devices was reviewed in ref.~\cite{stock86}, some of the more
recent particle production data obtained in the nineties at the SIS accelerator
and also at other
accelerators (AGS and SPS) covering higher energies have been summarized in
ref.~\cite{senger99}.

Our Collaboration has published pion data before~\cite{pelte97au},
\cite{pelte97ni}, \cite{hong97}, \cite{hong98}, \cite{hong05},
but these were more limited
in scope and, in particular, did not treat pion flow and isospin dependences.
Also, we call attention to the fact that the data of ref. \cite{pelte97au}
(for Au on Au at $1.06A$ GeV) have been revised and should be superseded
by the present data (see section \ref{experiment} for details).
The aim of the present work is to present a more complete systematics of
highly differential pion emission in heavy ion reactions obtained with the FOPI
device, varying the incident energy
(from 0.4 to $1.9A$ GeV), the system's size (from $A_p+A_t =40+40$ to
$197+197$, where $A_p$ and $A_t$, are the projectile and
target mass number, respectively) and the system's isospin, 
and, of course,  the event selection method i.e. the
centrality.

After describing the experimental methods we report on the
following observables for charged pions of both polarities:
\begin{itemize}
\item longitudinal and transverse rapidity distributions and stopping;
\item polar anisotropies;                                              
\item pion multiplicities;
\item transverse momentum spectra;
\item ratios $(\pi^+ /\pi^-)$ of average transverse momenta and of yields;
\item directed flow;
\item elliptic flow.
\end{itemize}

While discussing each of these subjects,
we shall refer more explicitly to relevant earlier
work and compare data where possible.
All along we shall also present the results of simulations with a transport
code showing the degree to which our data can be understood on a microscopic
level and assessing conditions for equilibration (stopping),
 sensitivities to assumptions on the EOS and the
propagation of pions in the medium and searching for signals that might give
information on the isospin dependence of the EOS.
While pions present a probe of hot and compressed matter of high interest
in its own right, it is also important to have this observable under firm
theoretical control as it is a link to understanding the production of
strangeness under subthreshold conditions where the pion emitting baryonic
resonances are thought to play an essential role in the collision sequences
leading to outgoing strange particles, such as kaons.
We will end with a summary.

\section{Experimental method} \label{exp}

The experiments were performed at the heavy ion accelerator SIS of
GSI/Darmstadt using the large acceptance FOPI detector \cite{gobbi93,ritman95}.
A total of 18 system-energies are analysed for this work (energies in $A$ GeV
are given in parentheses):
$^{40}$Ca+$^{40}$Ca (0.4, 0.6, 0.8, 1.0, 1.5, 1.93),
$^{96}$Ru+$^{96}$Ru (0.4, 1.0, 1.5),
$^{96}$Zr+$^{96}$Zr (0.4, 1.0, 1.5),
$^{197}$Au+$^{197}$Au (0.4, 0.6, 0.8, 1.0, 1.2, 1.5).
Particle tracking  and energy loss determination are
done using two drift chambers, the CDC (covering polar angles
between $35^\circ$ and $135^\circ$) and
the Helitron ($9^\circ-26^\circ$), both
located inside a superconducting solenoid operated at a magnetic field of
0.6T.
A set of scintillator arrays, Plastic Wall $(7^\circ-30^\circ)$,
Zero Degree Detector
$(1.2^\circ-7^\circ)$, and Barrel $(42^\circ-120^\circ)$,
 allow us to measure the time of flight
and, below $30^\circ$, also the energy loss.
The velocity resolution below $30^\circ$ was $(0.5-1.5)\%$,
the momentum resolution in the
CDC was $(4-12)\%$ for momenta of 0.5 to 2 GeV/c, respectively.
Use of CDC and Helitron allows the identification of pions, as well as
good isotope separation for  hydrogen and
helium clusters in a large part of momentum space.
Heavier clusters are separated  by nuclear charge.
More features of the experimental method, some of them specific to pions,
have been described in Ref.~\cite{pelte97au}.

\subsection{Pions: reconstructing $4\pi$ from FOPI} \label{fourpi}

The pion data presented in this work are limited to the CDC.
In one of two independent analyses methods, particle tracking was based on the
Hough transform method, HT.
Track quality cuts were varied systematically and the results extrapolated
to zero cuts to achieve estimates of the tracking efficiency.
The relative efficiency of positively and negatively charged pions was inferred
from studies of the isospin symmetric system $^{40}$Ca+$^{40}$Ca.
An alternative method of data analysis with a local tracker, LT, has also been
used allowing extensive cross checking.
This method is documented in ref.~\cite{stockmeier02} and we shall
briefly come back to it in section~\ref{IQMD}.

The measured 
momentum space distributions of pions do not
cover the complete $4\pi$ phase space and must be complemented by
interpolations and extrapolations.
In the HT method,
we  filter the data to eliminate regions of
distorted measurements (such as edge effects)
and correct for efficiency where necessary.
Since this study is limited 
to symmetric systems, we require
reflection symmetry in the center of momentum ($c.o.m.$).
Choosing the $c.o.m.$ as reference frame, orienting the z-axis in the beam
direction, and ignoring for the moment deviations from axial symmetry (see
section \ref{azimuthal})
 the two remaining dimensions are characterized by the longitudinal
rapidity $y\equiv y_z$,
given by $exp(2y)=(1+\beta_z)/(1-\beta_z)$ and the transverse
(spatial) component $t$ of the four-velocity $u$, given by $u_t=\beta_t\gamma$.
The 3-vector $\vec{\beta}$ is the velocity in units of the light
velocity and $\gamma=1/\sqrt{1-\beta^2}$.
In order to be able to compare longitudinal and transversal degrees of
freedom on a common basis,
we shall also use the {\em transverse} rapidity, $y_x$,
which is defined by
replacing $\beta_z$ by $\beta_x$ in the expression for the longitudinal
rapidity.
The $x$-axis is laboratory fixed and hence randomly oriented relative to the
reaction plane, i.e. we average over deviations from axial symmetry.
The transverse rapidities $y_x$ (or $y_y$) should not be confused with
$y_t$ which is defined by replacing $\beta_z$ by
$\beta_t\equiv\sqrt{\beta_x^2+\beta_y^2}$.

For thermally equilibrated systems $\beta_t=\sqrt{2}\beta_x$ and the local
rapidity distributions $dN/dy_x$ and $dN/dy_y$ (rather than $dN/dy_t$)
should have the same shape and height than the usual longitudinal
rapidity distribution $dN/dy_z$, where we will omit the subscript $z$ when no 
confusion is likely.
Throughout we use scaled units $y_0=y/y_p$ and $u_{t0}=u_t/u_p$,
with $u_p=\beta_p \gamma_p$, the index p referring to the incident projectile
in the $c.o.m.$.
In these units the initial target-projectile rapidity gap always extends from
$y_0=-1$ to $y_0=1$.
It is useful to recall that in non-viscous one-fluid hydrodynamics many
observables scale when the system's size and energy, but not the shape or
impact parameter, are varied. 

\begin{figure}[!t]
\begin{minipage}{79mm}
\begin{center}
\epsfig{file=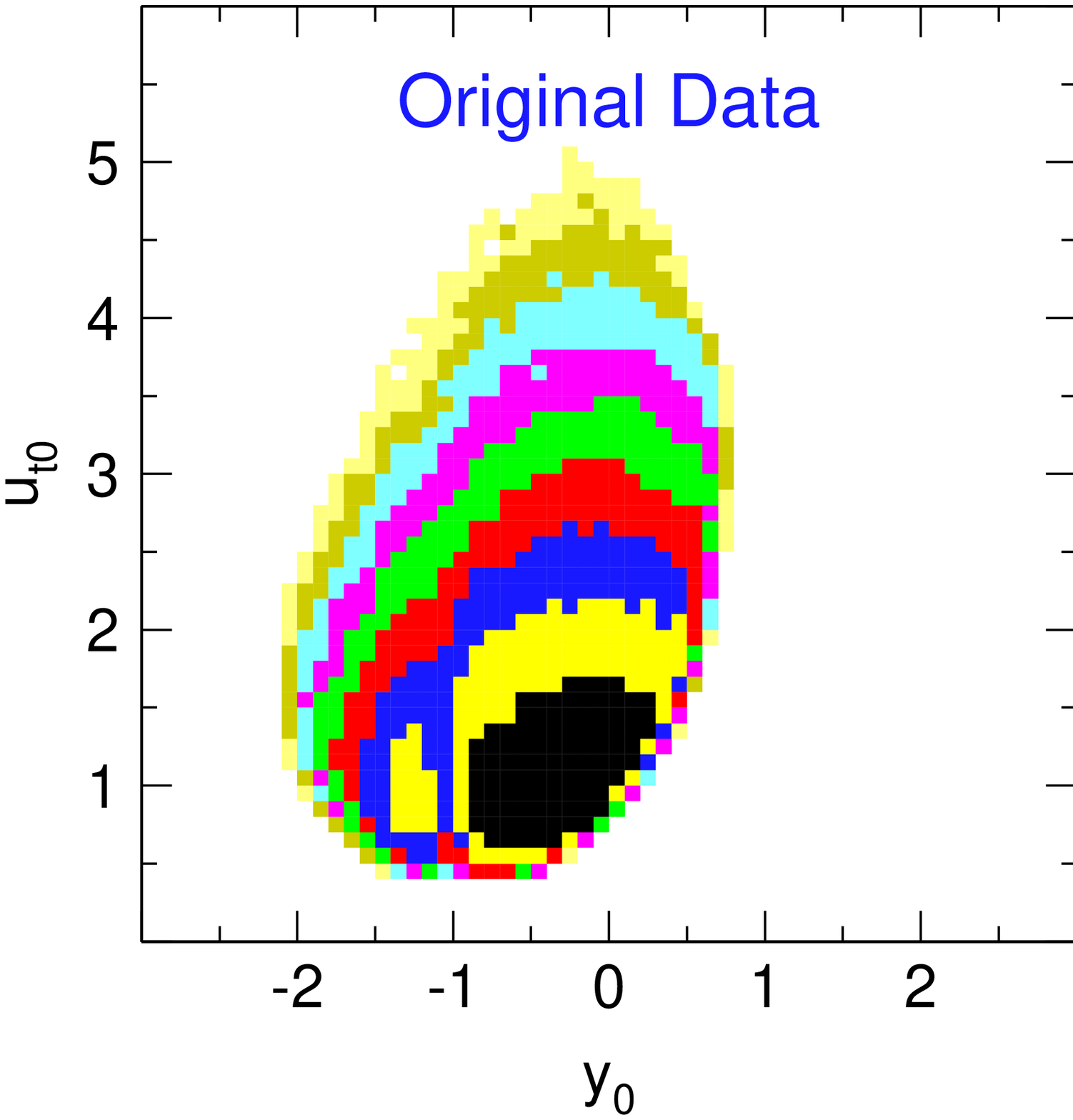,width=7.9cm}
\end{center}
\end{minipage}
\begin{minipage}{79mm}
\begin{center}
\epsfig{file=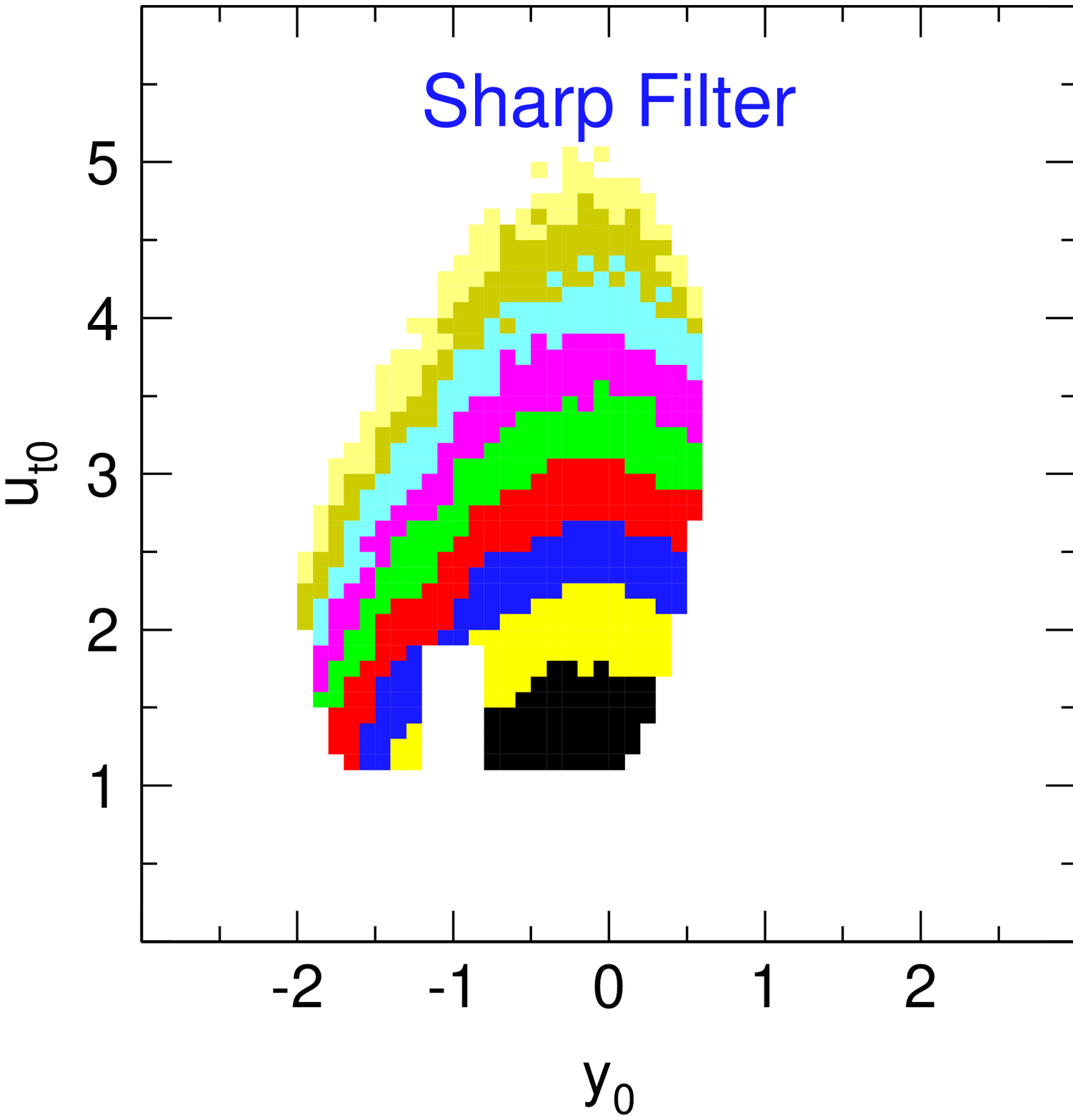,width=7.9cm}
\end{center}
\end{minipage}

%
\begin{minipage}{79mm}
\begin{center}
\epsfig{file=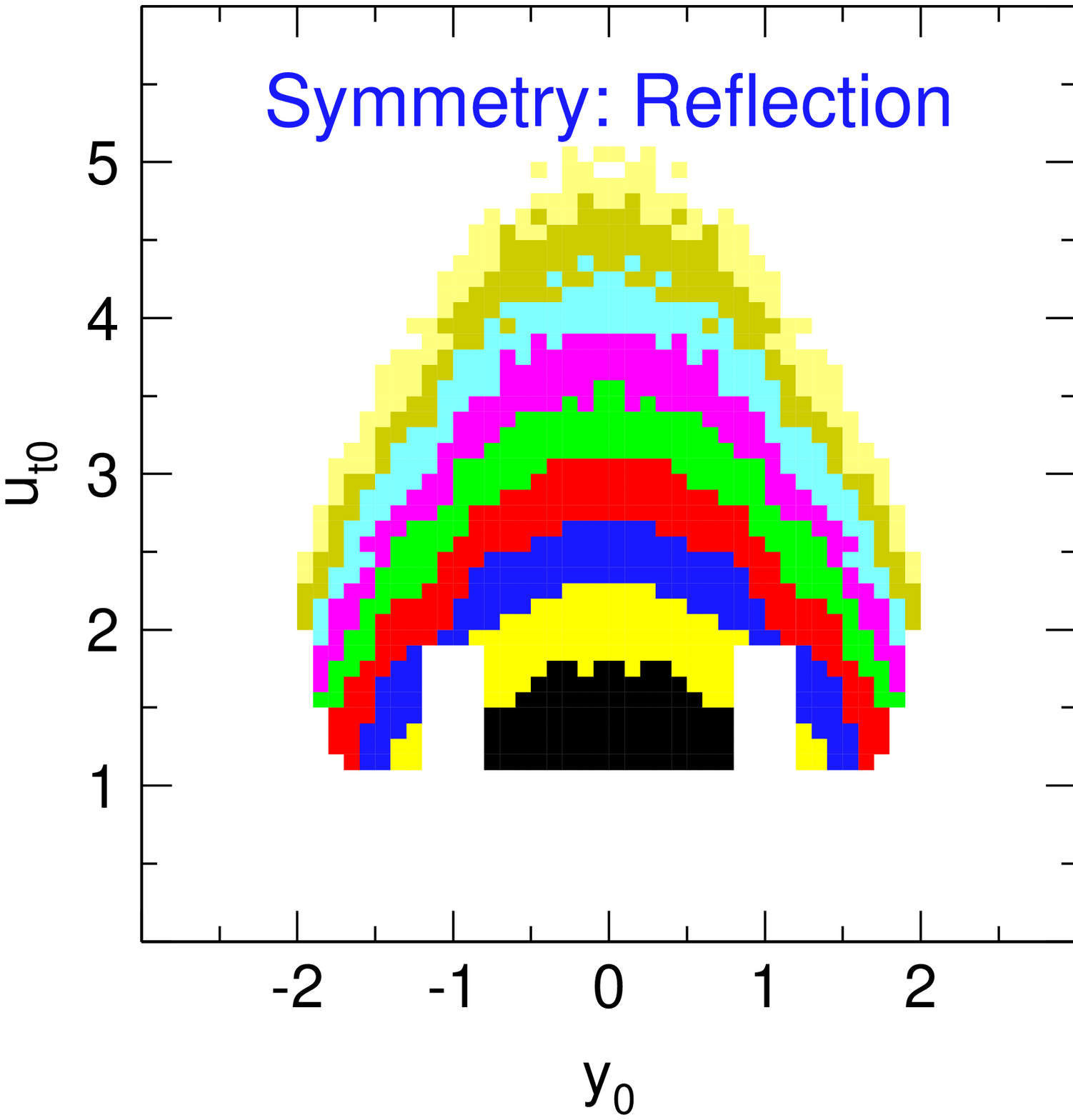,width=7.9cm}
\end{center}
\end{minipage}
\begin{minipage}{79mm}
\begin{center}
\epsfig{file=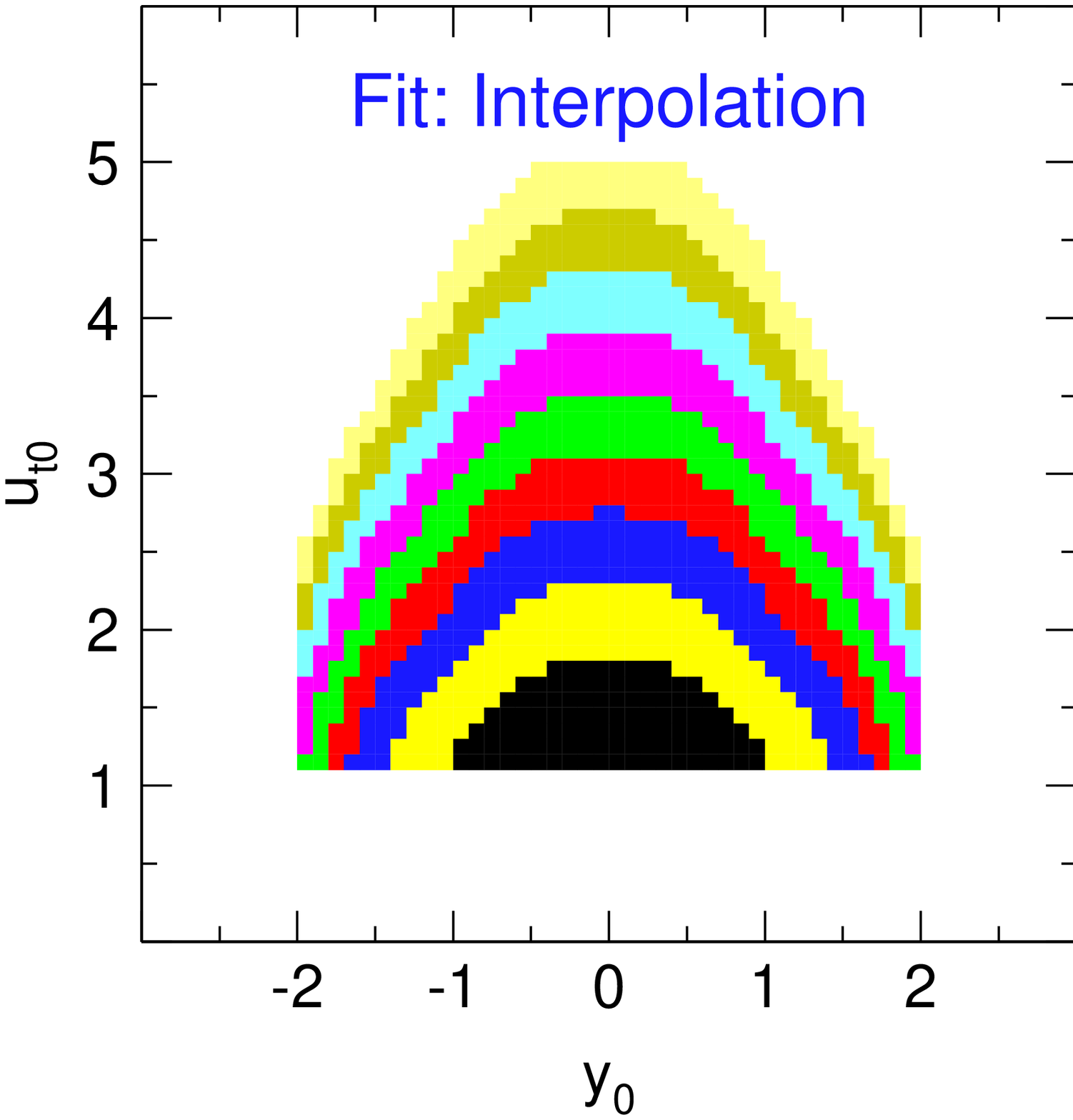,width=7.9cm}
\end{center}
\end{minipage}
%

\begin{minipage}{79mm}
\begin{center}
\epsfig{file=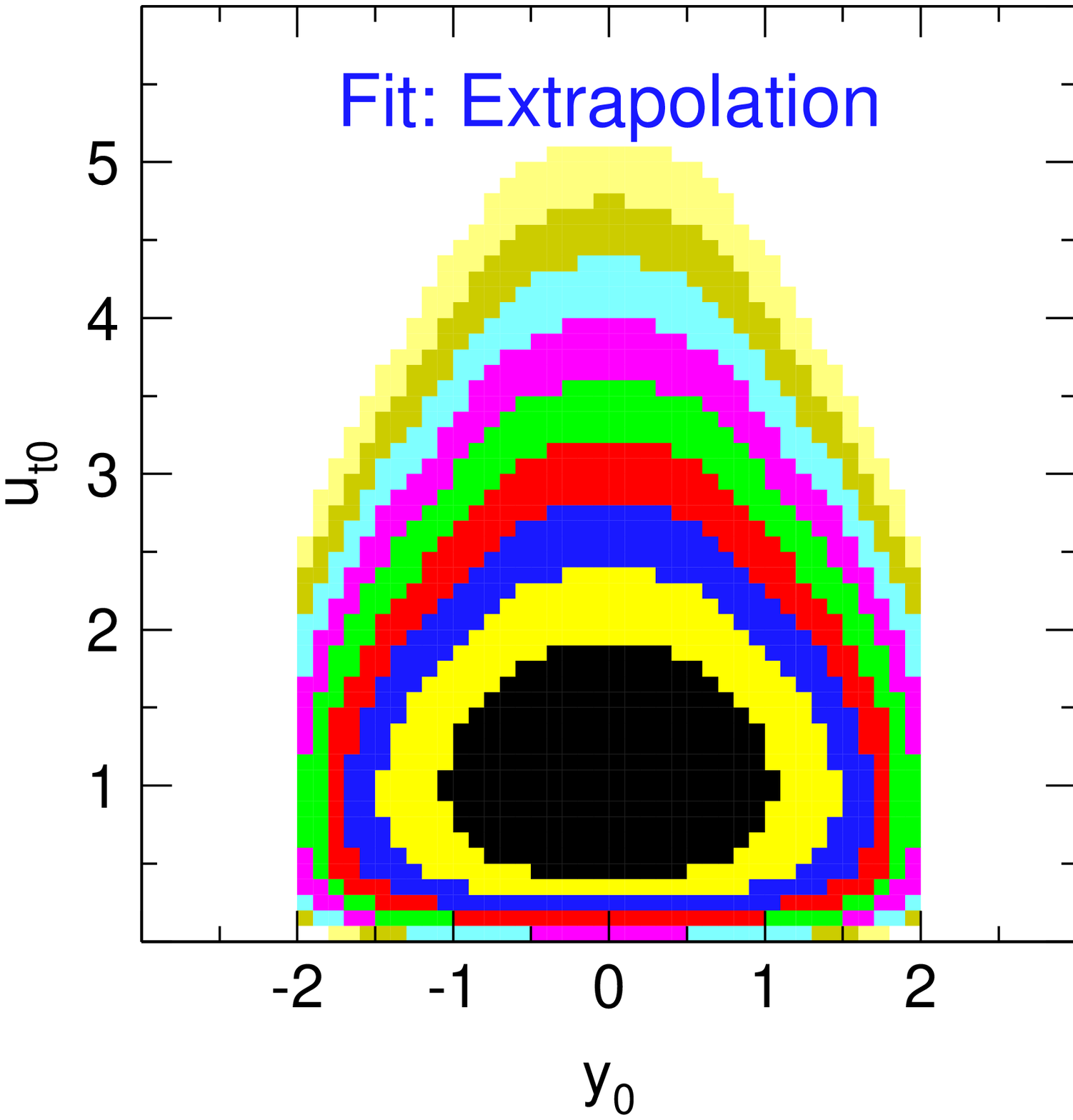,width=7.9cm}
\end{center}
\end{minipage}
\begin{minipage}{79mm}
\begin{center}
\epsfig{file=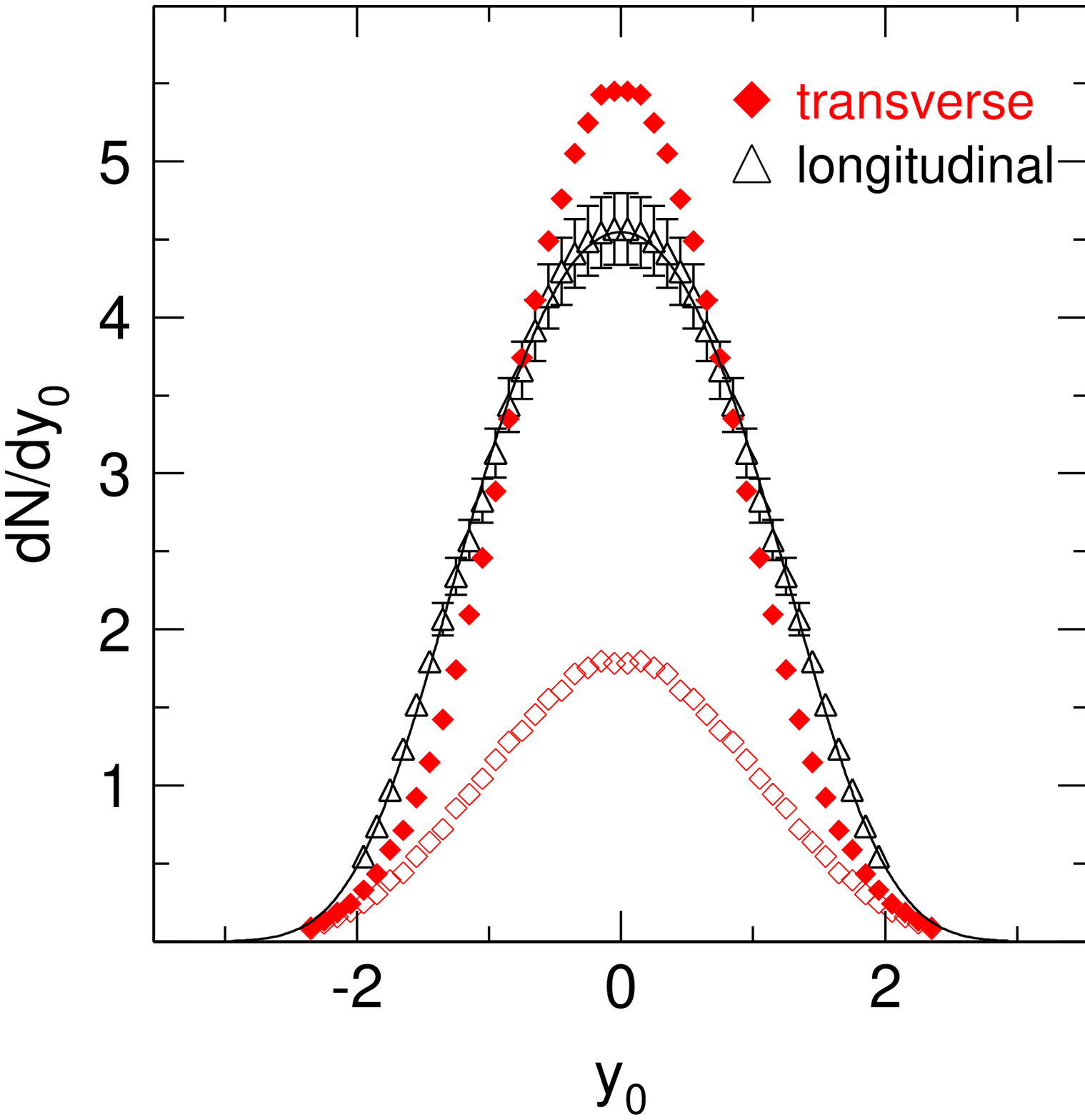,width=7.9cm}
\end{center}
\end{minipage}

\caption{
Distributions $dN/du_{t0} dy_0$ of pions ($\pi^-$)
emitted in central collisions
of Au on Au at $0.8A$ GeV. The various color tones correspond to cuts differing
by factors 1.5.
Five panels show the evolution of the data treatment (see text),
while the sixth panel (lower right) shows deduced longitudinal and transverse
rapidity distributions. 
The transverse rapidity distribution with a cut on longitudinal rapidity
$(|y_0|<0.1)$, is also shown. }
\label{uty6}
\end{figure}

Choosing central collisions of Au on Au at $0.8A$ GeV as a typical example, we
show in the upper left panel of Fig.~\ref{uty6} 
the original $\pi^-$ (CDC) data in the $(y_0$-$u_{t0})$ plane.
Proceeding from the upper left  to the lower right panel of the
figure, the next two panels show the data after application of a sharp filter
and the use of reflection symmetry.

\newpage

\bigskip
In the following step each measured phase-space cell
$dy_0*du_{t0}$ and its local surrounding $N_{yu}$ cells, with
$N_{yu} = (2n_y+1)(2n_u+1)$, is  least squares fitted using the ansatz
$$  \frac{1}{u_{t0}} \frac{d^2 N} {du_{t0} dy_0 }
 = \exp \, [f(y_0,u_{t0})] $$
where $f(x,y)=a_2 x^2 + b_2 y^2 + d |x| y + a_1 |x| +b_1 y +c_0$
is a five parameter function.

This procedure smoothens out statistical errors and allows
subsequently a well defined iterative extension to gaps in the data.
Within errors the smoothened representation of the data follows  the
topology of the original data:
typical deviations are $5\%$ i.e. of a magnitude that exceeds statistical
errors in most cases and is caused by local distortions of the apparatus
response, thus revealing typical systematic uncertainties.

The technical parameters of the procedure were chosen to be
$dy_0=0.1$, $du_{t0}=0.1$, $n_y=4$, $n_u=6$ 
(except for the data at 0.4A GeV where $n_y=6$ and $n_u=8$).
These choices are governed by the available statistics and the need to follow
the measured topology  within statistical and systematics errors.
Variations of these parameters were investigated and found to be uncritical
within reasonable limits.

The smoothened data (middle right panel)
are well suited for the final step: the extrapolation
to zero transverse momenta also shown in the figure (bottom left panel).

The low $p_t$ extrapolation procedure was guided by  microscopic event
simulations (to be described in more detail in section~\ref{IQMD}) 
that take into account the influence of Coulomb effects,
which differ for the two kinds of charged pions.

\begin{figure}[h]
\begin{minipage}{75mm}
\begin{center}
\epsfig{file=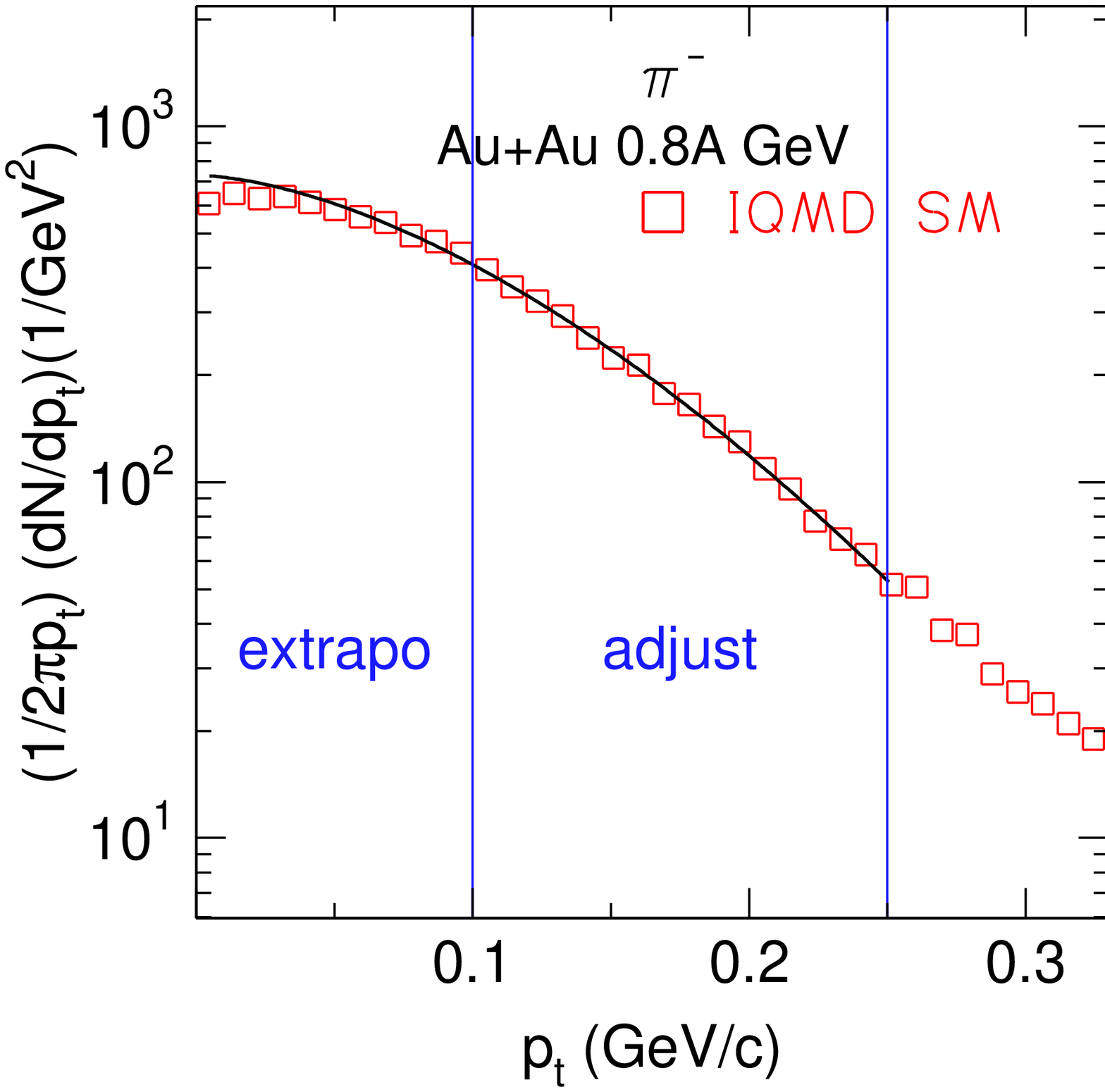,width=75mm}
\end{center}
\end{minipage}
\begin{minipage}{75mm}
\begin{center}
\epsfig{file=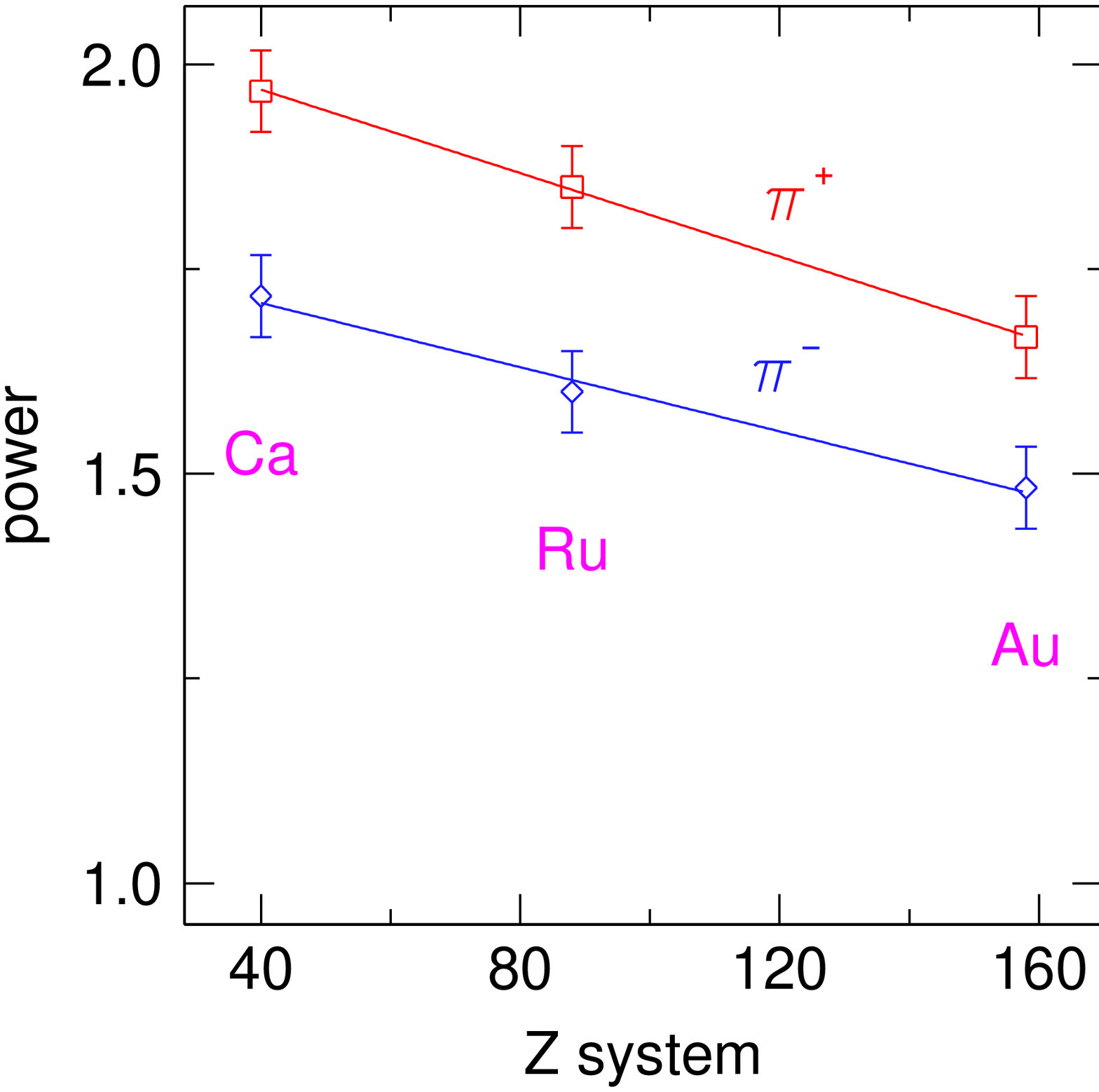,width=75mm}
\end{center}
\end{minipage}

\begin{minipage}[t]{75mm}
\caption{
Transverse momentum spectrum of negative pions in the reaction Au+Au at
$0.8A$GeV. 
The symbols represent data from a simulation with IQMD SM.
The solid line is a least squares fit (see text) using the data in
the momentum range (marked 'adjust') from 0.1 to 0.25 GeV/c.}
\label{extrapo}
\end{minipage}
\hspace{8mm}
\begin{minipage}[t]{75mm}
\caption{
{\small 
The power parameter used to reproduce the simulated data for
positive and negative pions as a function of the size (total charge) of
the system. }
}
\label{power}
\end{minipage}
\end{figure}
 
\bigskip
Fig.~\ref{extrapo} shows a simulated transverse momentum spectrum for $\pi ^{-}$
emitted in central Au on Au collisions at 0.8A GeV. It is integrated
over all longitudinal rapidities.
The range to be extrapolated for the experimental data ($p_t < 0.1$ GeV/c)
is indicated.
The simulated  data were fitted in the indicated adjustment range
(0.1-0.25 GeV/c) with the function 
$$ N \exp (-p_t^x/c)$$  
the power $x$ and the constant $c$ being two shape
 parameters, and $N$ a normalization parameter
(in the non-relativistic, no-Coulomb regime a 'thermal' fit would
mean $x=2.0$ and $c= 2mT$ in terms of a temperature $T$).
The fit function was accepted when it also gave a good reproduction of the
theoretical data in the
low-$p_t$ range (even though it was not included in the fitting procedure).
The best powers $x$, shown in Fig.~\ref{power}, turned out to be somewhat
lower than 2, depending on the system size and the pion charge polarity.
The dependence  on the incident energy was very weak and therefore was
ignored.

\begin{figure}[t]
\begin{minipage}{80mm}
\hspace{-9mm}
\epsfig{file=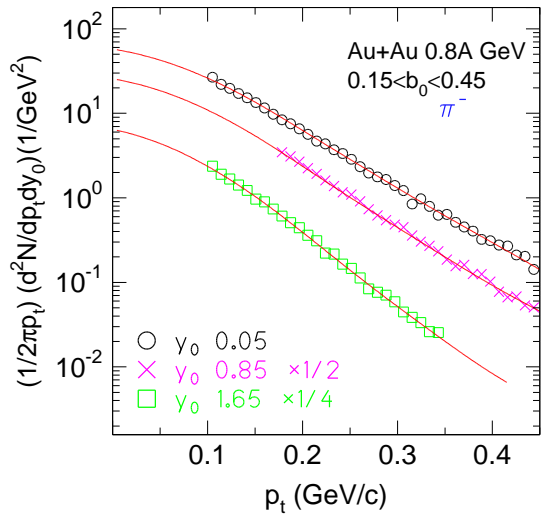,width=85mm}
\end{minipage}
\begin{minipage}{80mm}
\caption{
{\small
Measured
transverse momentum spectra of $\pi^-$ in the reaction Au+Au at $0.8A$ GeV
in various scaled rapidity bins of widths 0.1. 
Centroids and data scaling factors are given in the figure.
The data are given by symbols, the solid lines represent a smoothened version of
the data plus extrapolations.}
}
\label{ptspec}
\end{minipage}
\end{figure}

The experimental data were then extrapolated using the same (available)
adjustment range and $x$-values suggested by the simulation
but varying $N$ and {\em one} shape parameter only, $c$.
The typical outcome is shown in Fig.~\ref{ptspec} for three indicated
scaled rapidity intervals.

\subsection{Centrality selection} \label{centrality}

Collision centrality selection was obtained by binning
distributions of the ratio, $ERAT$~\cite{reisdorf97a}, of total transverse
and longitudinal kinetic energies.
In terms of the scaled impact
parameter, $b_0=b/b_{max}$, we choose the same centralities for all the
systems: $b_0<0.15$ (i.e. $2.25\%$ in terms of total cross sections),
$b_0<0.25$, $0.25<b_0<0.45$, $0.45<b_0<0.55$. 
We take $b_{max} = 1.15 (A_{P}^{1/3} + A_{T}^{1/3})$ fm
as effective sharp radius and estimate $b$
from the measured differential cross sections for the $ERAT$
distribution using a geometrical sharp-cut approximation.
$ERAT$ selections do not imply {\em a priori} a chemical bias.
Autocorrelations in high transverse momentum population, that are caused by
the selection of high $ERAT$ values, are avoided by not including identified
pions in the selection criterion.

\section{Simulation using the IQMD transport code.}\label{IQMD}

In the present work we are making extensive use of the code
IQMD~\cite{hartnack98}
which is based on Quantum Molecular Dynamics~\cite{aichelin91}.
Many interesting predictions and revealing interpretations of the mechanisms
involving pions have been made~\cite{bass93}-\cite{bass95b}, \cite{bass95a}
with this model
before our data were
available making it almost mandatory to confront it now with the present 
measured data.

One of the motivations for numerical simulations of heavy ion reactions with
transport codes is the possibility to investigate the effect of the underlying
equation of state, EOS, on the experimental observables without relying on the
restrictive assumption of local (and {\em a fortiori} global) equilibrium
followed by 'sudden' freeze-out of all elementary hadronic collisions.
In between collisions, nucleons are propagating in mean fields,
the nuclear parts of which correspond in the limit of infinite matter to well
defined zero temperature EOS.
These EOS can be chosen to be 'stiff' or 'soft', as characterized in terms of
incompressibilities $K = 380$ MeV, respectively 200 MeV, where
$K = 9\rho^2\partial ^2(E/A)/\partial\rho^2$ near $\rho=\rho_0$, the
saturation density.
We shall henceforth label these two options, HM, respectively SM.
The M in HM and SM stands for the momentum dependence of the nucleon-nucleon
interaction. 
IQMD incorporates a phenomenological Ansatz fitted to experimental data on the
real part of the nucleon-nucleus optical potential.
In the IQMD code pions are produced by the decay of the 1232 MeV $\Delta$
baryon resonance and may be reabsorbed exclusively by forming a $\Delta$ again.
The dominance of the lowest nucleonic excitation in the $1A$ GeV
energy regime has been directly demonstrated in proton-pion
correlation studies \cite{hjort97}, \cite{eskef98}, \cite{matulewicz00}.

While in between collisions, the $\Delta$ baryons are propagating 
in the same nuclear
and Coulomb fields as the nucleons, pions are feeling only the Coulomb
potential.
For simplicity, and in order to limit the number of input variables we
'shut off' the poorly known (isospin) symmetry potential in these
exploratory calculations.
Thus 'isospin effects', if any, would have to result either from the 
isospin dependent NN cross sections (implemented in the code), e.g.
a 'cascade' effect, and/or
the Coulomb fields.
As in the model pion production and absorption is strongly connected with
$\Delta$ baryon production and absorption, the physics of $\Delta$ propagation
in the medium becomes important both for the observed final number of pions and
the flow pion 'daughters' inherit from their 'parents'.
If not otherwise stated we have used in our calculations the scheme of
ref.~\cite{kitazoe86} for the $\Delta$ baryon mass distribution and width.

Some recent implementations of transport theoretical
codes for heavy ion collisions in the $1A$ GeV regime contain more advanced
 features than those just described.
Without attempting to be complete, we  mention the works of
Bao An Li and coworkers~\cite{bali04} and of the Catania group~\cite{baran05}
which  implement various isospin dependences of the {\em mean field}.
Other codes have made progress implementing non-equilibrium aspects of the
local densities~\cite{gaitanos},
better accounting for off-shell effects when particle (resonance) lifetimes
are short and/or the collision rate is high~\cite{cassing00},
reassessing the in-medium cross sections for the $NN \rightarrow N\Delta$
reaction~\cite{larionov03}, etc. .
The code of Danielewicz~\cite{danielewicz95}
includes nucleonic clusters up to mass three
and has been used in our context here to clarify the role of expansion
on the pion observables.
Very recently, the relativistic code UrQmd has also been used~\cite{qfli05}
to study SIS energy
reactions, although its original design~\cite{bass98} was directed towards
much higher energies. Most of these codes    include the influence of many
higher baryonic resonances.

\begin{figure}[h]
\begin{center}
\epsfig{file=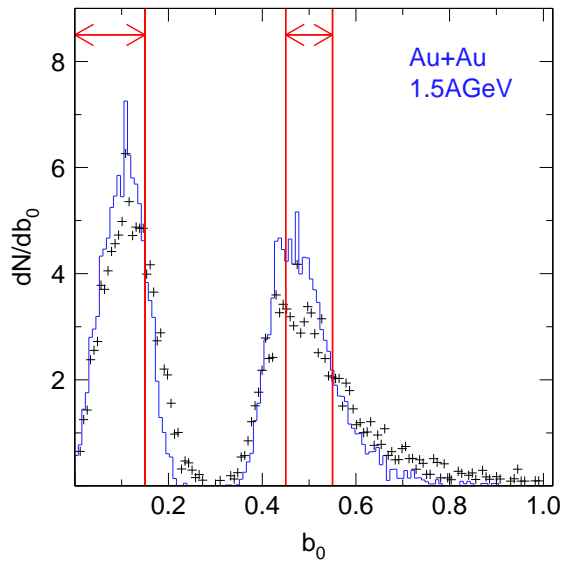,width=90mm}
\end{center}

\caption{
\small{
Simulated reduced impact parameter distributions 
for Au+Au collisions at $1.5A$ GeV using the global observable
$ERAT$ for event selection.
The two peaks correspond to nominal centralities $b_0 < 0.15$ and
$0.45 < b_0 < 0.55$, respectively, as indicated by the vertical lines and
the double arrows.
Histograms (crosses) correspond to unfiltered (filtered) data.}
}
\label{b-au1500}
\end{figure}

One of the advantages of a QMD type code over BUU 
(Boltzmann-Uehling-Uhlenbeck) implementations
might be its capability to induce finite (nucleon) number fluctuations
without resorting to additional assumptions or recipes.
For the simulation of the experimental situation it is of paramount importance
that there is an accurate 'centrality matching'.
i.e. the centrality criterion of the experiment (here the observable $ERAT$)
and its realistic fluctuations should be simulated.
We vary $b_0$ uniformly between 0 and 1 and add the events
(about 50000 per system-energy) weighed by $b_0$.
Fig.~\ref{b-au1500} shows $b_0$ distributions resulting from $ERAT$ selection in
two of our standard centrality intervals.
The nominal $b_0$ intervals are also shown, as well as the effect of applying a
filter that takes into account the geometrical and the threshold limits
of the apparatus.
As can be seen this filter does not have a dramatic influence on the
distributions, and hence its details  are not critical.
However, the effect of using $ERAT$, rather than the experimentally elusive
$b_0$, is not negligible, even with a perfect detector.
For the energy range of interest here, we find that the alternative selection
method, binning charged particle multiplicities, yields less sharp $b_0$
distributions as long as $b_0<0.40$, but is competitive for more peripheral
collisions. 
In particular, multiplicity binning avoids the high $b_0$ tails beyond $b_0=0.6$
visible in the figure.
Multiplicity selected data are available, but will not be shown here.
$ERAT$ has the advantage that it does not require the simulation to reproduce
the degree of nucleonic clusterization,
a difficult (still) task for transport codes (see ref.~\cite{reisdorf04b}). 
Finite number effects deteriorate the $b_0$
'resolution' when lighter systems, such as Ca+Ca, are studied, an additional
reason to try to be realistic when simulating the experiment.

We have also used the IQMD code with a more technical aim:
we have introduced it~\cite{stockmeier02}
as an event generator for a GEANT based~\cite{geant} Monte Carlo simulation
of our apparatus response to better assess tracking efficiencies
and losses due to geometrical limitations.
This alternative independent data analysis~\cite{stockmeier02}
 using a local tracking method, LT,
instead of the Hough-transform based tracker, HT, was found to
yield $4\pi$ pion multiplicities 
in good agreement ($10\%$ or better) with the method outlined above (e.g. see
Figs.~\ref{ruru} and \ref{sumpion} in section~\ref{production}).

\newpage
\section{Rapidity distributions and stopping} \label{rapidity}

\begin{figure}[h]
\begin{center}
\hspace{-15mm}
\begin{minipage}{67mm}
\epsfig{file=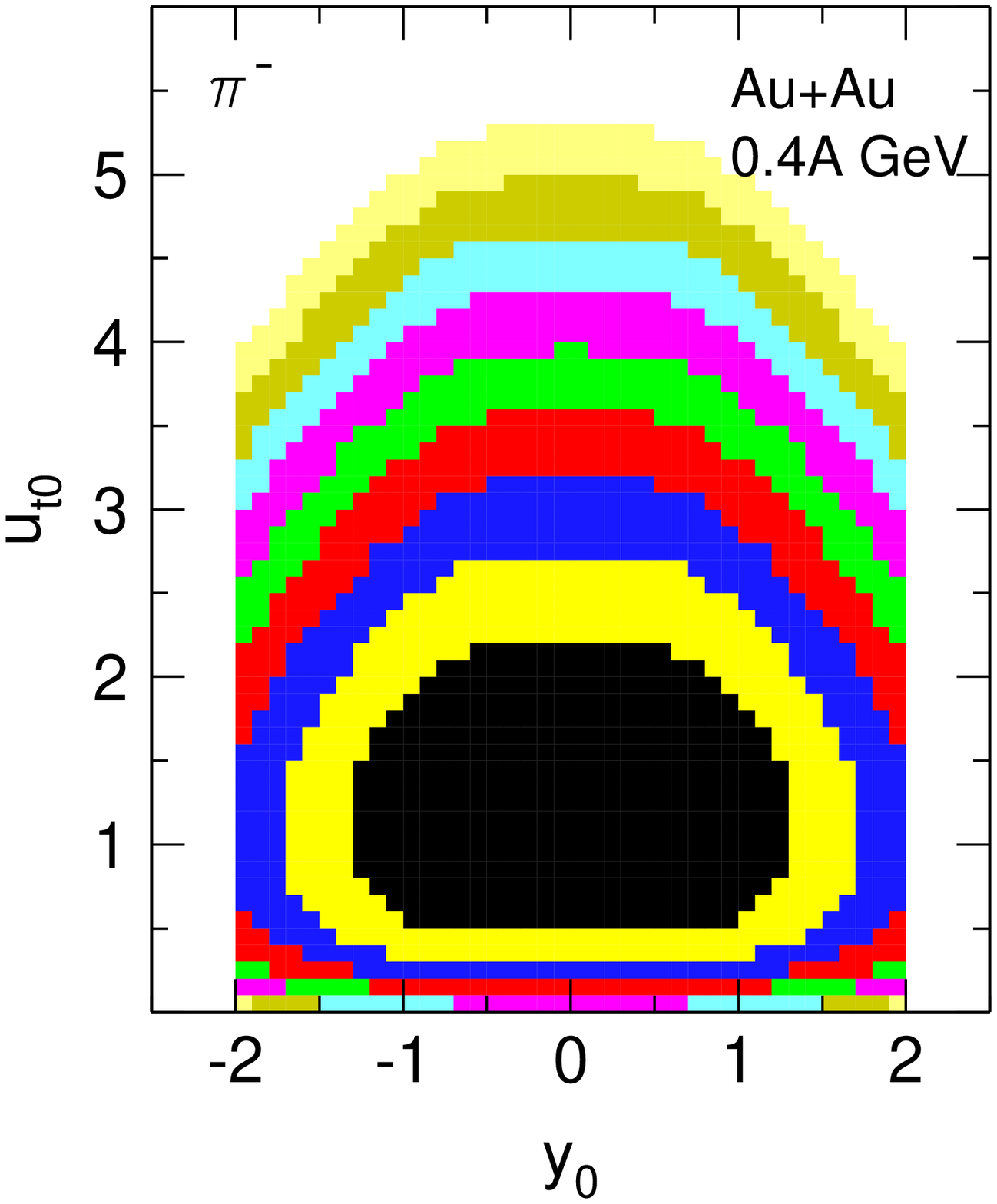,width=70mm}
\end{minipage}
\begin{minipage}{67mm}
\epsfig{file=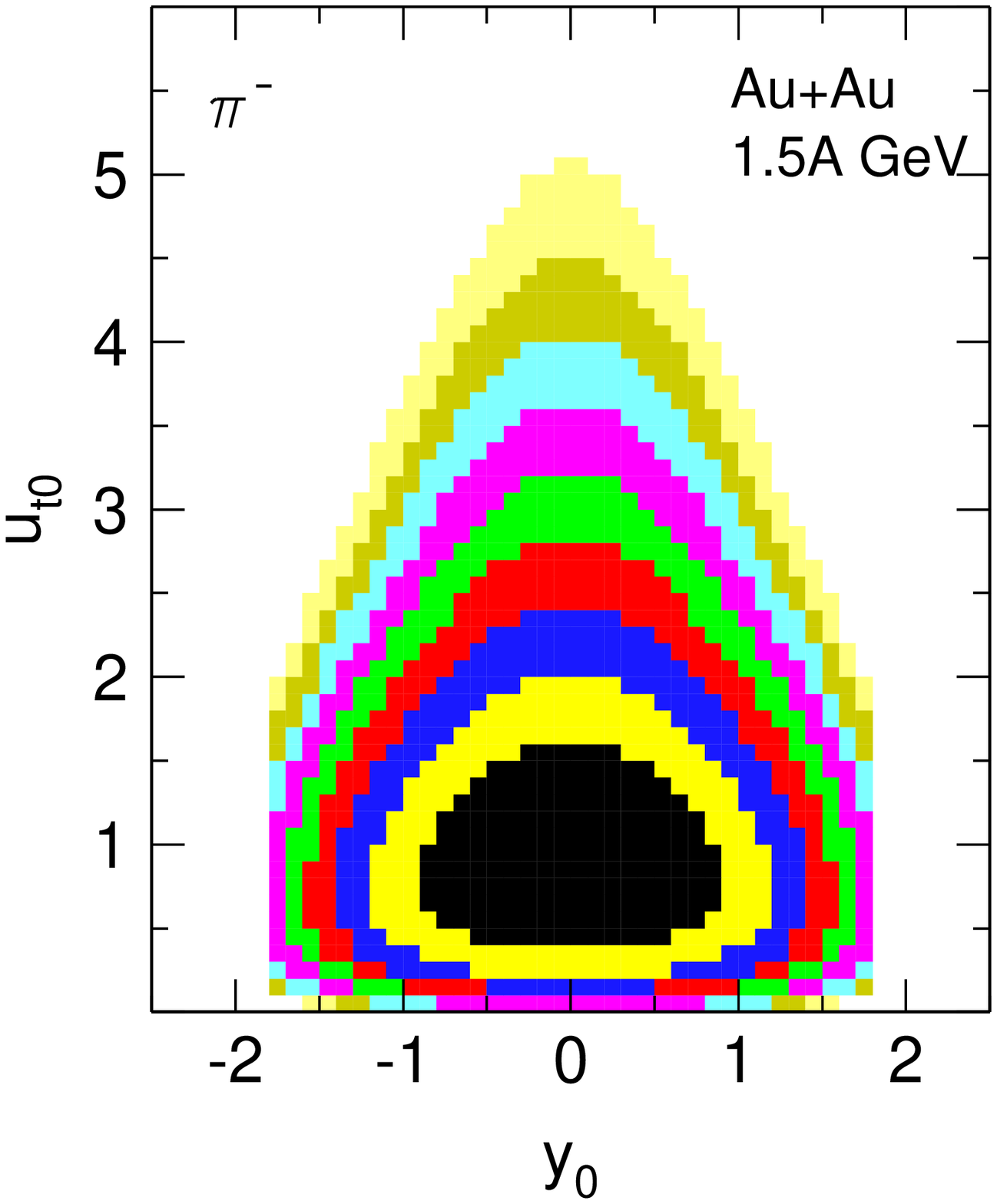,width=70mm}
\end{minipage}
\end{center}

\begin{center}
\hspace{-15mm}
\begin{minipage}{67mm}
\epsfig{file=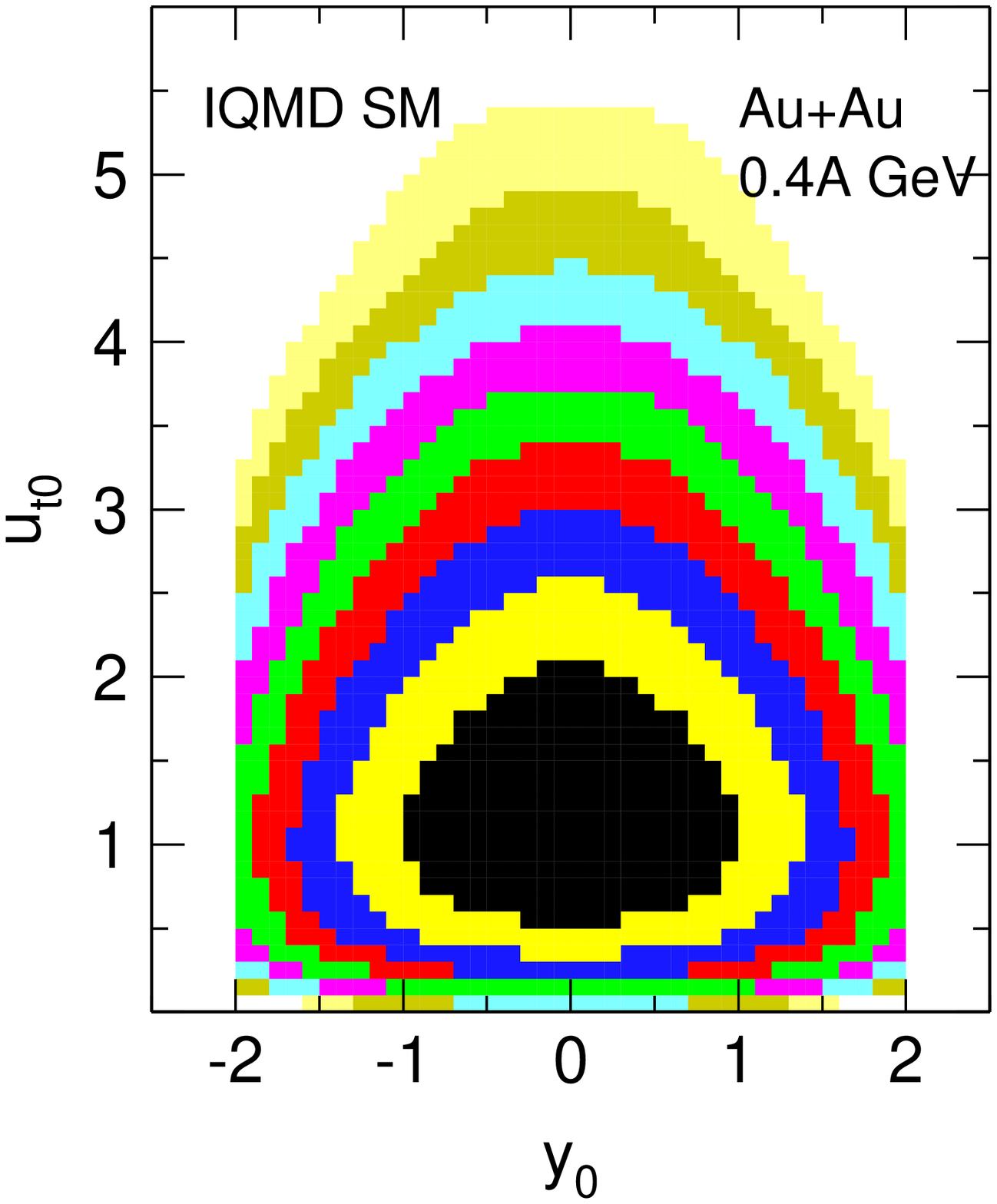,width=70mm}
\end{minipage}
\begin{minipage}{67mm}
\epsfig{file=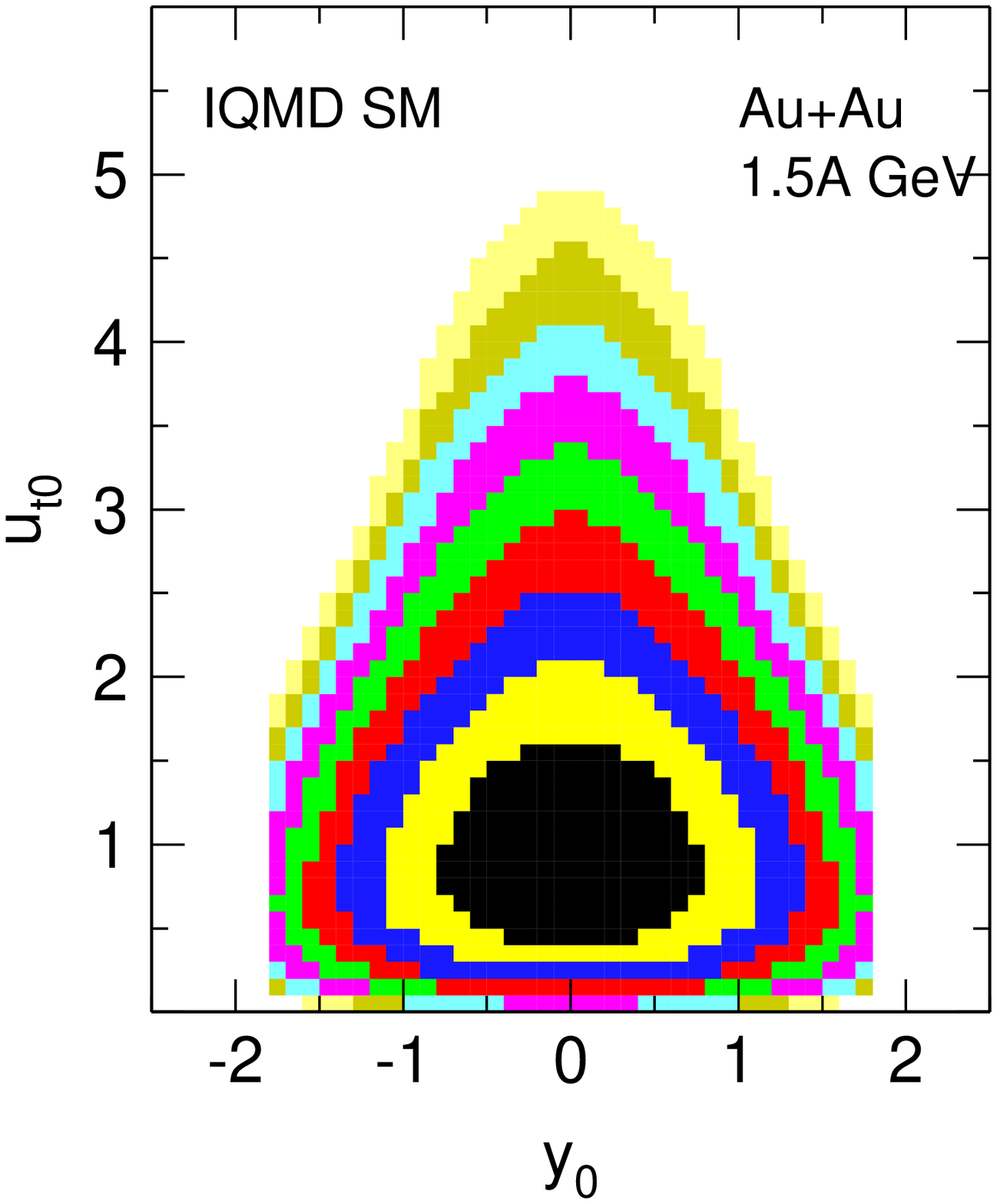,width=70mm}
\end{minipage}
\end{center}
\caption{
\small{
$dN/du_{t0} dy_0$ distributions for pions $(\pi^-)$ emitted in $b_0<0.15$
collisions of Au+Au at $0.4A$ GeV (left) and $1.5A$ GeV (right).
Top panels: data, bottom panels: simulation.
The grey (color) tones differ by factors 1.5.}
}
\label{uty}
\end{figure}

Before looking at special projections in momentum space it is useful to display
the full (2-dimensional) distributions $dN/du_{t0} dy_0$.
One such distribution was already shown in Fig.~\ref{uty6} for Au+Au at
$0.8A$ GeV.
To see the evolution of the topology with incident energy - i.e. from $0.4A$
to $1.5A$ GeV - two more such distributions are shown in the upper panels
of Fig.~\ref{uty}.

The pion (here $\pi^-$) sources are centered at midrapidity with no readily
evident memory of the initial conditions (i.e. around $|y_0|=1$).
Comparing the contours at $0.4A$ GeV with those at $1.5A$ GeV, one notices
that in the scaled units chosen for these plots, the distributions are
significantly wider in both dimensions at the lower energy.
The data at $0.4A$ GeV span more than twice the full rapidity gap.
At least two properties involved in pion production are not expected to scale:
a) the nucleonic Fermi motion which should be relatively more important
at the lower energy and 
b) the decay kinematics of the parent $\Delta$ (and other) baryonic resonances.

The simulation with IQMD (see the lower panels) qualitatively reproduces the
higher 'compactness' at $1.5A$ GeV, but is less expanded in the longitudinal
direction, especially at the lower energy.

\begin{figure}[htb]
\begin{center}
\epsfig{file=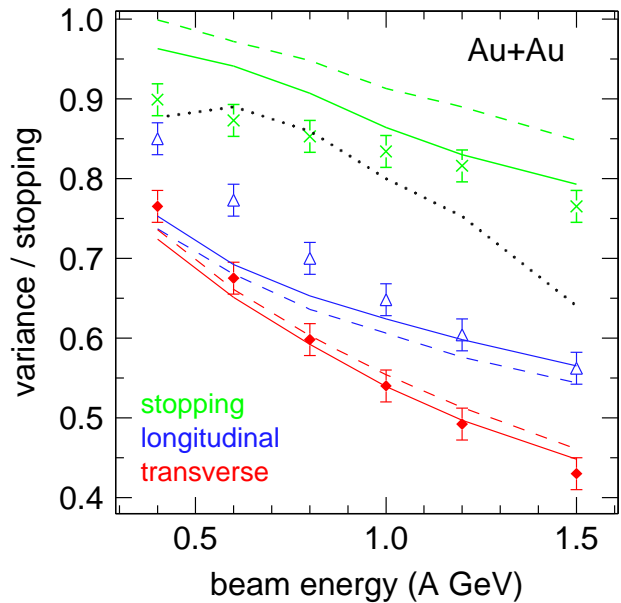,width=100mm}
\end{center}
\caption{
\small{
Various excitation functions for $b_0<0.15$ collisions of Au+Au.
The symbols represent from top to bottom the stopping observable
{\em vartl}, the variance of the longitudinal rapidity, the variance
of the transverse rapidity.
The corresponding observables from the IQMD simulation are represented by
solid (dashed) lines for SM (HM). 
The dotted curve is the measured \cite{reisdorf04a} stopping for nucleons.
}}
\label{vary}
\end{figure}

A more quantitative assessment of these features can be made in terms of
longitudinal, $dN/dy_{0z}$, and transverse, $dN/dy_{0x}$, rapidity distributions
deduced from the reconstructed $4\pi$ data.
Due to apparatus limitations, and in order to keep the technical definition of
the scaled variances of these distributions strictly constant over the full
range of system-energies and centralities, we define the variances 
$\sigma^2(y_{0z})$, resp. $\sigma^2(y_{0x})$, of these distributions in the
finite interval $|y_0|<1.8$ for pions.
Further we define the ratio $vartl=\sigma^2(y_{0x})/\sigma^2(y_{0z})$
which we shall loosely call 'degree of stopping' or just 'stopping'.
In a non-relativistic purely 'thermal' interpretation this ratio would
represent the ratio of transverse to longitudinal 'temperatures',
which, if different from one, would then imply non-equilibrium.
In Fig.~\ref{vary} we show measured excitation functions for 
$\sigma^2(y_{0z})$, $\sigma^2(y_{y0x})$ and of $vartl$ (symbols indicated
in the figure).

The scaled
variances are seen to decrease significantly and steadily with incident
energy, as expected from the qualitative discussion of Fig.~\ref{uty}.
The 'pionic stopping' is consistently below one, but decreases less rapidly
with energy.
We also include from our earlier work~\cite{reisdorf04a} 
the 'nucleonic stopping' (dotted) for
comparison: it decreases faster with energy than its pionic counterpart.
Some of the difference between pions and nucleons results from the fact that
the observed pionic momenta result from a convolution of the excited
nucleon momenta with the decay kinematics, a convolution that is expected
to lead to a more homogeneous (isotropic) final momentum distribution.
A more subtle effect would be
that different hadrons witness {\em on the average} different
collision histories in a non-equilibrium situation.

These effects should be taken care of in a microscopic simulation.
The solid lines in Fig.~\ref{vary} are the prediction of IQMD SM
which follow the measured transverse variances amazingly well, but
show a flatter trend for the longitudinal variances coming closer to the data
at the highest energy and underestimating the measured values at the
low energy end, confirming again
trends  seen qualitatively in Fig.~\ref{uty}.
As a consequence the 'stopping' is also overestimated especially at the low
energy end.
The mean field (SM versus HM) has a modest, but not negligible, influence: the
stiffer EOS leads to a general increase of the stopping (uppermost dashed line)
by roughly $5\%$ caused by the correlated decrease/increase of the
longitudinal/transverse variance.

\begin{figure}[htb]
\begin{center}
\epsfig{file=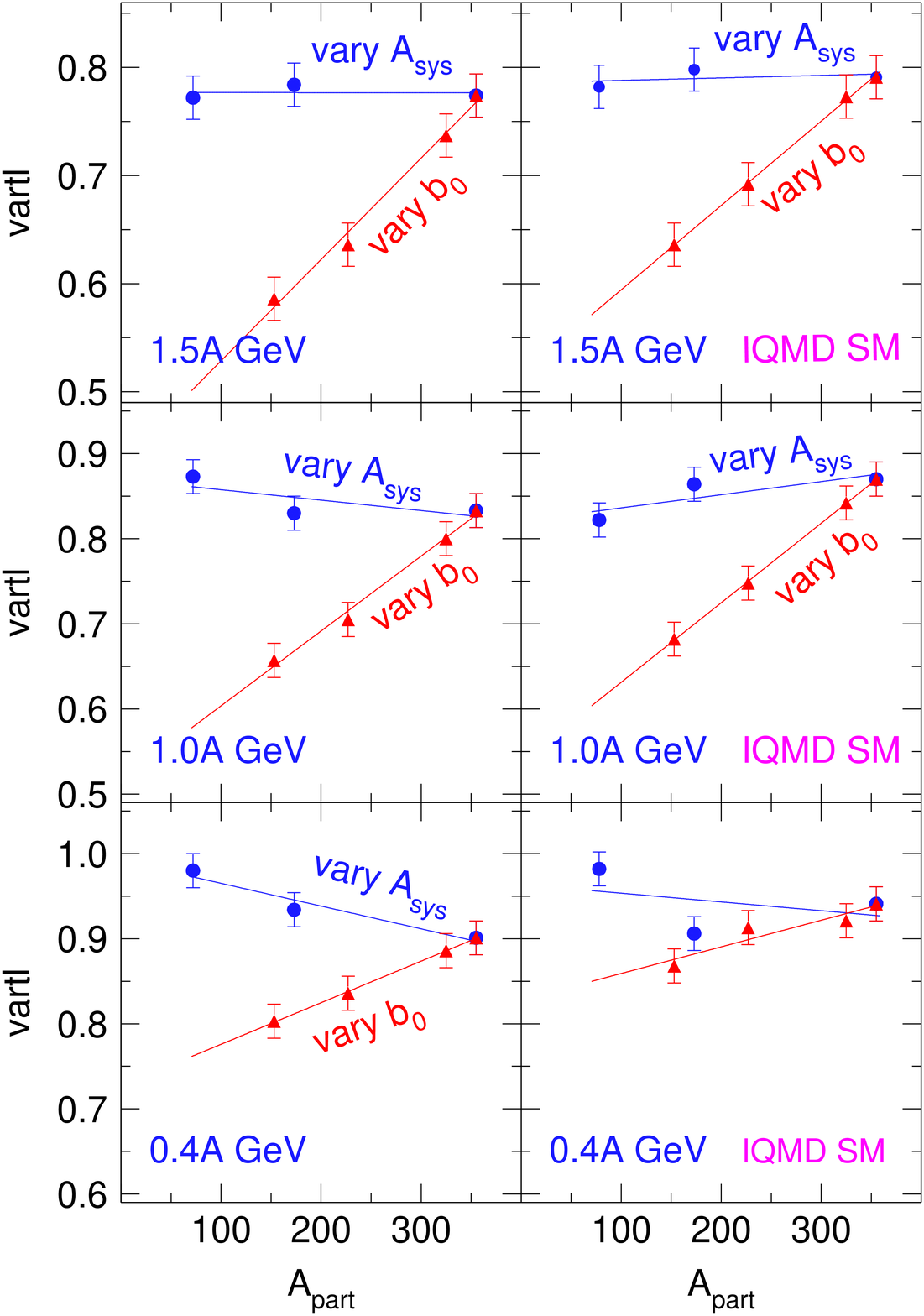,width=100mm}
\end{center}
\caption{
\small{
Pion stopping, {\em vartl}, as function of the number of participant nucleons,
$A_{part}$ for collisions at $1.5A$, $1.0A$ and $0.4A$ GeV (top to bottom).
Left panels: data, right panels: IQMD simulation.
The two branches correspond to two modes of varying $A_{part}$:
1) central ($b_0<0.15$) collisions varying the total system size, marked 'vary
$A_{sys}$', and 2) collisions of Au+Au varying centrality, marked 'vary $b_0$'.
}}
\label{vartlapart}
\end{figure}

Naively, if stopping is incomplete, the degree of stopping is expected to
depend on the system's size characterized by the total number of nucleons,
$A_{sys}$.
Our data comprise the systems Ca+Ca ($A_{sys}=80$), Ru+Ru ($A_{sys}=192$)
and Au+Au ($A_{sys}=394$) where we roughly double the nucleonic size
from one system to the next.
In order to compare, we keep the centrality in {\em scaled} impact parameters
constant ($b_0<0.15$).
In terms of 'participants' one can also vary the (participant) system
size $A_{part}$, by changing the impact parameter $b_0$, keeping $A_{sys}$
constant.
(In order not to interrupt the flow of arguments here, we refer the reader to
section~\ref{production} for a definition and discussion of $A_{part}$.)
With the concept of participant size, we are then in a position to establish
with our data a summary of the pion stopping systematics, based on
the two methods just discussed, see Fig.~\ref{vartlapart} and its caption.
In the figure we have plotted averages for both polarities, $\pi^+$ and
$\pi^-$, as no significant difference was observed.

The main message of this systematics
(which is reproduced semiquantitatively by the simulation)
is that there is no unique dependence on
$A_{part}$, but also a dependence on either the shape of the participant volume
(which depends on the collision geometry, hence $b_0$), or
on the presence of more or less 'spectator' matter in the
early stages of the reaction, an interpretation we favour in the present
energy regime.
The idea is that pions, being fast moving {\em light} hadrons (notice in
Fig.~\ref{uty} some pions are seen to freeze out with 4-5 times the incident
nucleon velocities $u_p$) are able to penetrate into the spectator matter
before the latter has left the neighbourhood of the participant zone.
These pions are then rescattered  experiencing a partial
longitudinal reacceleration.
This results in a smaller apparent stopping, an effect that increases with the
size of the spectator matter, as suggested by Fig.~\ref{vary}.
These pion (or $\Delta$ baryon) rescatterings in spectator matter
also influence the
anisotropies of azimuthal emissions~\cite{bass93,bass93a} to be discussed later.
\newpage
\section{Anisotropies} \label{aniso}

In thermal model analyses of pion data (see for example \cite{averbeck03})
it is often assumed that pion emission is isotropic in the c.o.m. in order to
extend to $4\pi$ the measured mid-rapidity data.
However, already early studies~\cite{wolf79,nagamiya81,mayer93}
of inclusive reactions have reported deviations from isotropy:
usually the polar angle distributions have a minimum near $90^{\circ}$ c.o.m..
More recently, the TAPS collaboration has
observed~\cite{mayer93,holzmann96,tyminska06} anisotropic  $\pi^0$ emission
in asymmetric heavy ion systems at subthreshold energies and interpreted
the inclusive data in terms of a 'primordial' pion emission followed by
final state interactions (rescattering and absorption).
The averaging over impact parameters and the presence of asymmetrically
distributed  spectator matter in the chosen reactions complicate 
the microscopic interpretation of such data. 
In the present work,
the fact that the stopping observable $vartl$ (see previous section) was found
to be less than one,  indicates that isotropy is not fulfilled even in
the most exclusive central collisions, despite minimal amounts of 
'shadowing' spectator matter.

\begin{figure}[h]
\begin{minipage}[b]{75mm}
\hspace{-10mm}
\epsfig{file=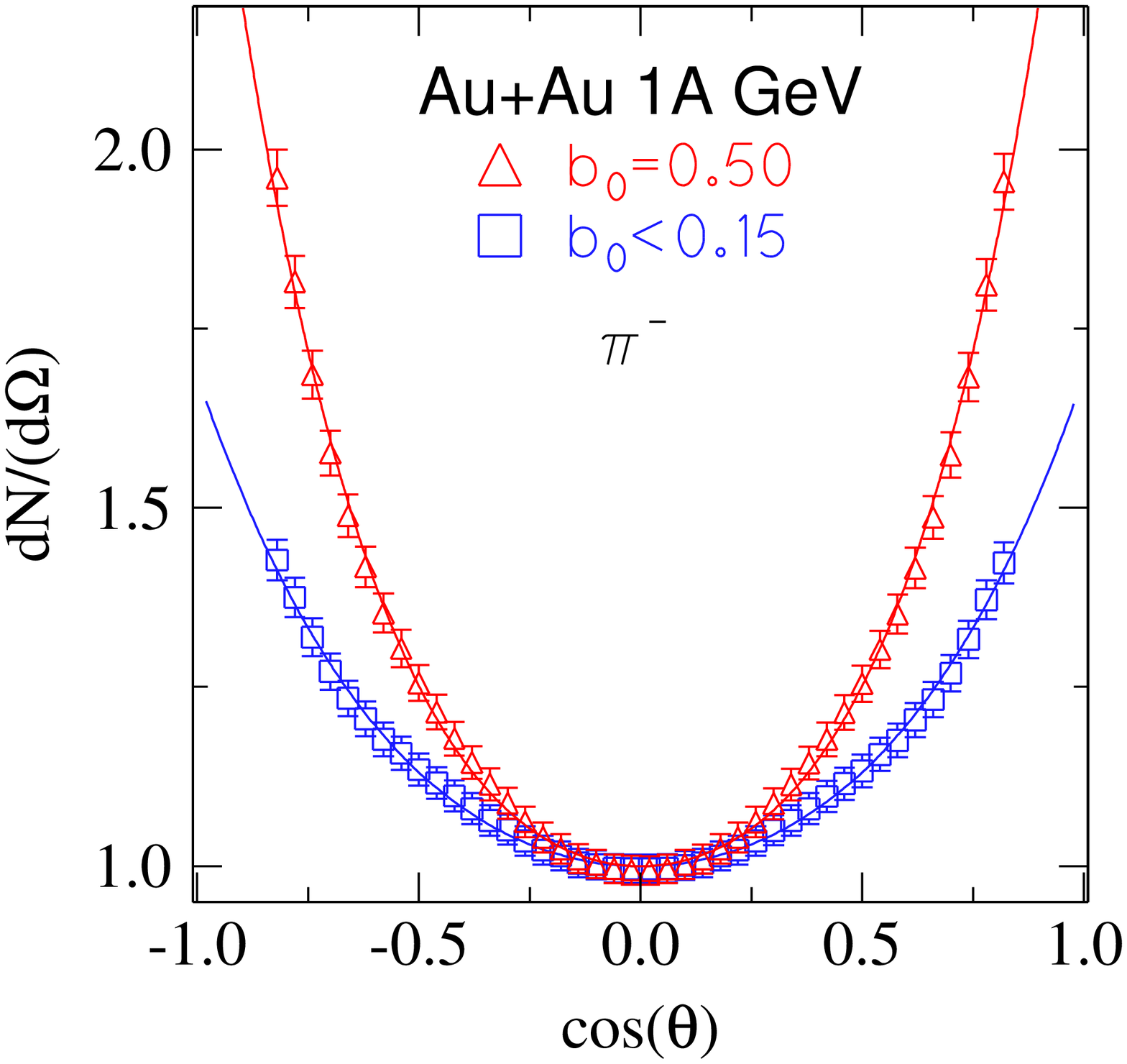,width=77mm}
\end{minipage}
\hspace{-10mm}
\begin{minipage}[b]{75mm}
\epsfig{file=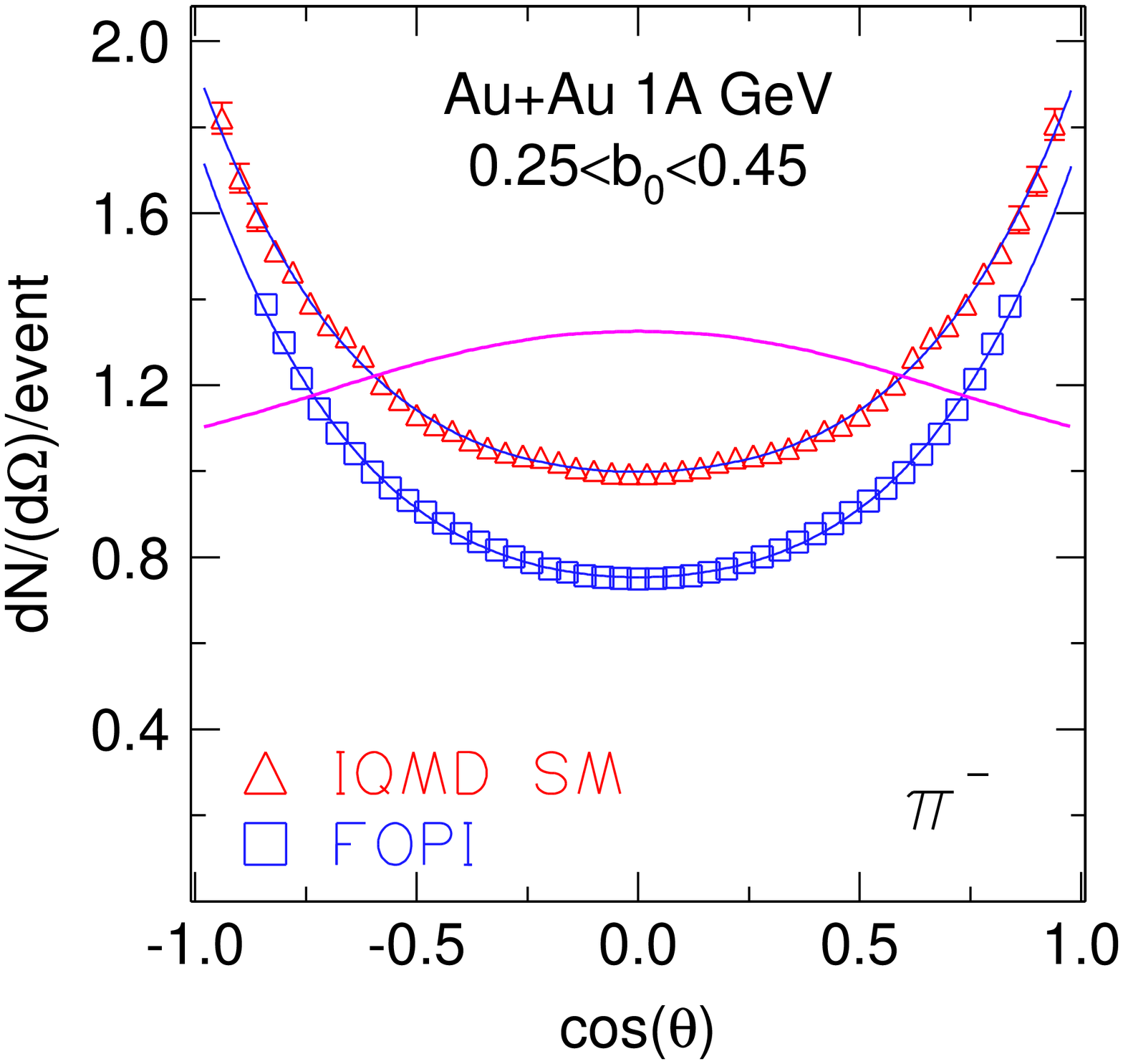,width=77mm}
\end{minipage}
\caption{
\small{
Polar (c.o.m.) angle distributions of $\pi^-$ mesons in the reaction Au+Au at
$1A$ GeV.
The solid lines are least squares fits of the three-parameter function
$a_0+a_2 \cos ^2  \theta +a_4 \cos ^4 \theta $.
Left panel: Experimental data.
Open squares (blue): $b_0 < 0.15$, triangles (red): $b_0=0.50\pm0.05$.
The data have been normalized to 1 at $90^{\circ}$ to emphasize the differences
in shape.
Right panel: Comparison with IQMD for $0.25<b_0<.45$. The ratio IQMD/FOPI
is also shown (Thick solid curve).}
}
\label{dndo}
\end{figure}

To quantify anisotropy it is useful to pass from the coordinates
$(y_z,p_t,\phi)$ to spherical coordinates $(p,\theta,\phi)$,
where $\theta$ is the polar and $\phi$ the azimuthal angle.
This coordinate change does not introduce new information, of course, but
allows to define observables that show deviations from isotropy more
directly and more sensitively, especially when $vartl$ is only slightly below 1.
The distributions $dN/du_{0t}dy_0$ such as those shown in Fig.~\ref{uty}
were transformed to spherical coordinates duly taking into account the
Jacobian.
Fig.~\ref{dndo} shows the projection onto the polar angle axis for Au+Au at
$1A$ GeV for various centralities.

We define as 'anisotropy factor' $A_f$ the ratio
$$ A_f = \int^{+1}_{-1}f(x)dx / 2 a_0$$
where
$$ f(\cos \theta) = a_0 + a_2 \cos^2 \theta + a_4 \cos^4 \theta$$
is least squares fitted to the data.
$A_f$ is the factor with which a measurement restricted to $90^{\circ}$
(c.o.m.) or to mid-rapidity, has to be multiplied to
correct the measured yields for deviations from isotropy.
For the cases shown in Fig.~\ref{dndo} we find $A_f=1.20\pm 0.02$ (for $b_0 <
0.15$) and $A_f = 1.44 \pm 0.04$ (for $b_0=0.5\pm 0.05$).

In terms of our $A_f$ the anisotropy measured~\cite{brockmann84} for 
central collisions
$^{40}$Ar + KCl at $1.8A$ GeV , $A_f=1.19$, compares well
with  our values of $1.21 \pm 0.03$ and $1.22 \pm 0.03$ for $6\%$
central collisions of $^{40}$Ca + $^{40}$Ca at 1.5 and $1.93A$ GeV,
respectively.

The polar angle anisotropies cannot be explained in the framework
of thermal models.
The simulation with IQMD, however, does a reasonable job for
this observable, aside from the absolute normalization to be
discussed later (see right panel of Fig.~\ref{dndo}).
High polar angle anisotropies suggest that a significant fraction of the
{\em observed} pions must be produced and emitted close to the system's
surface where they can escape without rescattering.
One could conjecture that some, potentially more thermalized,
pions (or nucleonic resonances)  created deeper inside
the system did not reach the detectors because of in-medium absorption.

\begin{figure}[!t]
\begin{center}
\epsfig{file=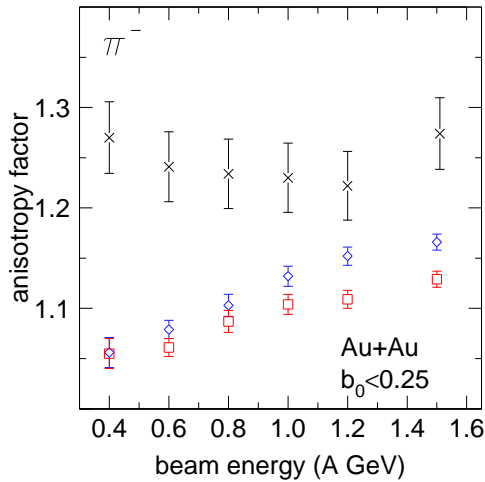,width=80mm}
\end{center}
\caption{
Excitation function of the anisotropy factor for Au+Au collisions with
$b_0<0.25$. The (black) crosses are the measured data (with systematic errors),
(blue) diamonds and (red) squares are predictions from IQMD SM and
HM, respectively (with statistical errors).
}
\label{aniso-e}
\end{figure}

More systematic details reveal differences between the data and the
simulations.
While stopping (Fig.~\ref{vary}) shows a weakly
decreasing trend with increasing energy,
the measured excitation function of $A_f$ looks rather flat (within the
indicated systematic errors), see
Fig.~\ref{aniso-e}, while the IQMD simulations always underestimate $A_f$
and predict a rise with beam energy which is slightly more pronounced with
the softer EOS.

\begin{figure}[!t]
\begin{center}
\epsfig{file=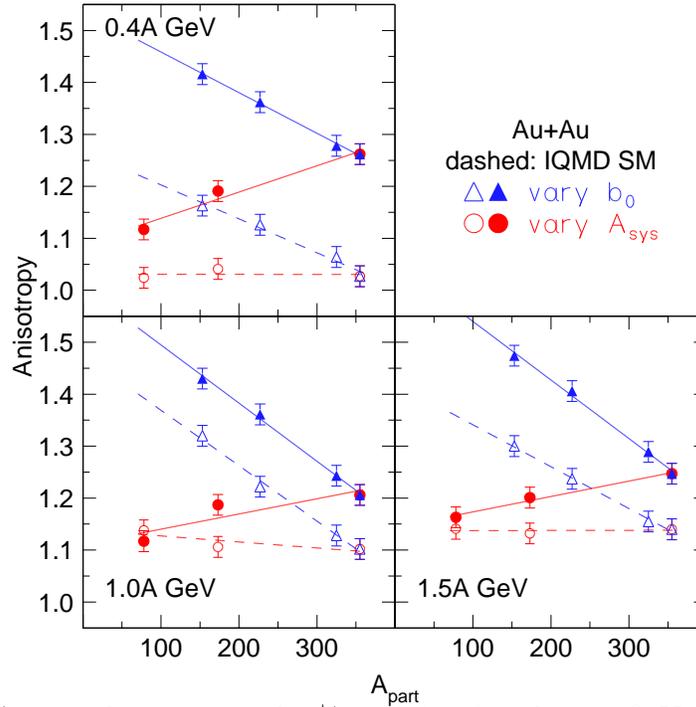,width=110mm}
\end{center}
\vspace{-8mm}
\caption{
\small{
Anisotropy (averaged over $\pi^-$ and $\pi^+$) at 0.4, 1.0 and $1.5A$ GeV
incident energy as a function of $A_{part}$.
Full symbols and solid lines represent the data, open symbols and dashed lines
represent the simulation IQMD SM.
The lines are linear least squares fits added to guide the eye.
As in Fig.~\ref{vartlapart}, there are two branches, one varying $A_{part}$
via centrality $(b_0)$ binning in Au+Au collisions, the other via changes of 
the system size $(A_{sys})$, keeping the centrality constant $(b_0<0.15)$.}
}
\label{anisoapart}
\end{figure}

The observables of anisotropy, $A_f$, and of stopping, $vartl$, are not
equivalent since they quantify different aspects of momentum space population.
However, they are related to the degree that high transparency that would yield
$vartl \ll 1$ should be characterized by $A_f \gg 1$.
The two branches seen in the stopping when varying either the size of the
system or the geometry, Fig.~\ref{vartlapart}, are also seen with the anisotropy
observable, Fig.~\ref{anisoapart}.
IQMD simulations reproduce these features qualitatively.
A somewhat puzzling feature is the fact that the measured data seem to suggest
that the anisotropy grows slightly with $A_{part}$ when the centrality is kept
high ($b_0<0.15$).

%

\begin{figure}[!b]
\begin{center}
\epsfig{file=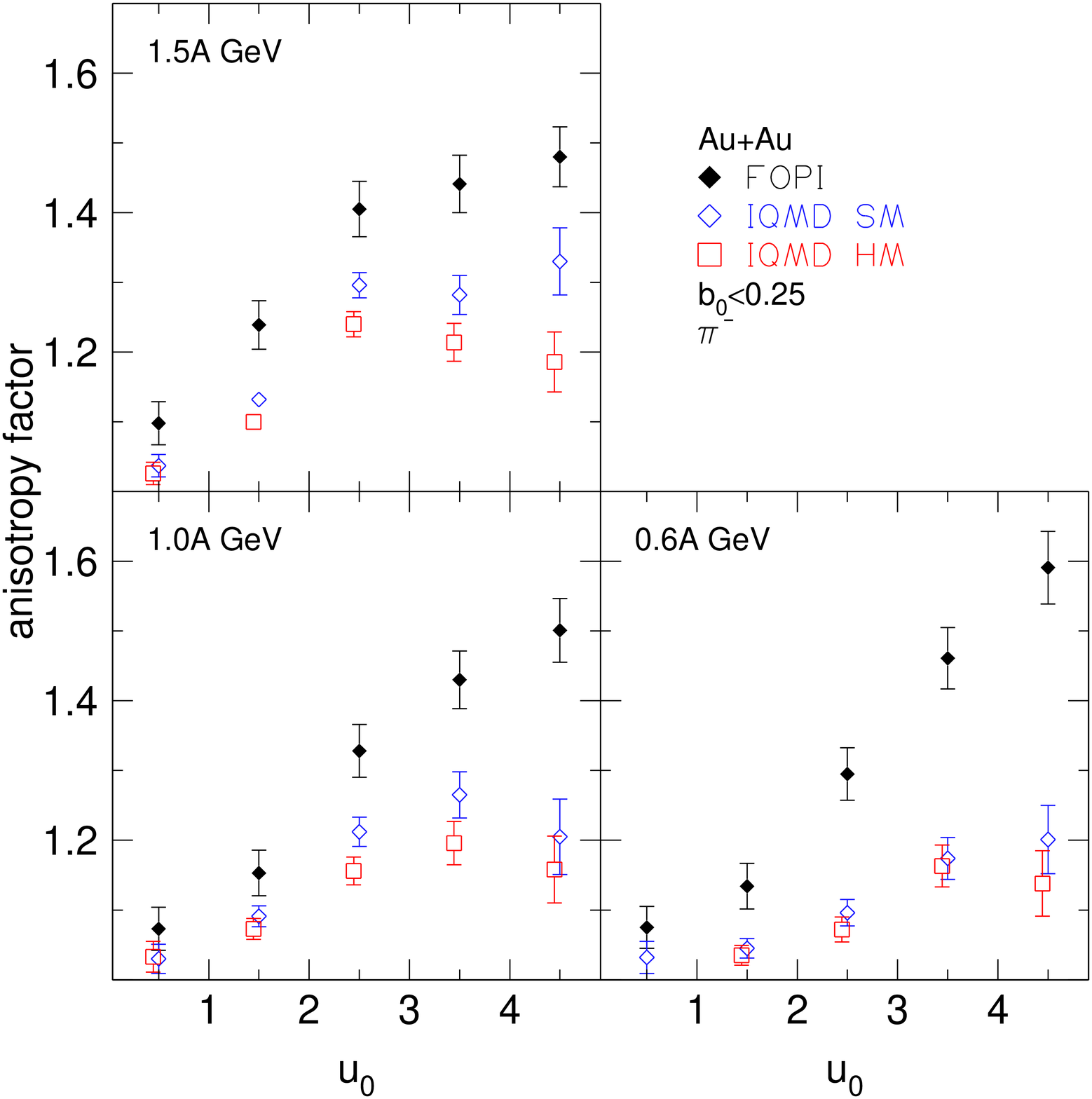,width=100mm}
\end{center}
\vspace{-8mm}
\caption{
\small{
Anisotropy of $\pi^-$ in Au+Au collisions $(b_0<0.25)$ at 1.5, 1.0 and $0.6A$
GeV as a function of the scaled momentum $u_0$.
Full (black) diamonds: data, open (blue) diamonds: IQMD SM, open (red) squares:
IQMD HM.}
}
\label{anisou0}
\end{figure}

In Fig.~\ref{anisou0} we take a more differential look at anisotropy
displaying for $b_0<0.25$ collisions the  dependence on the
scaled momentum $u_0 \equiv (p/m_{\pi})_0$ ($m_{\pi}$ pion mass)
for three different incident energies.
The simulations with IQMD SM and HM are also shown.
Again, IQMD underestimates $A_f$ systematically, shows a weak mean-field 
sensitivity and some tendency to saturate around $u_0 \geq 3$.
Except maybe at $1.5A$ GeV, the data do not show such saturation: $A_f$ rises
approximately linearly within the $u_0$ range shown and converges to zero at
zero momentum. 
This is expected on general grounds and indicates some consistency in our
extrapolation to low $p_t$ which affects primarily $A_f(u_0)$ values
below $u_0=1$.

\begin{figure}[htb]
\begin{minipage}{79mm}
\epsfig{file=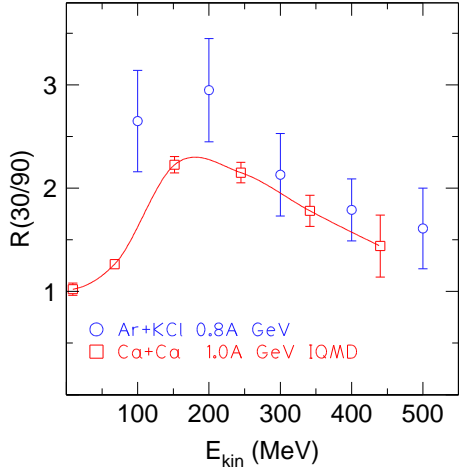,width=78mm}
\end{minipage}
\begin{minipage}{79mm}
\caption{
\small{
Ratio of pion yields at $30^{\circ}$ to that at $90^{\circ}$ for inclusive
reactions. Circles: Ar+KCl at $0.8A$ GeV~\cite{nagamiya81}.
Squares joined by a smooth curve: Ca+Ca at $1A$ GeV using IQMD SM.}
}
\label{nagamiya}
\end{minipage}
\end{figure}

For inclusive data it appears that the momentum (or kinetic energy $E_{kin}$)
dependence of $A_f$ shows a maximum near $u_0=2.8$ (or $E_{kin}$=150 MeV
at $E/A=0.8A$ GeV) as demonstrated in Fig.~\ref{nagamiya} which reproduces the
early data of Nagamiya et al.~\cite{nagamiya81} for Ar+KCl at $0.8A$ GeV.
It is interesting to note that the authors concluded (in 1981) '{\em
these features cannot be explained by any conventional theoretical model}'.
In the figure we have included a calculation with IQMD for a similar
system-energy, Ca+Ca at $1A$ GeV, which is seen to reproduce fairly well the
observed features.
Inclusive data are generally more difficult to interprete in a definite way.
Due to trigger biases used to enhance central collisions and minimize
background reactions~\cite{pelte97au} the present data cannot be directly
compared with inclusive data such as those shown in Fig.~\ref{nagamiya}.

\clearpage
\newpage

\section{Pion production } \label{production}
\subsection{Experimental trends} \label{experiment}
In this section we  present our pion multiplicity data and compare
the results with those of earlier work, as well as with those of 
transport model calculations.

In the earlier literature some authors reported on pion production in terms
of pion multiplicity per 'participant', $M(\pi)/A_{part}$.
The motivation for this choice came  from the observation, probably first
reported in ref.~\cite{fung78}, that in relativistic heavy ion reactions 
this reduced multiplicity did not appear
to depend on the size of the system if the incident energy was kept fixed.
Like impact parameters, 'participants' (or 'spectators') 
are not direct observables, they are not defined rigorously  and are
estimated using  recipes that may vary with the authors.
This limits the level of accuracy with which such 'reduced' data from different
experiments can be compared.
In experiments that measure multiplicities eventwise
there are two ways to determine the number of participants:
first, one defines all nucleons to be participants that have momenta  outside 
the Fermi spheres around target and projectile momenta, and, second,
one calculates from the 'impact parameter' the geometrical size
of a straight-trajectory, sharp-geometry overlap zone~\cite{gosset77}
where the 'impact
parameter' is estimated from some global event observable ($ERAT$, charged
particle multiplicity, etc) as described earlier.
In the present work, using the second method and the observable $ERAT$,
we estimate 
a $(90 \pm 5 \%)$ 'participation' for our most central, $b_0 < 0.15$, sample.

\begin{figure}[htb]
\begin{minipage}{79mm}
\epsfig{file=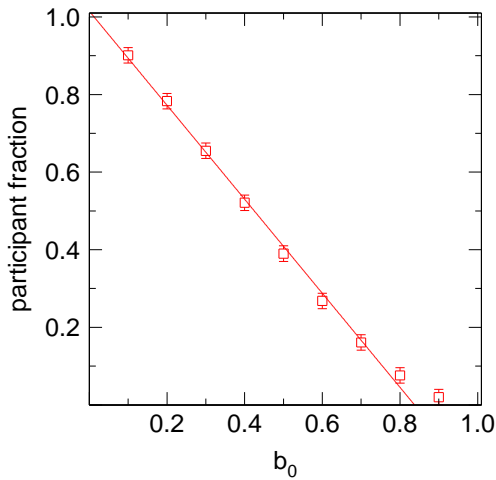,width=78mm}
\end{minipage}
\begin{minipage}{79mm}
\caption{
\small{
Dependence of the participant fraction on the scaled impact parameter 
$b_0$ in the sharp-cut geometrical model.}
}
\label{apart}
\end{minipage}
\end{figure}

Although the calculation of $A_{part}$ with the sharp-cut geometrical
model \cite{gosset77}
is a simple mathematical exercise, we show the result in 
Fig.~\ref{apart} to raise the awareness of an approximate fact:
the participant fraction $A_{part}/A_{sys}$
decreases linearly with
the reduced impact parameter $b_0$ if one confines oneself to $b_0 < 0.8$.
Within this limit, it also does not matter much if the geometrical model allows
for a diffuse~\cite{pelte97au},  rather than a sharp surface.

For inclusive measurements, i.e. pion multiplicity data taken without a
'centrality bias', $A_{part} = 1/4 A_{sys}$ is often assumed.

To our knowledge no single experimental apparatus measures all
three isospin components of the pion.
Assuming that Coulomb effects on pion production are very small,
isospin symmetry implies that $M(\pi^+) + M(\pi^o) + M(\pi^-) =
3*M(\pi^0) = 1.5*[M(\pi^+) + M(\pi^-)]$.
If only $M(\pi^-)$ has been measured, assumptions must be made
to obtain the full pion multiplicity, $M(\pi) = f*M(\pi^-)$.
Using for the following only the data for the most central collisions, 
our excitation functions for the total reduced pion multiplicity are  simply 
excitation functions of the quantity
$[M(\pi^-)+M(\pi^+)]*1.5/(0.9A_{sys})$.

\begin{figure}[htb]
\begin{center}
\epsfig{file=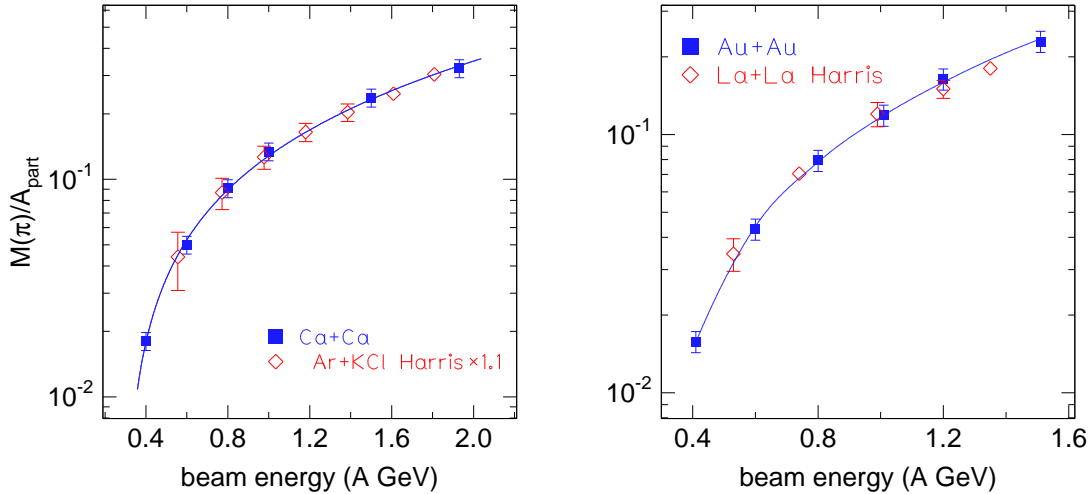,width=158mm}
\end{center}

\caption{
\small{
Excitation functions of the reduced pion multiplicity.
The left panel compares the present data for Ca+Ca with the data (rescaled by a
factor 1.1) of Harris et al.~\cite{harris85} for Ar+KCl.
The right panel compares the present data for Au+Au with the La+La
data~\cite{harris87}.}
}
\label{harris}
\end{figure}

%

The two-panel figure shows our measured pion-multiplicity excitation
functions for the systems Ca+Ca (left panel) and Au+Au.
They are compared with Streamer Chamber data for
Ar+KCl~\cite{sandoval80,stock82,harris85}
and for La+La~\cite{harris87}, respectively. 
The smooth curve (a second order polynomial) has been fitted
to the FOPI data only.
The agreement between the two sets seems to be excellent.

At this point we have to remark that the present new Au on Au data
are in conflict with our earlier publication~\cite{pelte97au} concerning
pion production at $1.06A$ GeV.
A reassessment of our older data has lead to the following conclusion:
due to a too low setting of the potential voltage of the CDC the Chamber was
not fully efficient for low ionizing particles (ref.~\cite{pelte97au}
had reported the very first application of the CDC).

A few further comments on the comparison with the Streamer
Chamber data are useful.     
In the La+La experiment~\cite{harris87} a 384 scintillator hodoscope    
covering angles $\theta_{lab} < 18^o$ was added to the Streamer Chamber
and the projectile-spectator angle window was defined
'to be the region centered about beam-velocity Z=1,2 fragments containing
$90\%$ of the charged particles in minimum bias'.
An efficiency factor, estimated from a cascade code simulation, was applied.
The authors also mention a correction (approximately $14\%$)  to the
pion multiplicities to account for track losses and misidentifications.
Finally, the negative pion multiplicities were multiplied by a factor f=2.35
to account for $\pi^+$ and $\pi^0$ emission.
From our $\pi^-/\pi^+$ systematics we know that the factor depends on the
incident
energy (we estimate f=2.06 at .4A GeV and f=2.43 at 1.5A GeV for La on La).

Concerning Ar+KCl,
the number of 'participating protons' and the $\pi ^-$
multiplicities are conveniently given in a Table~\cite{sandoval80}. 
Fragments with projectile velocity in a $4^o$ forward cone and
positive tracks (around target rapidity) with laboratory momenta
$p_{lab} < 200$ MeV/c were counted as 'spectators'.
All 'participant' tracks were assumed to be singly charged.
This is a good approximation  at the highest energy.
From the geometrical model we estimate a participant total
charge of 28 (for a 180mb trigger) in reasonable agreement with
the tabulated values~\cite{sandoval80} for beam energies above 1A GeV.
The system Ar+KCl $(A_{sys}=77.25$, $Z_{sys}=36$ and
 hence $N_{sys}/Z_{sys}=1.146$) is not
a strictly isospin-symmetric system. Using the simple-minded 'isobar'
formula~\cite{stock86} one obtains $M(\pi^-)/M(\pi^+) = 1.255$, i.e. a number
significantly different from 1 and implying f=2.70. 
Again, using our energy dependent systematics, we get factors f=2.55 at
0.4A GeV and f=2.74 at 1.5A GeV, rather than f=3 mentioned in~\cite{harris85}.
From the tabulated 'proton participants' and $M(\pi^-)$ in~\cite{sandoval80}
we get $M(\pi)/A_{part}$ values very close to those shown in~\cite{harris85}
if we apply the isobar model f factor to the $\pi^-$ multiplicities and
the factor $A_{sys}/Z_{sys}$ to the proton participant multiplicities.
In our figure we show that we get perfect agreement with our data if
we boost the Harris data by $10\%$.
No correction for the Streamer Chamber response is mentioned in
the three publications~\cite{sandoval80,stock82,harris85} which appear to
pertain to the same experiment. 

At the GSI/SIS accelerator site
pion data have also been obtained using two other experimental setups.
The pion multiplicity data of the TAPS collaboration have been
summarized in \cite{averbeck03}. 
Some (unpublished) data (for E/A=1GeV) from the thesis work of
Wagner~\cite{wagner96}
measured with the KaoS setup, are listed in
a Table of ref.~\cite{senger99}.

\begin{figure}[!h]
\begin{center}
\epsfig{file=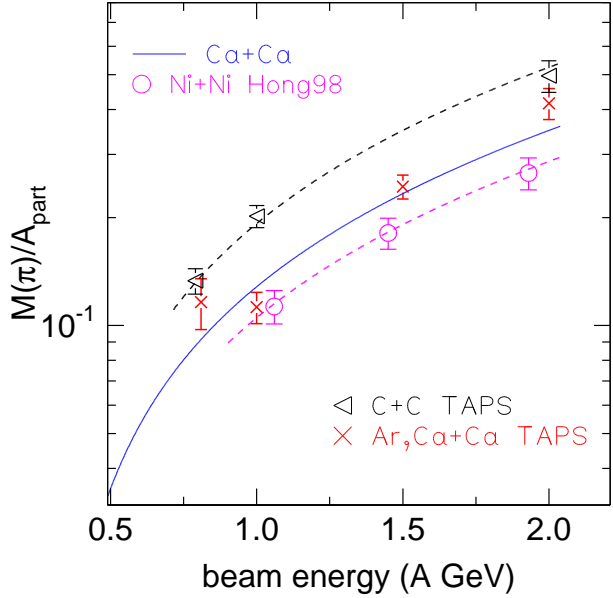,width=100mm}
\end{center}
\caption{
{\small
Excitation functions for the reduced pion multiplicities in reactions of
Ca+Ca (smooth solid line identical to the one shown in Fig.~\ref{harris}),
Ar+Ca and Ca+Ca \cite{averbeck03}, Ni+Ni \cite {hong97}, and C+C
\cite{averbeck03}. The respective symbols are indicated in the figure,
for further comments see text. }
}
\label{taps}
\end{figure}

A common feature of these experiments is that they are triggering
on the particle(s) of interest (charged pions in the KaoS case and
two-photon candidates for a $\pi^0$ decay in the TAPS case), within
an angular range that is substantially limited if compared to the
large acceptance devices (but can be varied by moving the apparatus).
As a consequence they are not measuring pion multiplicities event by event,
but rather differential cross sections in a restricted part of momentum space,
preferably 'around mid-rapidity', under more or less exclusive conditions.
To convert the, apparatus-response corrected, data to reduced $4\pi$ pion
multiplicities one must divide by a nuclear reaction cross section, $\sigma_r$
which in reference \cite{averbeck03} was taken to be 
$\sigma_r = \pi r_0^2 (A_p^{1/3} + A_t^{1/3})^2$ with $r_0 = 1.14$fm. 
Further, an extrapolation to  phase-space outside mid-rapidity has to be
made and possible biases in the triggering modes have to be
corrected for if inclusive data are wanted. 
One asset of the TAPS method is that momenta are measured all the way down to
zero.
In  Fig.~\ref{taps} we compare the present Ca+Ca data and some of our earlier
data~\cite{hong97}, \cite{hong98} with the inclusive
TAPS data for neutral pions (multiplied by f=3) \cite{averbeck03} for the
systems Ar+Ca, Ca+Ca, recalling that the FOPI data are high centrality data.   
For completeness we also mention here TAPS data taken at much lower energies
(Ar+Ca at $0.18A$ GeV \cite{martinez99}) well outside the range of the figure.

The published mid-rapidity TAPS data have been obtained under the assumption
of isotropic emission. 
Using our anisotropy information (previous section) we have boosted these
data by a factor 1.25. If this is done, a satisfactory agreement with the
FOPI data (and the old Streamer Chamber data) is obtained, the TAPS data
having somewhat more straggling around the smooth trend inferred from our data
(solid curve). 

The figure also shows inclusive data for C+C, \cite{averbeck97},
again, corrected
for anisotropy, and for central collisions of Ni+Ni~\cite{hong97,hong98}.
The two dashed curves represent the Ca+Ca curve (solid) rescaled by factors
0.82 and 1.58.
Clearly, there is some system size dependence of the {\it reduced} pion
multiplicity.
We note that the C+C system corresponds to $A_{part} = 6$ (inclusive data),
the Ar+Ca (TAPS) data to $A_{part} = 20$ and the FOPI data to
$A_{part}=72$ and $100$ for Ca+Ca and Ni+Ni, respectively.

\begin{figure}
\begin{center}
\epsfig{file=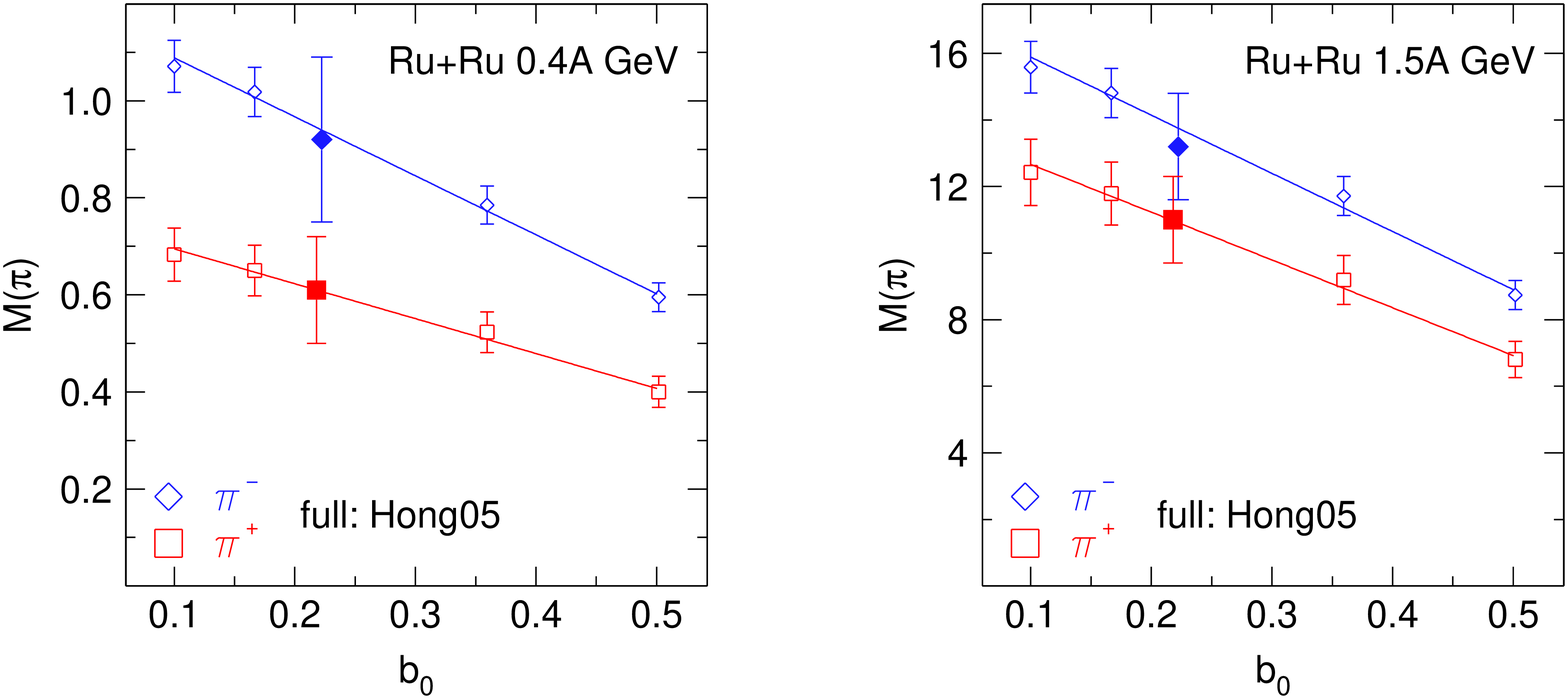,width=150mm}
\end{center}
\caption{
{\small
Multiplicities of $\pi^+$ and $\pi^-$ mesons in $^{96}$Ru + $^{96}$Ru
reactions at $0.4A$ (left) and $1.5A$ GeV versus the reduced impact parameter
$b_0$.
The data points marked with larger full symbols are from an earlier
publication~\cite{hong05}.}
}
\label{ruru}
\end{figure}

Participant size dependences can also be shown by varying the 
collision geometry keeping the system fixed, as demonstrated in
Fig.~\ref{ruru}.
The straight lines are linear least squares fits demonstrating (within error
bars) the approximate linear dependence on the (reduced) impact parameter, which
implies, as shown in Fig.~\ref{apart}, an approximate linearity also in terms
of $A_{part}$. 
This figure also contains data from an earlier publication~\cite{hong05}
of our Collaboration which used the alternate analysis method 
('local tracking') briefly described in subsection~\ref{IQMD}.

\begin{figure}
\begin{center}
\epsfig{file=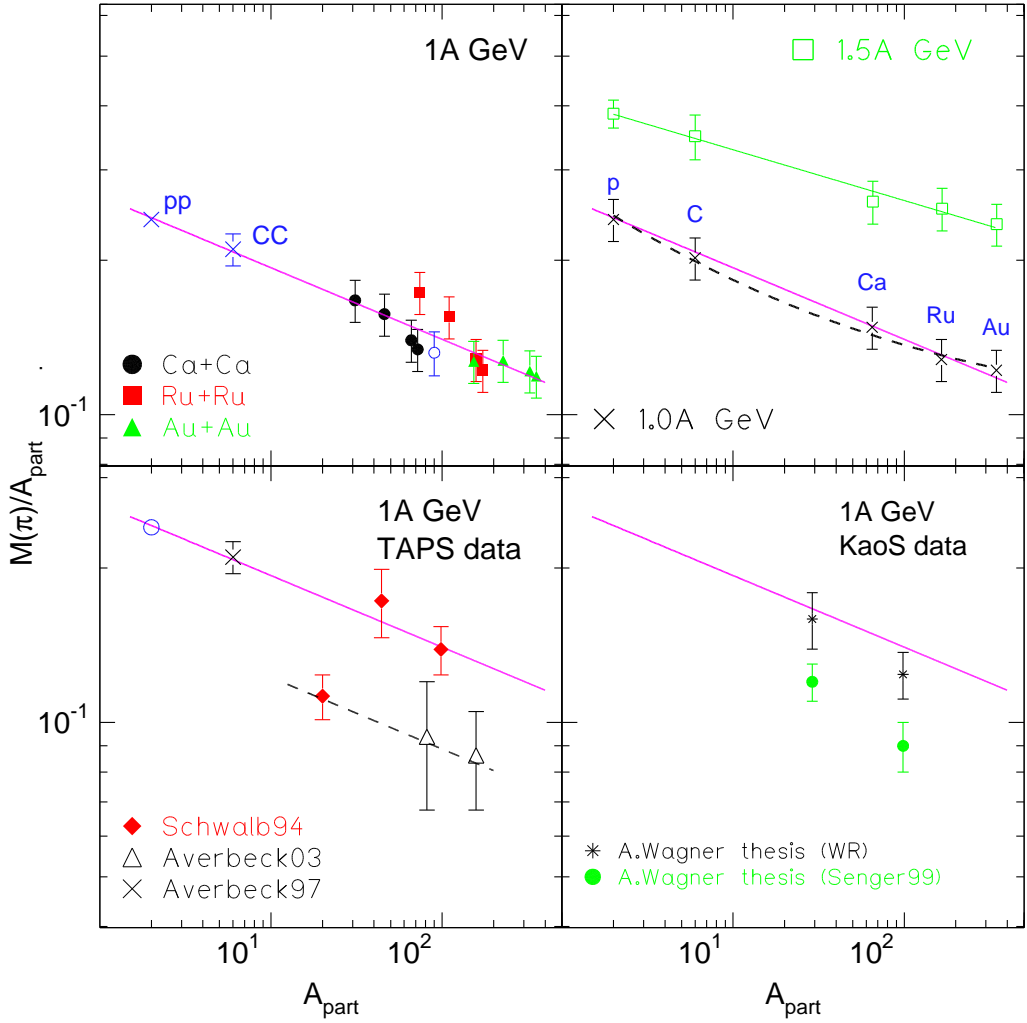,width=155mm}
\end{center}
\caption{ 
{\small
Reduced pion multiplicities and power law fits.
The data points marked pp (or p) and CC (or C) are from refs.
\cite{gazdzicki95} and \cite{averbeck97}, respectively.
The data in the left lower panel are TAPS data
\cite{averbeck97,averbeck03,schwalb94}.
KaoS data \cite{wagner96,senger99} are shown in the
right lower panel. See text for further details.}
}
\label{powerlaw}
\end{figure}

To obtain a quantitative evaluation of the $A_{part}$ dependence
we confine ourselves first to
the present data, thus avoiding systematic differences between various
experiments and hopefully profiting from the smaller point-to-point errors.
In the upper left panel of Fig.~\ref{powerlaw} 
we show the participant-size dependence of
 $M_{\pi}/A_{part}$ at an incident beam energy of 1A GeV.

It is of some interest to try to join up such data all the way down 
to the nucleon-nucleon system $A_{sys}=2$.
This raises the question~\cite{senger99} how to define the number of
'participants' in such a reaction.
Specifically for the $pp$ reaction, we follow ref.~\cite{senger99} and use
$(\sigma_{in}/\sigma_{tot}) [M(\pi^+)+M(\pi^-)]*1.5/2 $
where the multiplicities are obtained from the compilation (Table 4) of
\cite{gazdzicki95} and the ratio of inelastic to total cross section
is set at 0.4 (for a 1A GeV incident beam).
The resulting data point is marked $pp$ in the figure and plotted
at $A_{part}=2$, simply because 2 is the minimum number 
of 'participants' for any reaction.
The second 'non-FOPI' point, marked CC, has been determined by the TAPS
collaboration~\cite{averbeck97} for inclusive reactions ($A_{part}=6$) and
represents $1.25*3*M(\pi^0)$, the factor 1.25 accounting conservatively for
the (unknown) polar angle anisotropy, as discussed before. 
All other data are from the present work for various centralities
plus a Ni+Ni point after~\cite{pelte97ni} (for $A_{part}=90$, Fig.1).
In the Figure the  solid line is a power law fit,
 $M(\pi)/A_{part} =  c*A_{par}t^{\tau-1}$, 
{\it using   only the FOPI data} (i.e. $A_{part} > 20$).
We find $\tau = 0.86\pm 0.04$, significantly different from one.
(If it was one, the plotted {\em reduced} multiplicity would be constant.)
This fit, somewhat surprisingly, is nicely compatible with both the
$pp$ and the $CC$ data. Including all the data in the fit lowers the
uncertainty of the $\tau$ parameter  by a factor 2.

The power law curve is shown again in the other three panels.
In the upper right panel of Fig.~\ref{powerlaw} we show only the data points
for the most central collisions (except for the $pp$ and $CC$ data).
The power law fit made to the more extensive set of data in the upper left
panel is also optimal for this more limited set of data.
The advantage of a limitation to very central collisions is that one is less
dependent on the definition of $A_{part}$ since $A_{part} \approx A_{sys}$.
At the higher energy, $1.5A$ GeV, the same type of data,
also shown in the figure,  are again fully
compatible with an $A^{\tau}$ dependence, although $\tau=0.90\pm0.02$ is
somewhat closer to a pure 'volume' dependence.
In deriving power law descriptions of system-size dependences one hopes
to distinguish between 'volume' and 'surface' effects in a way reminiscent of
mass formulae, although the relative accuracy of mass measurements is multifold
better (and therefore the results more convincing).
Fitting the data with $c_v A + c_s A^{2/3}$ instead of $c A^{\tau}$,
i.e. again with a two-parameter Ansatz, one can describe the 
$1A$ GeV data equally well
as shown in the figure (dashed curve).
One finds $c_v=0.096\pm .011$ and $c_s=0.186\pm 0.033$.
With this description one can tentatively interprete the size dependence
 in terms of a decreasing surface to volume ratio: the fitted
coefficients suggest a $52\%$ surface effect for C+C decreasing to $21\%$
for Au+Au (both for $b_0<0.15$ collisions).

In principle, one could also add an isospin term, again following
mass formulae. The only unambiguous information in our data
comes from a comparison of $^{96}$Ru + $^{96}$Ru with $^{96}$Zr + $^{96}$Zr.
However, for the {\em total} multiplicity of pions (adding both charges)
we found no significant difference within errors.

From the smoothened trends described by the power law fits we can deduce that
the reduced pion yields in the reaction Au+Au around $1A$ GeV are lower by a
factor of about 0.85 with respect to the Ni+Ni system.
In Fig.~2 and Table~3 of ref.~\cite{pelte97ni} we had concluded 
that this factor was about 0.53. 
As mentioned earlier, the partially inadequate operation of the CDC in the
experiment leading to the results published in ref.~\cite{pelte97au}
forces us to retract this number and most of the quantitative aspects of the
conclusions in this early publication.

The TAPS data for 1A GeV~\cite{schwalb94,averbeck97,averbeck03} are
shown in the lower left panel together with the $pp$ point. 
We have applied a factor 1.25 to these mid-rapidity data
to  correct them for the polar anisotropies
discussed earlier.
The data from the earlier publication~\cite{schwalb94}, (red) full diamonds,
for the systems Ar+Ca, Kr+Zr and Au+Au, have been later
revised~\cite{averbeck03}, black open triangles, to account for a trigger 'bias
towards centrality'.
However, the Ar+Ca point was not changed.
These data were obtained in coincidence with the Forward Wall of FOPI.  
The C+C point was obtained in a different setup not involving the
FOPI apparatus.
As can be seen from the figure, after the revisal, the
three TAPS points (triangles) appear to be aligned, but are significantly below 
the solid power law line
(the dashed curve is our power law curve down-scaled
by a factor 1.58).
Note also that
if the correction is due to a centrality bias the revisal also leads to
a shift along the $A_{part}$ axis.
Trying to join up the three revised points to the $CC$ and $pp$ data,
one has to introduce two kinks in the overall TAPS curve.

Pion multiplicity data from the KaoS Collaboration are cited in~\cite{senger99}
and taken from an unpublished part of the thesis work of
Wagner~\cite{wagner96}.
The two data points for 1A GeV incident beams, cited as being inclusive,
for Ni+Ni, resp. Au+Au, are shown in the lower left panel as taken from
Table 2 of ref.~\cite{senger99}.
Also shown in the right lower panel of Fig.~\ref{powerlaw}
is a reassessment of these data points: first, a reaction cross
section in line with our geometrical scaling ($r_0 = 1.15$) and the TAPS
procedure ($r_0 = 1.14$) and, second, the application, again, of the
anisotropy factor $A_f = 1.25$.
When this is done, the Ni+Ni point completely, and the Au+Au point marginally,
agree with our power law fit.



It is of some interest to mention here that $\pi$-nucleus reaction data from the
meson factories~\cite{lee02} have also been described in terms of power laws.
In particular, 'true' absorption~\cite{ashery81} is characterized by
$\tau = 0.75\pm0.05$ for (200-400) MeV/c pion momenta.
However, the initial states of the 'inverse' (absorption) reaction cannot be
compared simply with the final states in the heavy ion induced pion emission.
Indeed, in the latter case, a 'bulk' or volume term for a strongly interacting
particle like the pion (or its parent $\Delta$ baryon)
can only be understood if one
invokes a {\em collective} expansion mechanism allowing pion absorption to
stop or 'freeze out' {\em globally} at some time in the evolution of the
reaction. This mechanism, presumably, is absent in $\pi$-nucleus reactions.

\begin{figure}[!h]
\begin{minipage}{78mm}
\epsfig{file=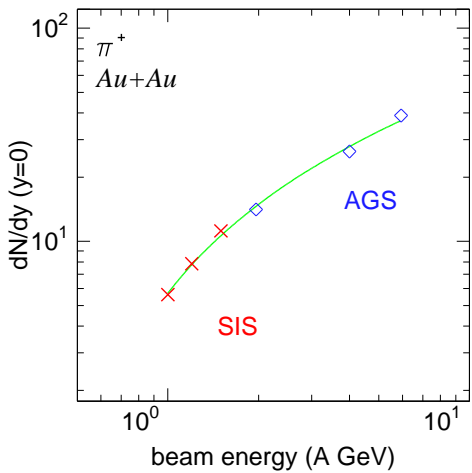,width=78mm}
\end{minipage}
\begin{minipage}{78mm}
\caption{
{\small                                                     
Excitation function for mid-rapidity $\pi^+$ mesons in central Au+Au
collisions.
The three data points at the higher energy are from ref.~\cite{ahle00}.
The data are for a centrality of $5\%$ (in terms of the reaction cross section)
within a rapidity interval $|y_0|<0.25$.}
}
\label{AGS}
\end{minipage}
\end{figure}

The data of our present work for central Au+Au reactions join up smoothly to
AGS data~\cite{ahle00} if taken under the same exclusive conditions.
This is shown in Fig.~\ref{AGS} for $\pi^+$ mid-rapidity data.           
Included in the figure is a least squares fitted second order polynomial,
in terms of $\ln (E/A)$, that reproduces the six data points with an
average accuracy of $6\%$, close to the point-to-point systematic errors of
both collaborations.
We conclude that there is no significant 'kink' in this excitation function, as
one might have expected to observe if the gradual passage to 'resonance' matter
(i.e. with a significant degree of nucleonic excitations) were to be associated
with a fast increase of entropy.
\subsection{Comparison with transport models} \label{transport}
The old
Ar+KCl and La+La data obtained at the BEVALAC accelerator
with a Streamer Chamber~\cite{harris85,harris87}
seemed to indicate within error bars that
the multiplicity per participant was independent
on the size of the fireball at a fixed incident energy.
The authors~\cite{harris87} concluded that
'{\em pion production is a bulk nuclear-matter probe
rather than a surface probe}' and '{\em is unaffected by the expansion phase}'.
As a consequence they deduced by comparison with thermal model expectations
that the nuclear EOS ('missing energy') '{\em is extremely stiff}'.
Probably the first publication putting into question the sensitivity of
pion multiplicities to the EOS was ref.~\cite{bertsch84} in which a transport
code based on the Boltzmann equation and including the mean field and
Pauli blocking effects was used.
Later, Kruse et al.~\cite{kruse85} using a different code confirmed that they
could not determine the EOS from the data of Harris 
et al.~\cite{harris85,harris87}, 
since they found only a $10\%$ effect when changing from a stiff to
a soft EOS.
In 1986 Kitazoe et al.~\cite{kitazoe86} reported that they were actually able
to reproduce the Harris data '{\em without introducing the nuclear compression
effect}'.

\begin{figure}
\begin{minipage}{158mm}
\epsfig{file=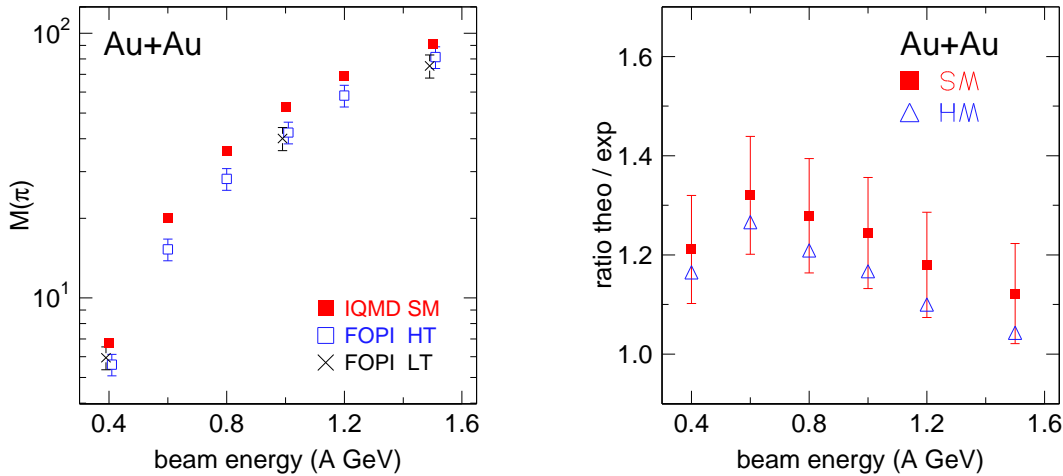,width=158mm}
\end{minipage}
\caption{
{\small
Left panel: excitation function for total pion multiplicity in central Au+Au
collisions. 
For the data at 0.4, 1.0 and $1.5A$ GeV two different trackers have
been used, HT and LT.
The measured data are compared to a simulation using IQMD SM
(full squares).
Right panel: ratio theory/experiment (HT). Full squares: IQMD SM with the
systematic uncertainties of the data;
open triangles: IQMD HM.}
}
\label{sumpion}
\end{figure}

Fig.~\ref{sumpion} shows a comparison of $4\pi$ integrated 
pion multiplicities ($1.5*(\pi^+ + \pi^-)$) with the predictions of IQMD
for central ($b^{(0)} < 0.15$) Au on Au collisions.
The ratio theory to experiment is found to be $1.22 \pm 0.08$ for the
soft EOS (SM), a number that holds independently of the tracker (LT or
HT) used.
The HM (stiff) version predicts at the highest energy ($1.5A$ GeV) a drop of
pion multiplicities by about $10\%$ and even less at lower energies.
Danielewicz~\cite{danielewicz95} showed that the 'missing energy' conjectured to
be the compression energy was actually taken up by the collective (radial) flow
energy generated during the expansion which has a high degree, but not perfect,
adiabaticity (and undergoes the associated cooling and memory loss).

In ref.~\cite{kolomeitsev05} eight different transport codes were compared:
the predictions for pion multiplicities in reactions of Au on Au at $1A$ GeV
differed by a factor 1.6 if the 'standard' versions were used. 
Some of this disturbing finding could be resolved by 'unifying' the treatment
of the $\Delta$ baryon lifetime in the medium which was found
to strongly influence pion production.
Performing an IQMD simulation with a 'phase shift prescription'
 \cite{wigner55,danielewicz96,kolomeitsev05} 
instead of the prescription proposed in
ref.~\cite{kitazoe86}, we found that the pion multiplicities for central Au+Au
collisions were reduced by $(26-30)\%$ to values somewhat lower than the
experimental values.
As we shall see later, the phase shift prescription also influences the
predicted pion flow, although not quite as significantly.

In ref.~\cite{larionov03} a realistic estimation of in-medium modifications was
performed.
Perhaps the most spectacular and interesting effect originates from a 
significant
modification of the 'elementary' free cross sections for the $NN \rightarrow
N\Delta$ reaction caused by the drop of the baryonic Dirac masses in the medium
predicted presently by a number of theoretical models (see for example
\cite{vandalen05}). 
Within the quasiparticle picture this affects the kinematical phase space
factors in front of the square of the transition matrix element determining the
cross sections.
Although, due to partial thermalisation and hence a setting in of the back
reaction via detailed balance, this does not translate linearly to the
observed pion multiplicities. The effects predicted \cite{larionov03} are still
significantly larger than the experimental uncertainties.

\begin{figure}
\epsfig{file=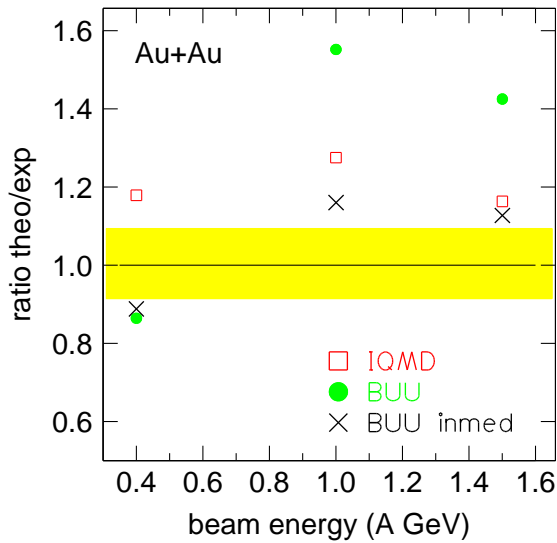,width=90mm}
\caption{
{\small
Ratio of theoretical and experimental pion multiplicities for central
Au on Au collisions versus beam energy. Crosses and full circles are
calculations
from ref.~\cite{larionov03} with and without in-medium corrected cross sections,
respectively. The open squares are IQMD calculations with free cross sections.
The shaded band reflects the uncertainty of the experimental data.}
}
\label{larionov}
\end{figure}

Medium effects are of special interest in this context since $\Delta$ baryon
resonances decay in the medium due to their very short lifetime
and undergo a cycle of several regenerations~\cite{bass95}.
At present it seems that the theoretical situation regarding these matters  is
not settled.
This is illustrated in Fig.~\ref{larionov} where we plot the ratio of
calculated pion multiplicities to experimental data
for Au on Au at various incident energies.
The ratios were evaluated using BUU results from ref.~\cite{larionov03} for the
so-called NL2-set2 combination of parameters (see the original paper for
details) which seemed to come closest to a preliminary version of our data.
Compared to the experimental uncertainty band indicated in the figure the
difference between the calculation fully accounting for in-medium modifications
and the 'standard' calculation with free cross sections is significant at the
two higher beam energies.
For comparison we also plotted the 'standard' (in the present work) IQMD
predictions.
The standard versions of IQMD and BUU make different predictions. 
At $1.5A$ GeV
the standard IQMD value is close to the in-medium corrected value of the BUU
code.

\section{Transverse momentum spectra} \label{transverse}

The main characteristics of pion emission transverse to the beam axis have
been described in section~\ref{rapidity} in terms of the variance of
the transverse rapidity distribution which was compared with the variance
of the longitudinal distribution.
In this section we briefly compare transverse momentum spectra in more detail
with, first, data from the KaoS Collaboration~\cite{kaos} and, second, with the
output of IQMD calculations.

\subsection{Comparison with KaoS}

\begin{figure}[!t]
\begin{center}
\epsfig{file=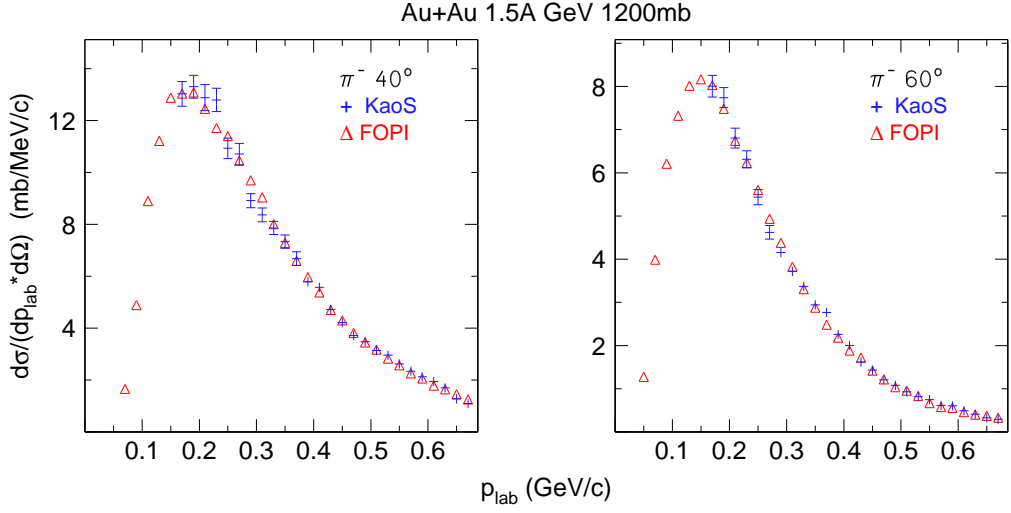,width=150mm}
\end{center}
\caption{ 
{\small
Laboratory momentum spectra, $d\sigma/(dp_{lab} d\Omega)$,
of emitted pions at $40^{\circ}$ (left)
and $60^{\circ}$ (right) for Au+Au at $1.5A$ GeV.
The KaoS and the FOPI (triangles) spectra are compared for the same centrality
selection.}
}
\label{KaoS}
\end{figure}

To convince ourselves that the filtered part of our data (upper right
panel in Fig.~\ref{uty6}) that we choose to analyse is not affected 
by  $p_t$ dependent distortions in the $p_t$ range that we are interested
in (i.e. excluding both very low $p_t$ which we cannot measure, as well as
very high $p_t$ which are rare),   
we have made a comparison with KaoS data~\cite{kaos} for
the reaction Au+Au at $1.5A$ GeV which is shown in Fig.~\ref{KaoS}.
Notice the comparison is shown on a {\em linear} ordinate scale.
To avoid the influence of assumptions and possible problems of aligning the
centralities, we have chosen a relatively large sample (the 1200mb
innermost centrality) at fixed laboratory angles indicated in the figure.
The agreement is good as far as the shape of the spectra is concerned
(the KaoS data have been rescaled to agree with the FOPI data on an 
integral basis).
Below 0.1 GeV/c the FOPI data are affected by apparatus response.

\subsection{Comparison with IQMD}

\begin{figure}[!b]
\begin{center}
\epsfig{file=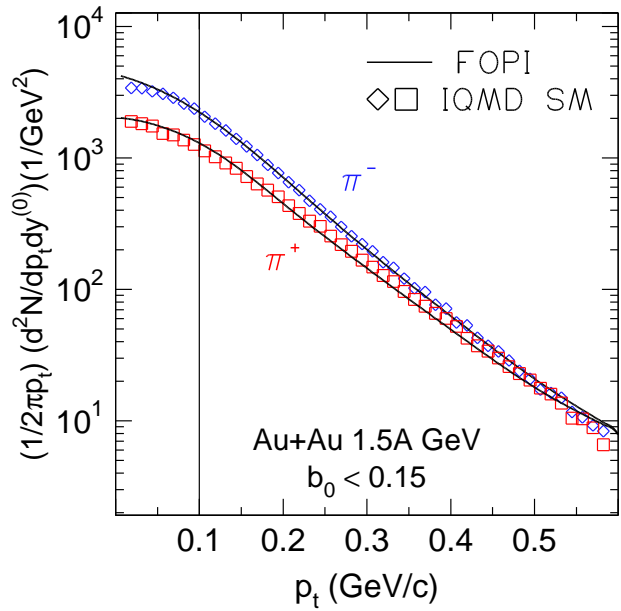,width=100mm}
\end{center}
\caption{
{\small
Transverse momentum spectra of $\pi^-$ and $\pi^+$ mesons in central
collisions ($b_0 < 0.15$)  of Au+Au at $1.5A$ GeV.
The symbols represent IQMD simulations, the solid lines are generated from
the smoothened representation of the measured data. 
The part to the left of the line at $p_t=0.1$ GeV/c is extrapolated.
The spectra are integrated over the longitudinal rapidity $|y_0|<1.8$.}
}
\label{pt}
\end{figure}

Attempts to describe transverse momentum spectra in terms of
thermal model parameters  are generally successful, see for
instance~\cite{hong97,hong98,hong05}.
Here we shall not go through the thermal exercise, but are rather
interested how well microscopic calculations that are free of ad hoc adjustable
parameters are reproducing our data.
In Fig.~\ref{vary} we have  shown that the variances of the 
{\em transverse} rapidities were rather well reproduced in Au+Au collisions.
We normalize the calculations to the experimental data, since we have already
discussed (see Fig.~\ref{sumpion}) the integrated pion multiplicities
and are now interested in the shapes of the spectra.
The outcome of the comparison is shown in Fig.~\ref{pt} for the reaction
Au+Au at $1.5A$ GeV and $b_0 < 0.15$. 
The, somewhat different, shapes of both the $\pi^+$ and $\pi^-$ mesons
are well reproduced in the $p_t$ range shown.
The level of accuracy on which the comparison is meaningful in the
figure is determined by the systematic errors in the data and
corresponds approximately to the size of the symbols.
We recall that IQMD takes into account the effect of the Coulomb fields
on the pion trajectories, but does not include (yet) higher resonances
which might influence the spectra at higher momenta.
With this partial success in hand, we can go back to the details of the
collision history that a microscopic code is able to furnish.
Rather than being a spectrum resulting from the decay of resonances moving
locally around with a single temperature, the model tells us~\cite{bass94}
that the measured
spectrum is a superposition of pions from various $\Delta$ baryon generations
representing different stages of the collision with spectra varying in
'hardness' as time proceeds.
This is a key to memorize {\em some} effect of the early compression stage
in contrast to a completely equilibrated freeze out scenario and
is probably  the reason why an, albeit small, effect of the mean field on
both the yields and the stopping can be seen.
As we shall find out later, the mean field effect also influences the small
azimuthal anisotropies.

%

\clearpage
\newpage
\section{Pion isospin dependences} \label{isospin}

The $\pi^-$ and the $\pi^+$ mesons are members of an isospin triplet.
A comparison of observables connected with differences between the $\pi^-$
and the $\pi^+$ mesons offers therefore, in principle, the possibility to 
explore isospin effects on the reaction dynamics and perhaps also on the more
fundamental issue, the isospin dependence of the EOS.

In this section we show our results for the ratio of average transverse momenta
and the ratio of yields.
Another important observable, isospin differences in the pion flow, will be
treated separately in section~\ref{azimuthal}.

\subsection{Average transverse momenta}

\begin{figure}
\begin{center}
\epsfig{file=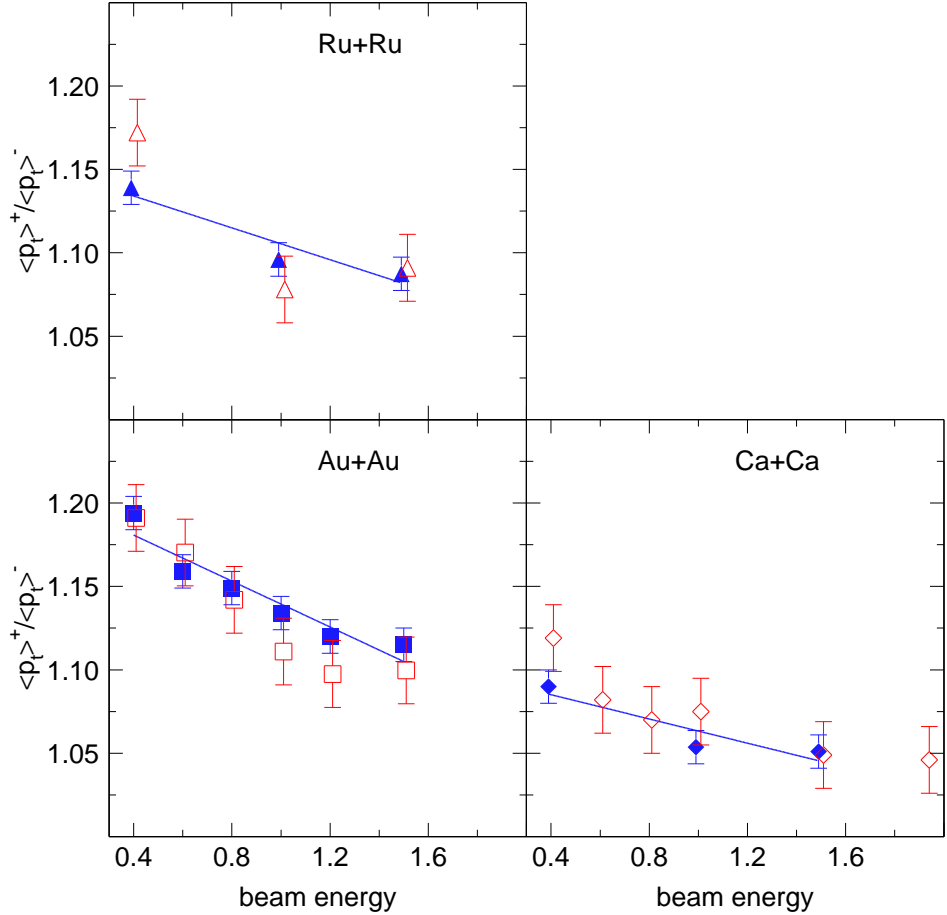,width=140mm}
\end{center}
\caption{
{\small
Ratio of the average transverse momenta of $\pi^+$ and $\pi^-$ mesons.
Shown are the excitation functions for the systems Ru+Ru (top panel),
Au+Au and Ca+Ca (lower right).
The IQMD SM predictions (full symbols) are joined by linear least squares fits
to guide the eye, the open symbols represent values inferred from our $4\pi$
extrapolated data.}
}
\label{ptratio}
\end{figure}

In the simulation with IQMD, pions are assumed to propagate in Coulomb fields.
As these change sign with the charge of the meson, higher average transverse
momenta, $<p_t>$, are predicted for $\pi^+$ relative to $\pi^-$ mesons.
This is shown in Fig.~\ref{ptratio} for central collisions of the systems 
$^{40}$Ca+$^{40}$Ca, $^{96}$Ru+$^{96}$Ru, and Au+Au as a function of beam
energy.
The calculations predict an increase of the ratio $<p_t>^+/<p_t>^-$ with
the total charge of the system, as one would expect, and an approximately
linear decrease with the beam energy.
The experimental data follow these trends, confirming, for this observable, the
dominant influence of Coulomb fields, presumably after the strong interactions
have frozen out.
In principle this effect can be used to infer the size of the freeze-out volume.
For an application of this idea using analytical formulae we refer
to~\cite{wagner98,hong05}.
Alternatively, since the microscopic simulation describes the transverse
momentum data well, one
could derive the freeze out conditions from the detailed intermediate output
of the code.
Here we rather invert the argument and conclude that the code is realistic as
far as freeze out densities are concerned.

A closer look into Fig.~\ref{ptratio} suggests that the decrease with beam
energy is somewhat faster at the lower energy end and has a tendency to saturate
at the higher energies.
This effect is marginal however, in view of the experimental uncertainties,
which are primarily caused by the necessity to extrapolate the measured data to
zero $p_t$.

\subsection{Ratio $\pi^-/\pi^+$}


The usefulness of the $\pi^-/\pi^+$ yield ratio for investigating the isospin
dependence of the EOS has been advocated recently~\cite{bali03,gaitanos04}.
The lessons learned in connection with total pion multiplicities
(see section~\ref{production}) leave room only for a cautious optimism.

It is argued~\cite{bali03} that both extremes of modelling, first chance
collisions dominated by the $\Delta$-resonance mechanism (the
$\pi^-/\pi^+$ ratio is then $\approx  (N/Z)^2$~\cite{stock86}), 
as well as equilibrium
statistical models, via the difference $\mu_n-\mu_p$ in the neutron and proton
chemical potentials, all depend on the isospin asymmetry densities 
quantified by $\rho_n/\rho_p$ or $\rho_n - \rho_p$
($\rho_n$, $\rho_p$ are the neutron, resp. proton densities).
Under the influence of the heavy ion dynamics changes in local asymmetries may
be induced that in turn create different $\pi^-/\pi^+$ generations.
In particular, if the EOS is stiff against isospin asymmetry at high densities,
the local energy density will tend to minimize by pushing the neutrons out
of high density areas (in neutron rich systems) leading to a 
local lowering of the $\pi^-$ production mechanism. 

\begin{figure}
\begin{center}
\epsfig{file=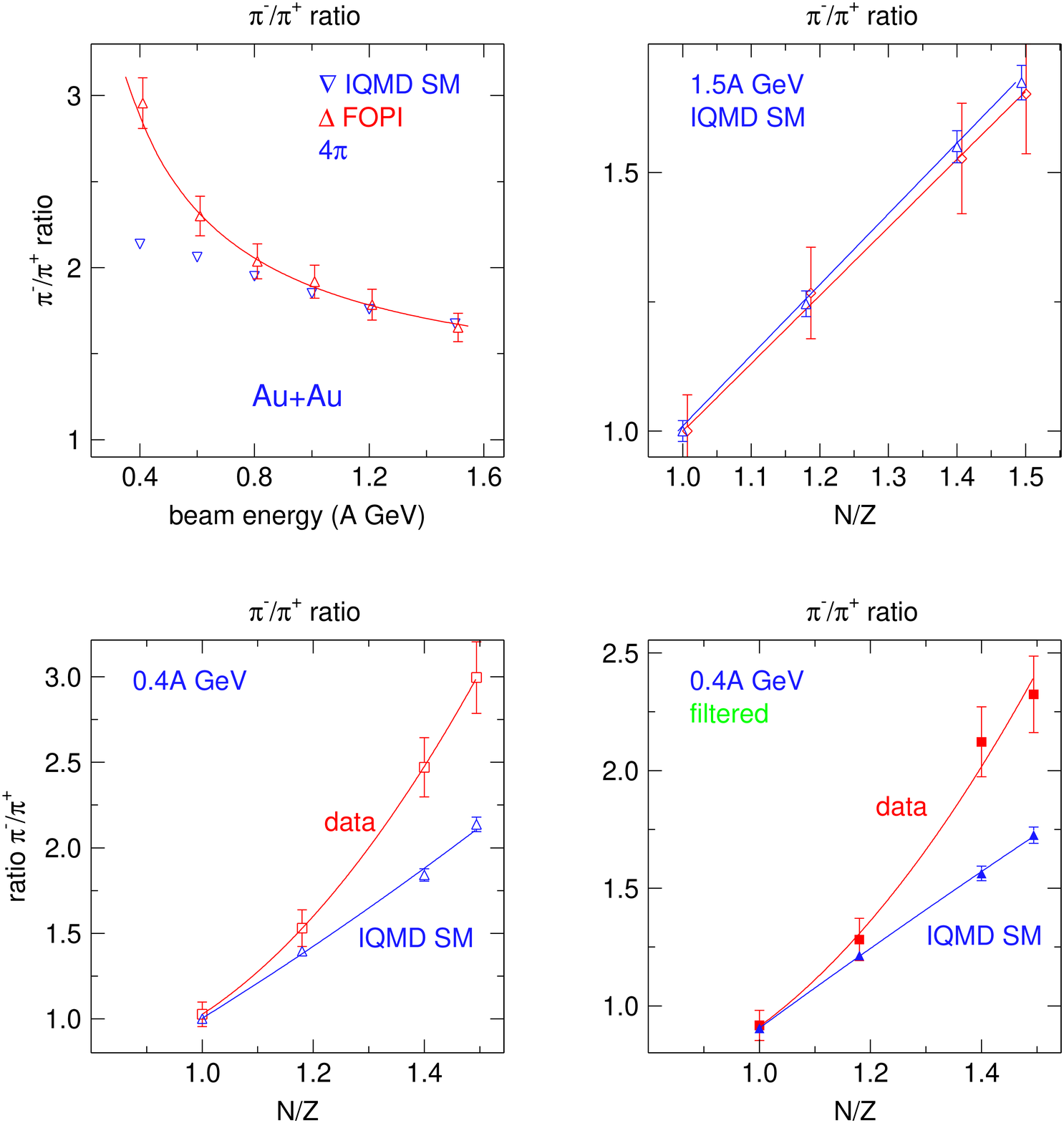,width=150mm}
\end{center}
\caption{
{\small 
Upper left panel: Excitation function of the $4\pi$-integrated ratio
of $\pi^-/\pi^+$ yields in central Au+Au collisions. 
The experimental data are joined by a least squares fit of the function
$c_0  + c_{-1}(E/A)^{-1}$ excluding the lowest energy point.
The IQMD SM prediction (triangles) is also given.
Upper right and lower left panels: the $N/Z$ dependence at $1.5A$, respectively
$0.4A$ GeV of the $\pi^-/\pi^+$ ratio.
The solid lines are least squares fits of linear or quadratic $(N/Z)$
dependence.
Lower right panel: same as lower left panel,but for filtered data.}
}
\label{piratio}
\end{figure}

In the various panels of Fig.~\ref{piratio} we show a summary of our measured
$\pi^-/\pi^+$ ratios and compare with IQMD predictions.
Briefly, one observes a decrease of the $\pi^-/\pi^+$ ratio with incident energy
(upper left panel) which is qualitatively also predicted by IQMD.
However, in this panel, and more so in the other three panels, it is seen that
while IQMD is doing a perfect job at $1.5A$ GeV, also when $(N/Z)$ is varied,
it clearly underestimates the pion ratio for large $(N/Z)$ at $0.4A$ GeV.
The right lower panel repeats the comparison at $0.4A$ GeV, but for the
{\em filtered} data, leading to the same conclusion and thus showing that the
extrapolations to $4\pi$ are not responsible for the discrepancy.

The linear $(N/Z)$ dependence at the higher energy instead of the expected
$(N/Z)^2$ dependence of the $\Delta$-resonance model (the model used in IQMD)
can be partially understood when realizing that the copious pion production
will move the system towards chemical equilibrium by lowering the $(N/Z)$
of the daughter system constrained by total charge 
conservation~\cite{kapusta77}.
However, the linear behaviour of IQMD at $0.4A$ GeV, where pion emission is a
modest perturbation, is less trivial, all the more as it disagrees with the
non-linear behaviour of the data.
 
\begin{figure}
\begin{center}
\epsfig{file=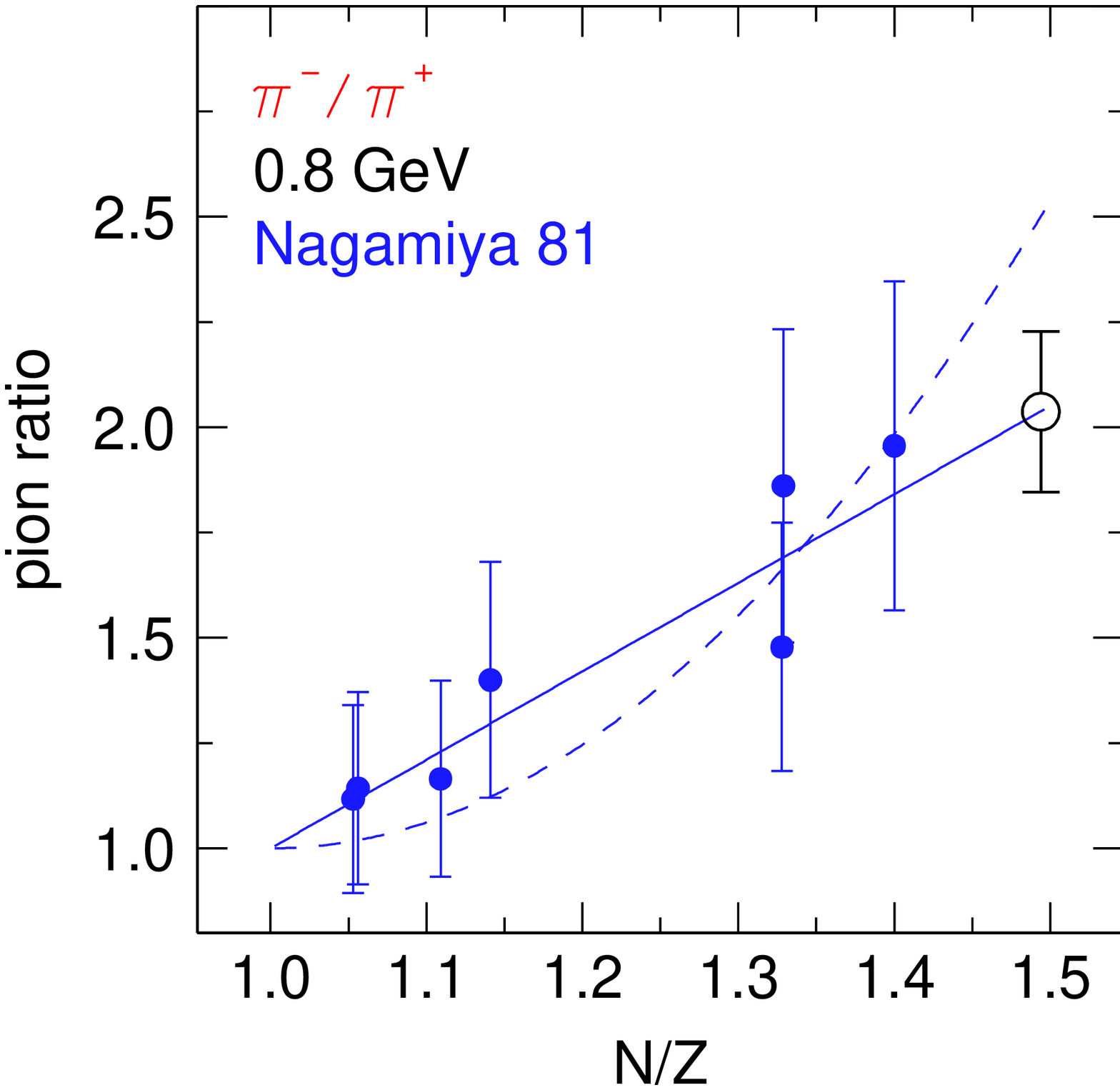,width=78mm}
%
\epsfig{file=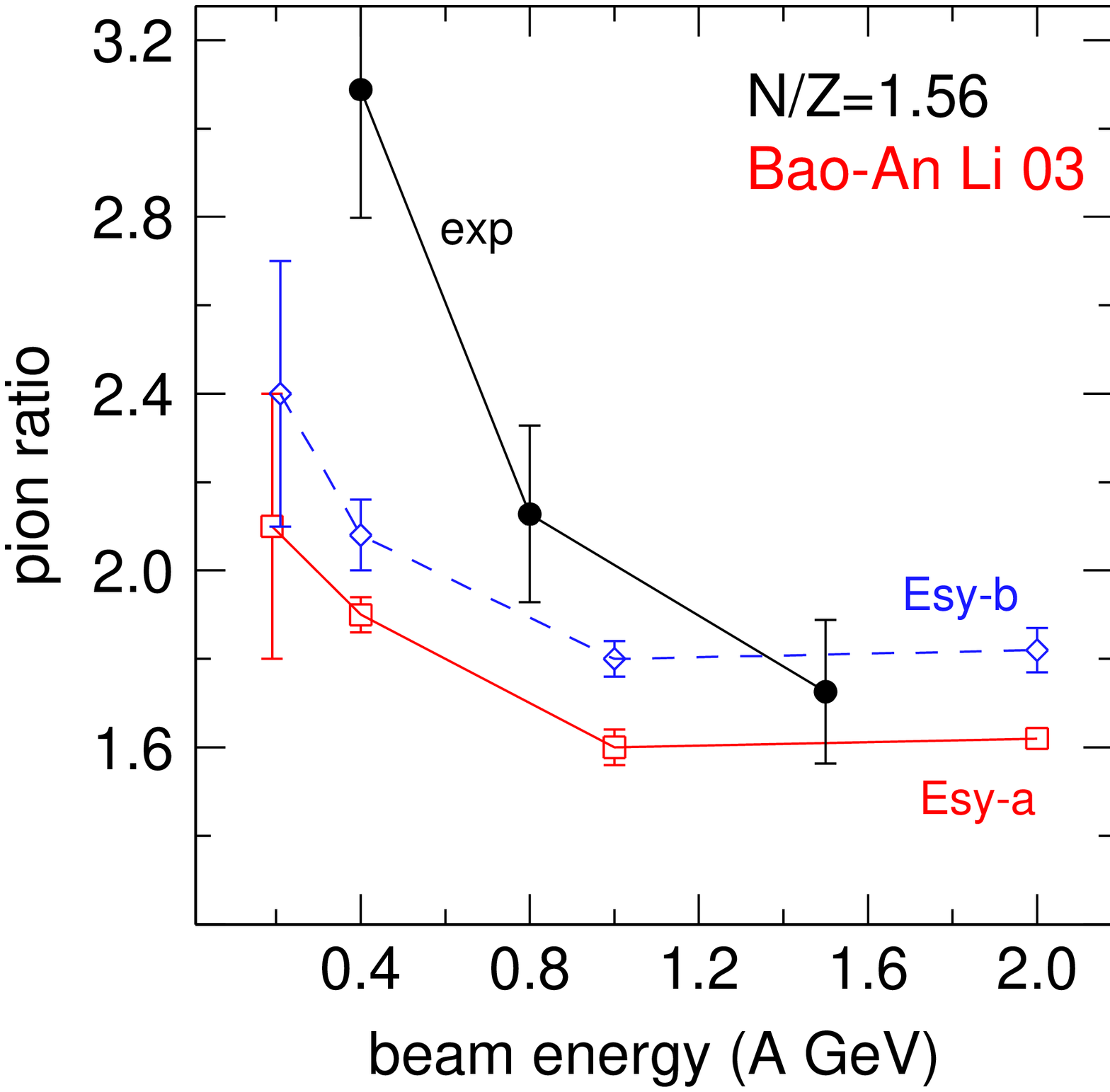,width=78mm}
\end{center}
\caption{
{\small
Left panel: $\pi^-/\pi^+$ ratio versus $N/Z$ of the 'fireball'
measured~\cite{nagamiya81} in various inclusive heavy ion reactions at $0.8A$
GeV (full circles).
The solid (dashed) curve is a linear (quadratic) least squares fit to the data
constrained to be one at $N/Z=1$.
The data point (open circle) from the present work holds for Au+Au and was not
included in the fit.
Right panel: $\pi^-/\pi^+$ ratios versus beam energy obtained in transport
calculations~\cite{bali03} for the system $^{132}$Sn+$^{124}$Sn ($N/Z=1.56$),
using two options for the symmetry energy, Esy-b, dashed, and Esy-a.
The ratios obtained from the present Au+Au data (solid circles) by linear
extrapolation (from $N/Z=1.494$) are shown for comparison.}
}
\label{bali}
\end{figure}

A systematics of $\pi^-/\pi^+$ ratios was first established for inclusive
reactions~\cite{nagamiya81} at $0.8A$ GeV beam energy using various, also
asymmetric systems.
In Fig.~\ref{bali} we reproduce these older data which were plotted as a
function of an estimated~\cite{nagamiya81} 'fireball' $(N/Z)$ composition.
Due to the limited accuracy,  both  linear and  quadratic $(N/Z)$
dependences are compatible with these inclusive data.
Our data point for Au+Au at the same energy, but for a central collision
selection, is also shown in the figure and is perfectly compatible with the
linear extrapolation.

In ref.~\cite{bali03} Bao An Li has performed more advanced calculations for
$^{132}$Sn+$^{124}$Sn at $0.4A$ GeV using two different options for the
asymmetry energy, Esy-a, a version that increases linearly with the density,
and Esy-b, an exotic variant that after an increase at lower density bends back
down to cross zero again at $\rho=3\rho_0$.
The changes in the $\pi^-/\pi^+$ ratio that these very different alternatives
induce are shown in Fig.~\ref{bali} which reproduces results from Fig.2 of
ref.~\cite{bali03}.

The $N/Z$ ratio of the rare isotope beam combination, $N/Z=1.56$ is not
very different from that of $^{208}$Pb+$^{208}$Pb, $N/Z=1.536$
(which would be a more readily available bigger system).
For the reaction $^{197}$Au+$^{197}$Au studied here, $N/Z=1.494$, we
anticipate,
by linear extrapolation to $N/Z=1.56$, $\pi^-/\pi^+$ ratios that have been
added to the figure. 
With this addition the problematics already known from the total pion
multiplicity studies show up.
First, the difference predicted from the calculation~\cite{bali03} between two
rather extreme options  is on the $10\%$ level and hence on the order
of the present experimental accuracy, and, second, none of the two predictions
follows the data.
Similar conclusions follow if we take the calculations in
ref.~\cite{gaitanos04}.

\section{Azimuthal correlations ('flow') of pions}\label{azimuthal}

Owing to collective flow phenomena, discovered experimentally in 
1984~\cite{gustafsson84,renfordt84},
it is possible to reconstruct the reaction plane 
event-by-event and hence to study
azimuthal correlations relative to that plane.
We have used the transverse momentum method~\cite{danielewicz85}
including all particles
identified outside the midrapidity interval $|y_0|<0.3$ and excluding
identified pions to avoid autocorrelation effects.
 
We use the well established parameterization

\[ u = (\gamma , \vec{\beta}\gamma) \ ; \ \  u_t = \beta_t\gamma \]
\[ \frac{dN}{u_t du_t dy d\phi} = v_0 [1 + 2v_1 \cos(\phi) + 2v_2
\cos(2\phi)] \]
\[ v_0 = v_0(y,u_t) \  ; \ \  v_1 = v_1(y,u_t) \ ; \ \  v_2 = v_2(y,u_t) \]
\[ v_1 = \left<\frac{p_x}{p_t}\right> = <cos(\phi)>\ ;
 \ \  v_2 = \left <\left( {\frac{p_x}{p_t}}\right)^2 -
  {\left(\frac{p_y}{p_t}\right)}^2 \right > = <cos(2\phi)>\]
\textBlack
where $\phi$ is the azimuth with respect to the reaction plane
and where angle brackets indicate averaging over events (of a specific class).
The Fourier expansion is truncated , so that only three
parameters, $v_0,v_1$ and $v_2$, are used to describe the 
'third dimension' for  fixed intervals of rapidity and transverse
momentum.
Due to finite-number fluctuations the apparent reaction plane determined
experimentally does not coincide event-wise with the true reaction plane,
causing an underestimation of the deduced coefficients $v_1$ and $v_2$
which, however, can be corrected by studying sub-events: we have used the 
method of Ollitrault \cite{ollitrault98} to achieve this.
The finite resolution of the azimuth determination is also the prime
reason why the measured higher Fourier components turn out to be rather small.

\begin{figure}[!h]
\begin{minipage}{8cm}
\hspace{-10mm}
\epsfig{file=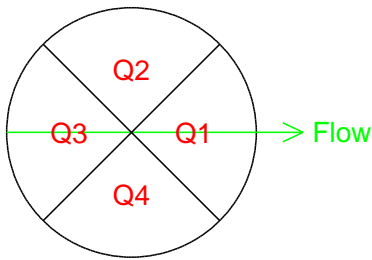,width=80mm}
\end{minipage}
\begin{minipage}{8cm}
\caption{
{\small
The four quadrants in a plane, $xy$, transverse to the beam direction, $z$.
The flow direction, $x$, is indicated.}
}
\label{quadrant}
\end{minipage}
\end{figure}

Alternatively to the three Fourier coefficients, one can introduce the
yields $Q_1$, $Q_2$, $Q_3$, $Q_4$ in the four azimuthal quadrants,
see Fig.~\ref{quadrant},
of which only three are independent (on the average over many events)
due to symmetry requirements

\begin{eqnarray*}
 Q_2 & = &  Q_4 \\
 Q_{0} &= & Q_1 + Q_2 + Q_3 + Q_4 \\
 Q_{24} & = & Q_2 + Q_4 \\
\end{eqnarray*}
\Black{The two equivalent triplets}
\begin{center}
$v_0$, $v_1$, $v_2$ $\longleftrightarrow$   $Q_{0}$, $Q_1$, $Q_{24}$
\end{center}
\Black{are related by}
\begin{eqnarray*}
2\pi\  v_0 & &= Q_0\\
\frac{2 \sqrt {2}}{\pi}\  v_1 & = 0.900 v_1 & = \frac{Q_1-Q_3}{Q_0} \\ 
-\frac{2}{\pi}\  v_2 & = -0.637 v_2 & = \frac{Q_{24}}{Q_0} - \frac{1}{2}\\
\end{eqnarray*}
\textBlack
These relations show that $v_1$ is a dipole, while $v_2$ is a quadrupole
strength. Statistical (count rate)
errors can be deduced with use of elementary algebra.
The quadrant formulation also has some advantage in assessing small
apparatus distortions leading to systematic errors, as we shall see.
The fact that $v_1$, as well as $v_2$, are found to be non-zero,
is generally called 'flow' in the literature and in particular
the first Fourier coefficient is  taken to be a measure of 'directed flow',
while the second Fourier coefficient has been dubbed 'elliptic flow'.

Recently more general methods based on Lee-Yang zeros have been
proposed~\cite{bhalerao03} and used~\cite{bastid05} to isolate 'true'
collective flow from other effects that might influence azimuthal anisotropies.
An analysis of some of our pion data with this method will be published
elsewhere~\cite{bastid05}.
Here we take the point of view that the most likely 'non-flow'
correlation, the $\Delta$-correlation between nucleons and pions is
part of the physics relating the pion flow to the nucleon flow, a physics that
is implemented in the IQMD code that we use to try to understand the
'flow' observables.
Removing this correlation from the data would require removing it from the
simulation as well when comparing.
In view of the high statistics required to do this with sufficient accuracy we
prefer in this survey to stick to the simpler 'standard' method.
This allows us also to compare with older data where possible.

Azimuthal anisotropies of pion emission in heavy ion reactions were first
reported in the refereed literature by Gosset {\it et al.}~\cite{gosset89}
using the DIOGENE setup in the reaction Ne+Pb at $0.8A$ GeV.
The authors attributed their observations to a target (Pb) shadowing
effect. This study concerned primarily what is now termed the $v_1$ component.
This $v_1$ component was studied in more detail in ref.~\cite{kintner97} for
the reaction of Au+Au at $1.15A$ GeV.
The fact that in sufficiently non-central geometries the $\pi^+$ 
 mesons had the opposite directed flow than the protons (termed
'antiflow') was reported supporting earlier theoretical
predictions~\cite{bali94,bass95}.
It was also observed that $\pi^-$ mesons had a different flow and it was
concluded~\cite{kintner97} that {\em 'the differences between the behavior of
the $\pi^+$ and the $\pi^-$ suggest that further consideration of this
phenomenon is needed'}.

The observation of a $v_2$ component was reported by the KaoS~\cite{brill93}
and the TAPS~\cite{venema93} collaborations in 1993.
A first shot at a $v_2$ systematics (Bi+Bi at 0.4, 0.7 and $1.5A$ GeV) was
made in ref.~\cite{brill97}.
A qualitative interpretation of the anisotropies making 
use of the transient vicinity
of spectator matter close to the participant fireball as a 'clock'
was presented in ref.~\cite{wagner00}.
In terms of the azimuthal quadrants just introduced, the authors considered
the ratios $Q_1/Q_2$, $Q_3/Q_2$ and $Q_1/Q_3$ which mix the two
Fourier components $v_1$ and $v_2$. 

A direct {\em quantitative} comparison of our present data 
and simulations with all these
earlier observations is not straight-forward, as we shall see.

For completeness we mention here that pion flow has also been studied at much
higher beam energies, $11A$ GeV (AGS \cite{E877}) and $40A$ GeV, as well as
$158A$ GeV (SPS \cite{NA49}).
These incident energies are rather far apart, making it
difficult to assess at present the gradual evolution of the pion flow observable
and the associated physics changes.

As pion 'flow' turns out to be small, when compared to nucleonic flow, its
observation requires a high degree of systematic and statistical accuracy.
Presently the latter is difficult to achieve in transport code simulations
if one has the ambition to predict $v_1(y,u_t)$ and $v_2(y,u_t)$ in their full
two-dimensional glory.
Despite our  computational efforts the simulations are still plagued
with statistical errors that are a factor 3-5 larger than the experimental
ones.
Therefore, in the sequel we shall use several flow characterizers, $<v_1>$,
$<v_1 u_{t0}> \equiv <u_{x0}>$, $<v_2>$ and $<-v_2 u_{t0}^2> \equiv <u_{yx}^2>
\equiv <u_{y0}^2 - u_{x0}^2>$
averaged ($< >$) over more or less large regions of phase space (besides the
averaging over events of a specific centrality class).
Although we present dipole and quadrupole flow in two separate subsections,
they should be considered as two sides of
the same (rather complex) phenomenon.
In the sequel averaging brackets will be omitted, but will be implied.

\subsection{Directed flow $v_1$}\label{v1}
\begin{figure}[!h]
\begin{center}         
\epsfig{file=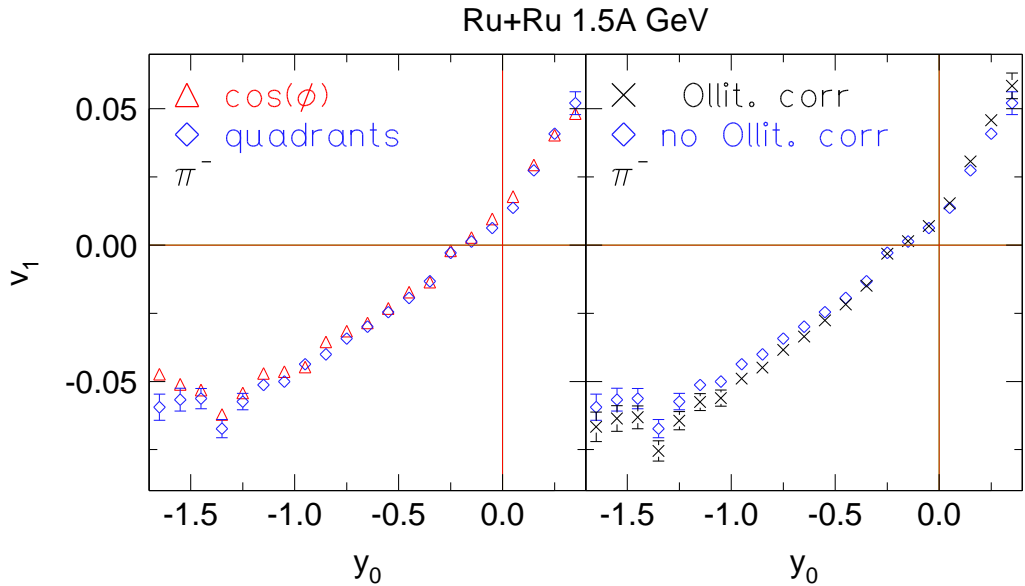,width=150mm}
\end{center}
\caption{
\small{
Rapidity dependence of the directed $\pi^-$ flow in the reaction $^{96}$Ru +
$^{96}$Ru at $1.5A$ GeV.
The centrality is $0.25 < b_0 < 0.45$.
Left panel: Comparison of the '$\cos(\phi)$' and the 'quadrants' methods.
Right panel: Comparison of the resolution corrected data with the uncorrected
data.}
}
\label{v1ruru}
\end{figure}

First, a few technical points will be mentioned.
The  near-equivalence of
using the '$\cos (\phi)$' method, eq. , or alternatively the
'quadrant' method, as well as the effect of the resolution correction
are shown in Fig.~\ref{v1ruru}.

Following \cite{ollitrault98} the correction factor is given by
$$v_n = v'_n/<\cos(n\Delta\Phi)>$$
where $v'_n$ is the (uncorrected) measured value and
$\Delta\Phi = \Phi_R - \Phi_R'$ is the azimuthal angle between the true
and the measured reaction planes, $\Phi_R$ and $\Phi_R'$, respectively.
Due to the large acceptance of FOPI the correction factors are essentially
given by the 'natural' finite number fluctuations rather than by apparatus
limitations.
Typical values for different average centralities at an incident beam energy of
$1A$ GeV are shown in Table~\ref{vcorr}.
The effect of the correction is smallest when nucleonic flow is large, i.e.
in the third centrality bin of the heaviest system.
Small inhomogeneities in the laboratory azimuthal acceptances were found to have
negligible influence on the flow results.

\begin{center}

\begin{table}
\caption{Reaction plane resolution correction factors for $E/A=1A$ GeV.}               
\label{vcorr}
\begin{tabular}{|c|ccc|ccc|}
\hline
$<b_0>$ & Au+Au & Ru+Ru & Ca+Ca & Au+Au & Ru+Ru & Ca+Ca \\
        &       &  $\cos(\Delta\Phi)$ &       &       & $\cos(2\Delta\Phi)$ & \\
\hline
0.100 & 0.852  & 0.707  & 0.561 & 0.583  & 0.365  & 0.216 \\
0.167 & 0.910  & 0.786  & 0.609 & 0.708  & 0.471  & 0.259 \\
0.360 & 0.963  & 0.892  & 0.698 & 0.860  & 0.663  & 0.353 \\
0.502 & 0.958  & 0.894  & 0.699 & 0.845  & 0.670  & 0.354 \\
\hline
\end{tabular}
\end{table}

\end{center}

-------------------------------------------------------------------------

For the symmetric systems that we study here, $v_1(y)$ should be asymmetric
with respect to midrapidity ($y_0=0$), in particular $v_1(0)$ should be
zero.
As can be seen in Fig.~\ref{v1ruru}, $v_1(y)$ does not cross the origin of the
axes. 
A systematic study of this mid-rapidity offset showed that it depended
on particle type, centrality and system size in a way suggesting that
it was correlated with the track density difference in the
'flow' quadrant $Q_1$ and the 'antiflow' quadrant $Q_3$.
While this could be simulated using our GEANT based implementation
of the apparatus response, a sufficiently accurate quantitative
reproduction of the offset at mid-rapidity was not achieved.
We  therefore opted for an empirical method to correct for
the distortion, using the very sensitive requirement of antisymmetry
with respect to midrapidity.

One can show that a good  first order 
correction simply consists in shifting $v_1(y)$ down by a rapidity 
independent correction $\Delta v_1$ until it crosses zero exactly at
mid-rapidity.
Essentially $v_1 \sim (Q_1-Q_3)/Q_0$ ; assume we correct $Q_1$ by replacing
it by $Q_1+cQ_1$ where $c$ is a constant not depending on transverse
momentum or rapidity
(some global loss in the flow quadrant).
The correction to $v_1$ then is $\Delta v_1 \sim c Q_1/Q_0 \sim c/4$
where we use the  approximation
$Q_1 \sim 1/4 Q_0$.
After trying  different  Ansatzes for the correction
$c=c(y,u_t)$ which could be in principle a  2-dimensional function,
we ended up using the rapidity independent Ansatz
$c=v_{0c}+v_{1c}*u_t$ which grows linearly with ut (or pt); 
$v_{0c}$ and$v_{1c}$ are parameters.  
The track density distortions are assumed to grow linearly with the density.
The quadrant $Q_3$ is left unchanged (only {\em relative} corrections matter),
 $Q_1$ is replaced by $Q_1(1+c)$
and $Q_{24}$ by $Q_{24}*(1+0.55c)$ implying that
in first approximation it is sufficient to use $Q_2 \approx 0.55*(Q_1+Q_3)$.
Using the factor 0.55, ($\pm 0.05$), in the $Q_{24}$ correction,
rather than 0.50, the
value for isotropic emission, allows roughly for globally enhanced out-of-plane
emission in the present energy regime.
The values of $v_{0c}$ and $v_{1c}$ are fixed from the two
mid-rapidity conditions $v_1(y_0=0)=0$ and $u_x(y_0=0)=0$.
This is done separately for each particle,centrality and beam energy.
The systematic error of $v_1$ after the distortion correction is assessed to be
0.007 or $7\%$ (whichever is larger)
from the uncertainty of the mid-rapidity offset determination.
The systematic error of $v_2$ (and $u^2_{yx})$) is $(7-10\%$) and was estimated
by varying the coefficient for the $Q_{24}$ correction between 0.5 and 0.6.
These systematic errors should be kept in mind when inspecting the
figures presenting our flow data, as these contain only the statistical errors.
In comparative cases, such as assessing isospin differences between the
systems Zr+Zr and Ru+Ru the systematic errors are expected to cancel to some
degree.

\begin{figure}
\begin{center}
\epsfig{file=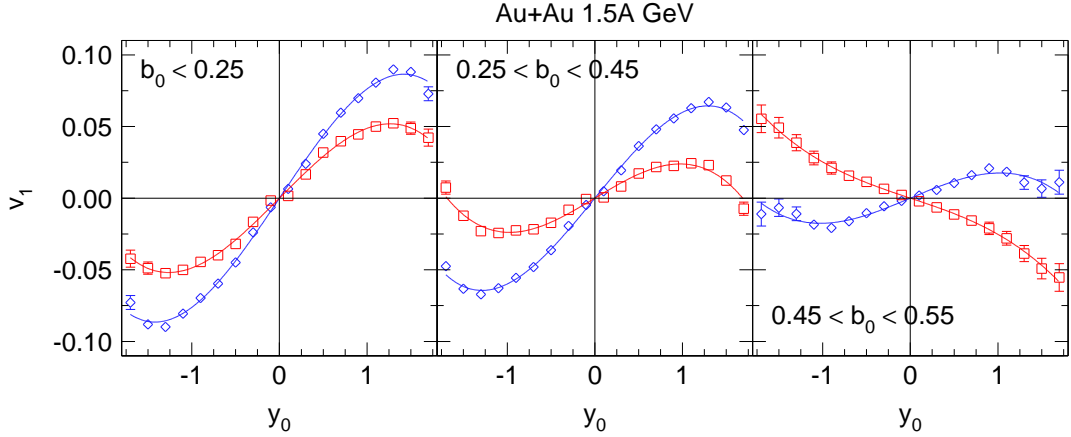,width=154mm}
\end{center}
\caption{
\small{
Rapidity dependence of the directed flow, $v_1(y_{0})$,
for Au+Au collisions at $1.5A$ GeV with centralities
$b_0<0.25$, $0.25<b_0<0.45$, $0.45<b_0<0.55$ (from left to right).
The data are taken in the interval $1.0<u_{t0}<4.2$.
(Red) squares: $\pi^+$, (blue) diamonds: $\pi^-$.
The solid curves are least-squares fits, see text.}
}
\label{v1-au1500}
\end{figure}

We are now in a position to review a representative sample of the rich flow
data that we obtained in the present work.
We start with the well known 'S-shaped' curves for the rapidity dependence of flow,
$v_1(y_0)$.
Fig.~\ref{v1-au1500} shows the data for Au+Au at $1.5A$ GeV and for three of
our 'standard' centrality intervals.
Several remarks can be made.
Only the data for $y_0<0$ were actually measured, (anti)symmetry was used to
infer the $y_0>0$ behaviour.
As $b_0$ is increased, the diagrams 'rotate' clockwise, the $\pi^+$ data
always 'preceding' the $\pi^-$ data.
In the interval around $b_0=0.5$ this 'rotation' has moved into a new
'quadrant', the antiflow side, for the $\pi^+$.
It is clear from this observation that pion flow cannot be simply derived from
a 'parent' $\Delta$ baryon flow assuming the latter to be equal
to that of single
protons, as was attempted in ref.~\cite{brill97} (for the elliptic part of the
flow).
A key question of theoretical analysis will be 
whether the $\pi^+/\pi^-$ difference
('isospin differential flow') is just a Coulomb effect or rather necessitates
a (nuclear) isospin effect in order to be reproduced.

\begin{figure}
\begin{center}
\epsfig{file=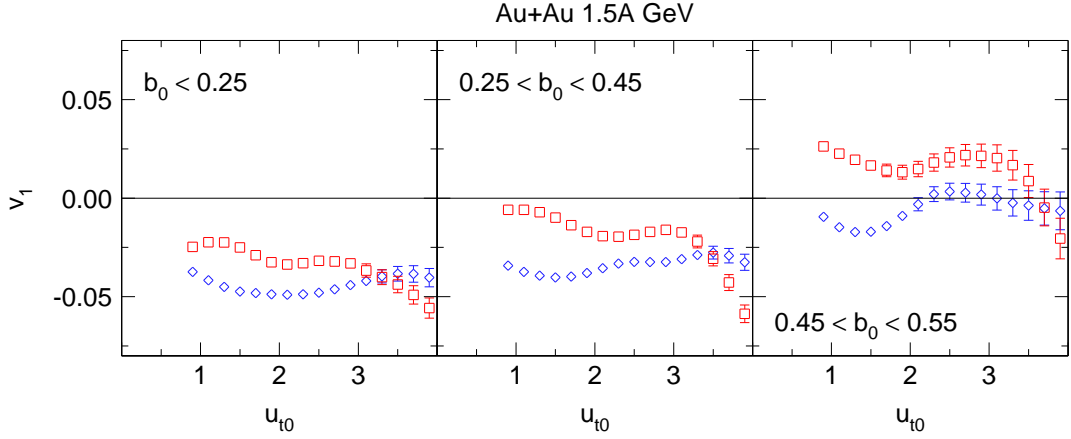,width=154mm}
\end{center}
\caption{
\small{
Transverse momentum dependence of the directed flow, $v_1(u_{t0})$,
for Au+Au collisions at $1.5A$ GeV with centralities
$b_0<0.25$, $0.25<b_0<0.45$, $0.45<b_0<0.55$ (from left to right).
The data are taken in the rapidity bin $-1.8<y_0<0.0$.
(Red) squares: $\pi^+$, (blue) diamonds: $\pi^-$.}
}
\label{v1ut-au1500}
\end{figure}

In Fig.~\ref{v1ut-au1500} we present the transverse momentum dependence,
$v_1(u_{t0})$, for data integrated over the backward hemisphere.
The switch of sign for $\pi^+$ near $b_0=0.5$ is again visible.
There is a puzzling weak wavy aspect, especially in the $\pi^+$ data that is at
the limit of apparatus distortions that we cannot completely exclude.
However, a simulation with IQMD, not shown here, predicts qualitatively similar,
even more pronounced, structures.


\begin{figure}
\begin{center}
\epsfig{file=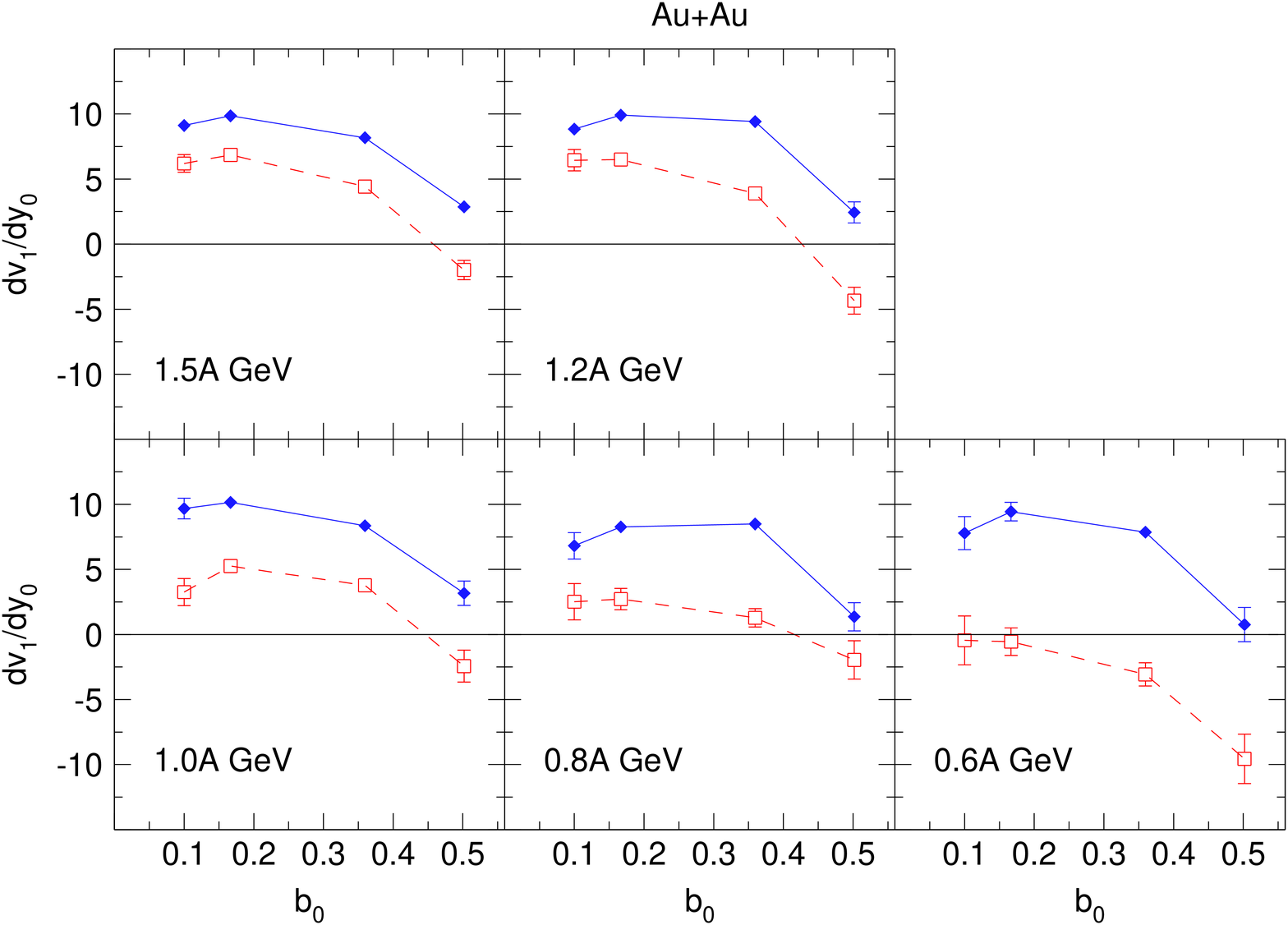,width=154mm}
\end{center}
\caption{
\small{
Centrality dependence of the midrapidity slope $dv_1/dy_0$ at various indicated
beam energies in the reaction Au+Au. 
Open squares joined by dashed lines: $\pi^+$, full diamonds: $\pi^-$.}
}
\label{v1slop-au}
\end{figure}

Although the detailed shapes of the flow data are seen to be complex, it is of
some interest to try to characterize flow with just one parameter, thus easing
the task of obtaining a survey of many system-energy-centralities.
One such parameter, the midrapidity slope $dv_1(y_0)/dy_0 |_{y0=0}$,
stresses more the midrapidity region, while the alternative parameter, which we
term 'large acceptance' flow is $v_1$ (or $u_{x0}$)
averaged over a large region of momentum
space and stresses more the regions closer to target (or projectile) rapidity
since $v_1$ is zero at $y_0=0$.

\begin{figure}
\begin{center}
\epsfig{file=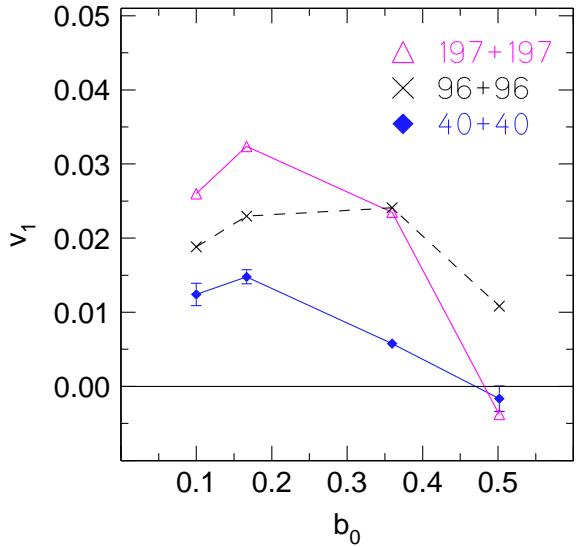,width=90mm}
\end{center}
\caption{
\small{
Centrality dependence of directed flow, $v_1$, for various indicated system
masses and incident beam energy $1.5A$ GeV.
The flow is averaged over $\pi^+$ and $\pi^-$.
The values for $A_p+A_t=96+96$ are an average between Zr+Zr and Ru+Ru.}
}
\label{v1-asys}
\end{figure}

A survey of midrapidity slopes in Au+Au reactions as a function of centrality
is shown for various indicated beam energies in Fig.~\ref{v1slop-au}.
The slopes were determined by least squares fitting with the polynomial
$v_{10}+v_{11}y_0+v_{13}y_0^3$ in the rapidity range $-1.8<y_0<0$
(the data are averaged over the interval $1<u_{to}<4.2$).
The quality of the fits can be visualized in Fig.~\ref{v1-au1500}
(smooth curves).
The constant $v_{10}$, which should be zero, accounts for remaining
uncertainties of the offset correction described earlier, $v_{13}$ is necessary
because, obviously, $v_1(y_0)$ is not linear over the extended rapidity range
and, finally, $v_{11}$ is identical to the slope $dv_1(y_0)/dy_0$ at $y_0=0$.
The $\pi^+$ slope data are seen to vary moderately with energy, the gap to the
$\pi^-$ slopes widens at the lower energies.

\begin{figure}
\begin{center}
\epsfig{file=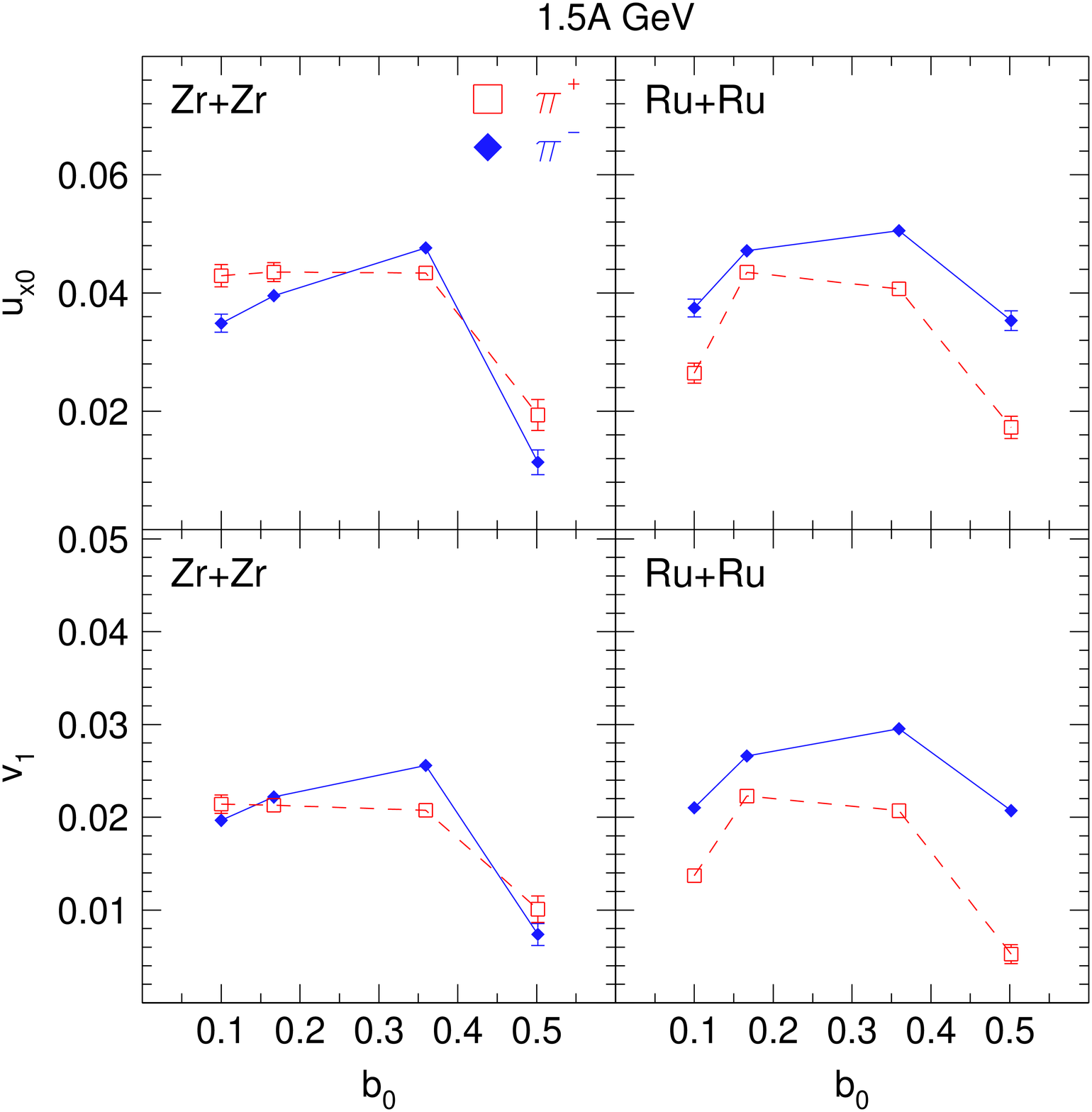,width=130mm}
\end{center}
\caption{
\small{
Comparison of the centrality dependence of directed flow ($v_1$ bottom and
$u_{x0}$ top) in the systems $^{96}$Zr+$^{96}$Zr (left) and $^{96}$Ru+$^{96}$Ru
(right). Open squares joined by dashed lines: $\pi^+$, full diamonds: $\pi^-$.}
}
\label{v1-iso}
\end{figure}

System size dependences and the system isospin dependences are shown in terms
of the large acceptance flow in Figs.~\ref{v1-asys} and \ref{v1-iso}.
For the averaging we choose the intervals $-1.8<y_0<0$ and $0.8 < u_{t0} < 4.2$
which are well covered by our setup for {\em all} measured system-energies.
Although the data were obtained in the backward hemisphere we give the values
in the forward hemisphere (which are opposite in sign) so that positive
(negative) values mean 'flow' ('antiflow') analogue to the midrapidity slopes.
For the size dependence study we have removed the isospin difference by
averaging over $\pi^-$ and $\pi^+$ data and also, for mass $96+96$, by
averaging over the two systems $^{96}$Zr + $^{96}$Zr  and $^{96}$Ru +
$^{96}$Ru.
The system size dependence of directed flow is seen to be complex, in
contrast to elliptic flow (see next subsection).

The last two mentioned systems were also the basis for an isospin dependence
study, see Fig.~\ref{v1-iso}.
We find for $v_1$, as well as for $u_{x0}$ (both in large acceptance),
an indication that the $\pi^- - \pi^+$ difference is slightly larger for Ru+Ru
than for Zr+Zr.
The Coulomb potential difference between the two systems is expected to be
close to $10\%$, a difference that does not seem to account quantitatively
for the observations.
Since $N/Z$ is larger for the Zr+Zr system, this can only be understood if there
is a nuclear isospin effect opposite in sign to the Coulomb effect.
This seems to be also suggested by the theoretical investigations of
Qingfeng Li {\em et al.}~\cite{qfli05}.
We shall come back to this when adding later also information on
isospin differential {\em elliptic} flow (Fig.~\ref{v2-iso}) which is easier
to parameterize.

\begin{figure}
\begin{center}
\epsfig{file=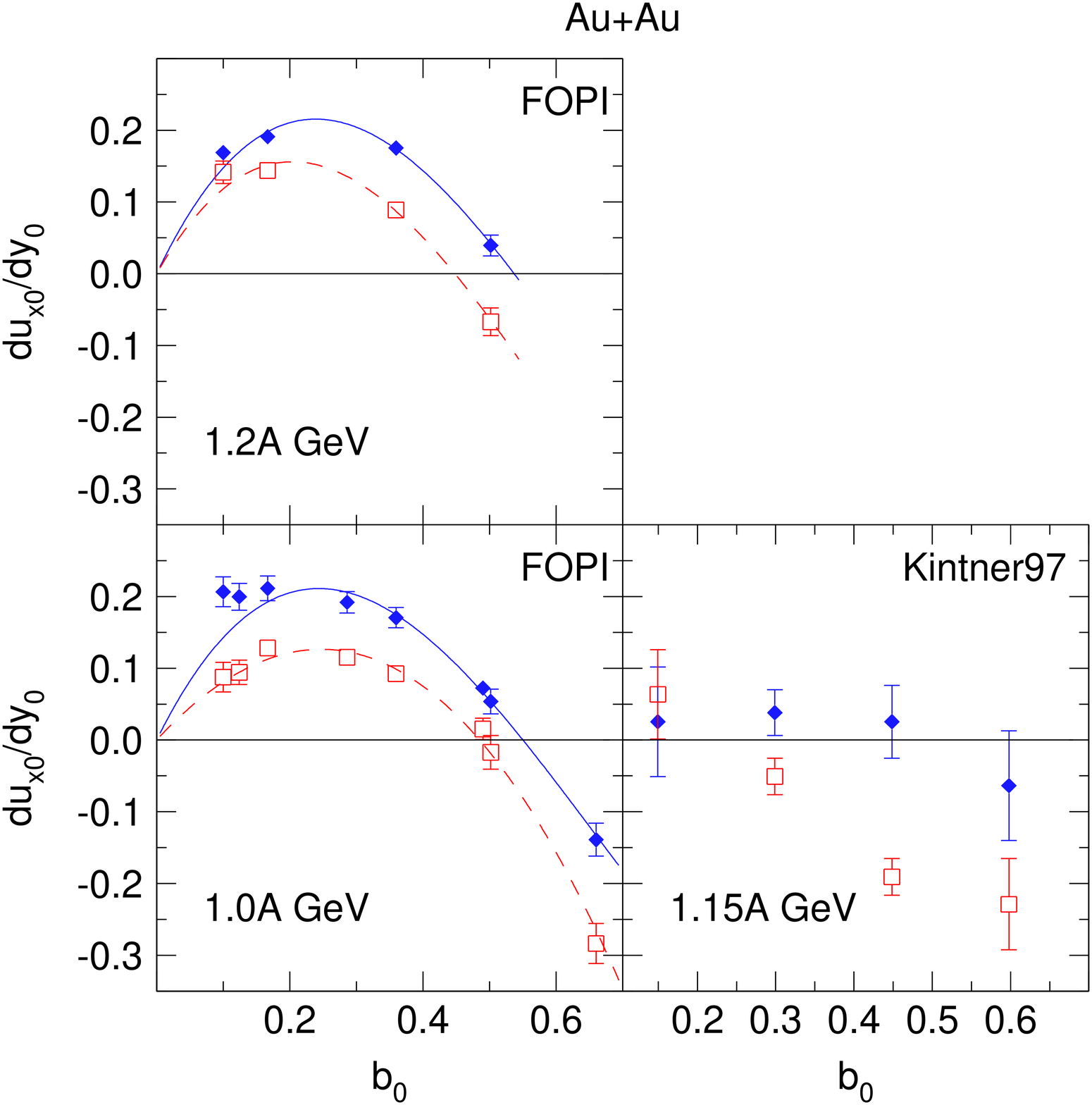,width=154mm}
\end{center}
\caption{
\small{
Centrality dependence of the midrapidity slope $du_{x0}/dy_0$ in the 
reaction Au+Au  (squares and dashed lines: $\pi^+$, full diamonds:
$\pi^-$).
The left panels represent the data of the present work ('FOPI') 
for $1.0A$ and $1.2A$ GeV beam energy with polynomial fits and 
a cut $1.0<u_{t0}<4.2$.
The right panel is adapted from ref.~\cite{kintner97} for $1.15A$ GeV.  }
}
\label{kintner}
\end{figure}

Before trying to assess the sensitivities of pion flow data to
theoretical input, we shall compare our data with those of
ref.~\cite{kintner97} which were taken at 1.15A GeV for the system Au+Au.
Such comparisons are not trivial, as the system-energy, the centrality,
the chosen specific flow-describing observable, and, last but not least,
the phase space covered, must be 'aligned'.
Fig.~\ref{kintner} presents in three panels the data relevant for this
comparison in terms of the centrality $(b_0)$ dependence of the mid-rapidity
slope $d u_{x0}/dy_0$.
As we have no measurement at $1.15A$ GeV, we show in two of the panels our data
at $1.0A$ and $1.2A$ GeV.
The analysis of the $1A$ GeV data has been extended to higher $b_0$ to allow
for a more complete comparison, but one can say that there is no dramatic
difference between the two energies.
The Kintner data have been converted to the same, scaled, axes
using the information from Fig.~3 of ref.~\cite{kintner97}.
Similar to the slopes of $v_1(y_0)$, 
our mid-rapidity slopes were determined from
a least squares fit of $u_{x0}(y_0)$ to the polynomial
$u_{x00}+u_{x01}y_0+u_{x03}y_0^3$ in the backward hemisphere $-1.8<y_0<0$.
At midrapidity $du_{x0}/dy_0 = u_{x01}$, while the constants $u_{x00}$ and
$u_{x03}$ take care of the uncertainty of the offset correction and of
non-linearities in the rapidity dependence, respectively. 
The quality of these fits is generally excellent and similar to those shown
in Fig.~\ref{v1-au1500} for $v_1(y_0)$.

Interpolating our data at $1A$ GeV, we find compatibility with the data of
Kintner {\em et al.} at $b_0=0.6$,
 but for lower $b_0$ the evolution is different.
The crossing point to 'antiflow' of the earlier data is seen to occur roughly
around $b_0=0.2$ for $\pi^+$, whereas our data suggest this to happen for
$b_0=0.45-0.55$.
This difference exceeds the uncertainty of the centrality determination in both
experiments.
For $\pi^-$ the data of ref.~\cite{kintner97} are very close to the no-flow
axis (given the error bars) preventing a reliable determination of the crossing
point, which appears in our data to be shifted by about 0.1 units relative to
the $\pi^+$ curve.
Also, our data show more $\pi^-$ flow in the maximum near $b_0=0.25$.
The midrapidity observable $du_{x0}/dy_0$ depends on the transverse momentum
range covered: in our case sharp cuts were applied ($1.0<u_{t0}<4.2$, or
$0.10<p_t<0.43$ GeV/c at $E/A=1A$ GeV).
The low $p_t$ limitations of the data were not discussed in
ref.~\cite{kintner97}. 

\begin{figure}
\begin{center}
\epsfig{file=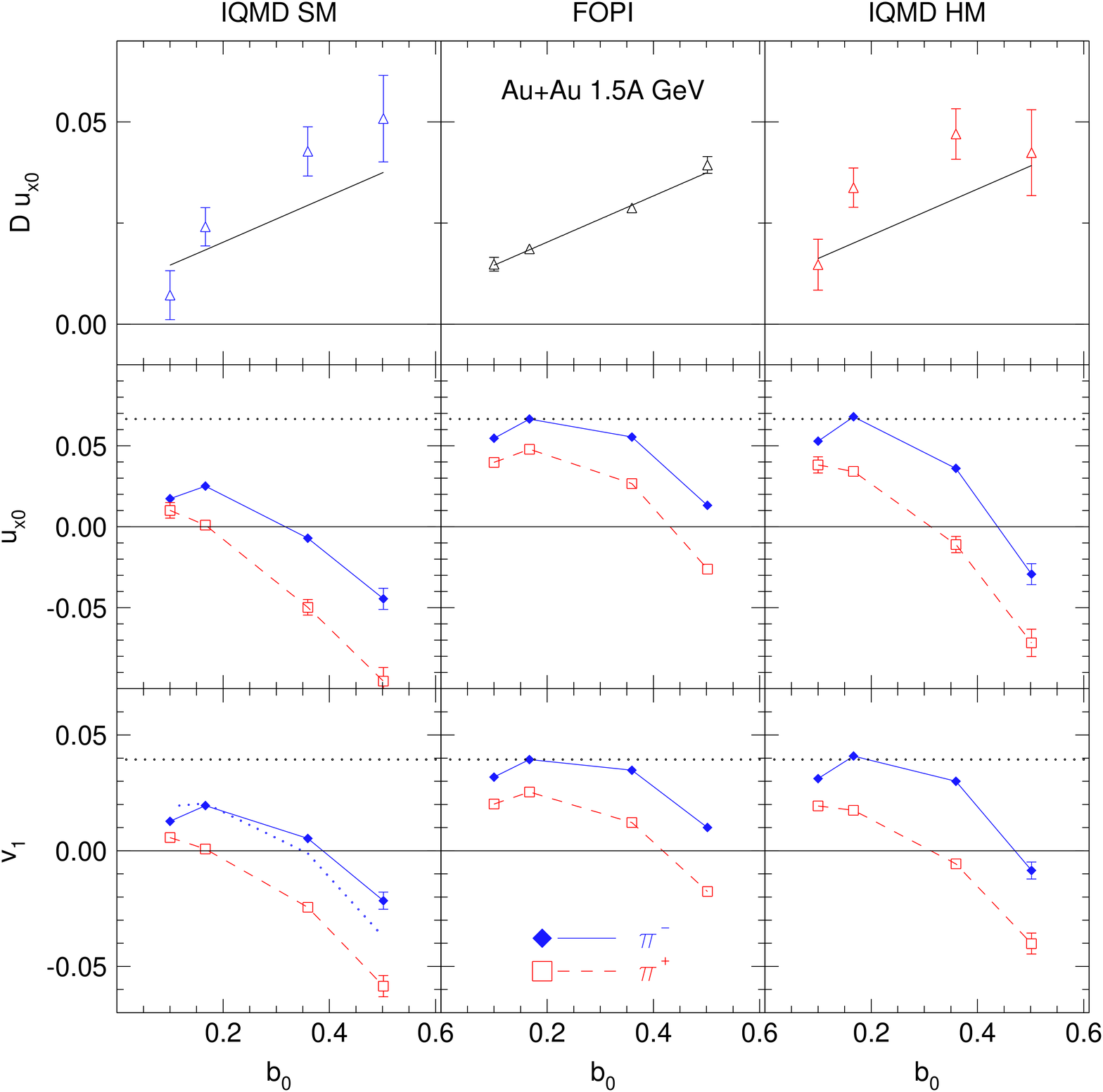,width=154mm}
\end{center}
\caption{
\small{
Centrality dependence of large acceptance (see text) directed flow in the 
reaction Au+Au at $1.5A$ GeV (squares and dashed lines: $\pi^+$, full diamonds:
$\pi^-$).
Going from left to right, the middle panels ('FOPI') represent the measured
data for $v_1$, $u_{x0}$ and the isospin differential flow $D u_{x0}$.
The left (right) hand panels are the simulated data using a soft (stiff)
equation of state.
The dotted horizontal lines are merely references to guide the eye.
The result of a linear least squares fit to the experimental
isospin-differential flow (top middle row) is repeated in the adjacent upper
panels. 
The dotted curve in the lower left panel results from a calculation
(for $\pi^-$) with the phase shift prescription.}
}
\label{flow-v1}
\end{figure}

Moving now to a comparison with simulations, we have chosen the large acceptance
flow, $|y_0|<1.8$ and $0.8<u_{t0}<4.2$,
for this purpose as it requires less events for a given statistical
accuracy.
The comparison is presented in Fig.~\ref{flow-v1} 
for the reaction Au + Au at $1.5A$ GeV where the experimental
data for $v_1(b_0)$ and $u_{x0}(b_0)$ and the isospin differential flow
$D u_{x0} = u_{x0}(\pi^-)-u_{x0}(\pi^+)$ are framed by calculations with a soft
EOS (left panel) and a stiff EOS (right panels).
The stiff EOS predicts statistically significantly higher values of both $v_1$
and $u_{x0}$ and a transition to antiflow in more peripheral collisions.
This sensitivity was noted earlier~\cite{bass95a}.
In the 'standard' version of IQMD we use, the stiff EOS is closer to the data,
especially for central ($b_0<0.2$) collisions.
However, even the stiff EOS predicts a transition to antiflow at smaller $b_0$
than our data.
This conclusion is not changed if we shift to the phase shift prescription
for the $\Delta$ baryon lifetime although this option changed the pion
multiplicities significantly (section \ref{transport}).
In the lower left panel of Fig.~\ref{flow-v1} we show also a calculation
(dotted curve) with the phase shift prescription for $\pi^-$.
The statistical errors (not shown) are similar to those of the standard
calculation.

The systematic difference between $\pi^-$ and $\pi^+$ flow is reproduced and
seems to be unaffected by the EOS.
Since we have not used an isospin dependent EOS, any isospin effect in the
calculation, besides Coulomb fields, would have to result from the 'cascade'
(i.e. collision) part of the model.
Past experience has shown that cascade effects influence directed flow only
moderately.
The upper middle panel shows that $D u_{x0}(b_0)$ varies linearly: the fit to
the difference data is also plotted in the panels showing $D u_{x0}$ from the
calculations.
The calculated $D u_{0}$ are similar to to the experimental values but, despite the statistical
limitations, seem to indicate a small surplus at intermediate $b_0$,
possibly a consequence of the missing nuclear isospin mean field in the
simulation.
The effects are small and hence difficult to assess quantitatively in a
convincing way.

\subsection{Elliptic flow $v_2$}\label{v2}

The presentation of our elliptic flow data follows in many ways the scheme of
the previous subsection on directed flow.
We start with aspects of rapidity and transverse momentum differential flow in
the reaction Au+Au at $1.5A$ GeV, then switch to a systematics of beam energy,
system size and system isospin dependences.
Then we compare our data to earlier measurements where possible and finally
compare a subset of the data to theoretical simulations.

\begin{figure}
\begin{center}
\epsfig{file=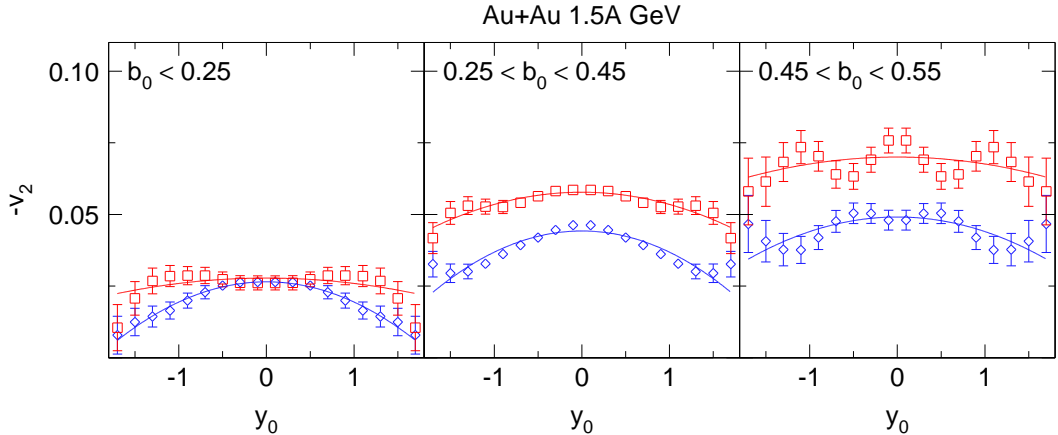,width=154mm}
\end{center}
\caption{
\small{
Rapidity dependence of the elliptic flow, $v_2(y_0)$,
for Au+Au collisions at $1.5A$ GeV with the indicated centralities.
The data are taken in the interval $1.0 < u_{t0} < 4.2$.
The solid curves are least squares fits of the two-parameter function
$c_0 + c_2 y_{0}^2 $.
(Red) squares: $\pi^+$, (blue) diamonds: $\pi^-$.}
}
\label{v2-au1500}
\end{figure}

In Fig.~\ref{v2-au1500} we show the rapidity dependence of elliptic flow,
$v_2(y_0)$, for Au+Au collisions at 1.5A GeV for various centralities.
Again, the flow of $\pi^+$ is different from that of the $\pi^-$, the
difference increasing with impact parameter.
The rapidity dependence is relatively flat: elliptic pion flow is not limited
to a narrow mid-rapidity interval.
There seem to be some weak structures (we remind however that only the backward
hemisphere was covered by our analysis). 
These marginal structures can be evened out by the two-parameter polynomial
fits (see the caption) on a level that is close to the statistical freedom and
therefore will not be discussed further.
The transverse momentum range is again sharply cut in the limits
$1<u_{t0}<4.2$.
Note that we always plot $(-v_2)$, rather than $(+v_2)$.
The sign of $v_2$ tells us that there is a surplus of emitted pions in the
direction ($y$) perpendicular to the reaction plane (zx).

\begin{figure}
\begin{minipage}{155mm}
\epsfig{file=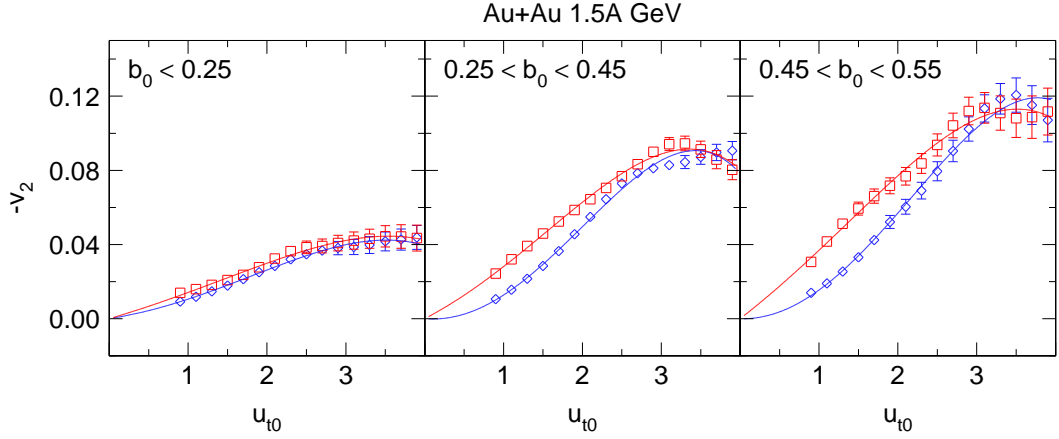,width=154mm}
\end{minipage}
\caption{
\small{
Transverse momentum dependence of the elliptic flow, $v_2(u_{t0})$,
for Au+Au collisions with indicated centralities.
The rapidity interval is $-1.8 < y_0 < 0$.
The solid curves are least squares fits of the polynomial               
$c_1 u_{t0} + c_2 u_{t0}^2 + c_4 u_{t0}^4$
to guide the eye.
(Red) squares: $\pi^+$, (blue) diamonds: $\pi^-$.}
}
\label{v2ut-au1500}
\end{figure}

The transverse momentum dependence of elliptic flow, $v_2(u_{t0})$,
integrated over the backward hemisphere,
is shown in Fig.~\ref{v2ut-au1500}.
It rises with transverse momentum but seems to saturate beyond $u_{t0}=3$
(or $p_t=0.375$ GeV/c at $1.5A$ GeV).
The difference between $\pi^+$ and $\pi^-$ flow is small but statistically
significant for intermediate centralities and momenta.

\begin{figure}
\begin{center}
\epsfig{file=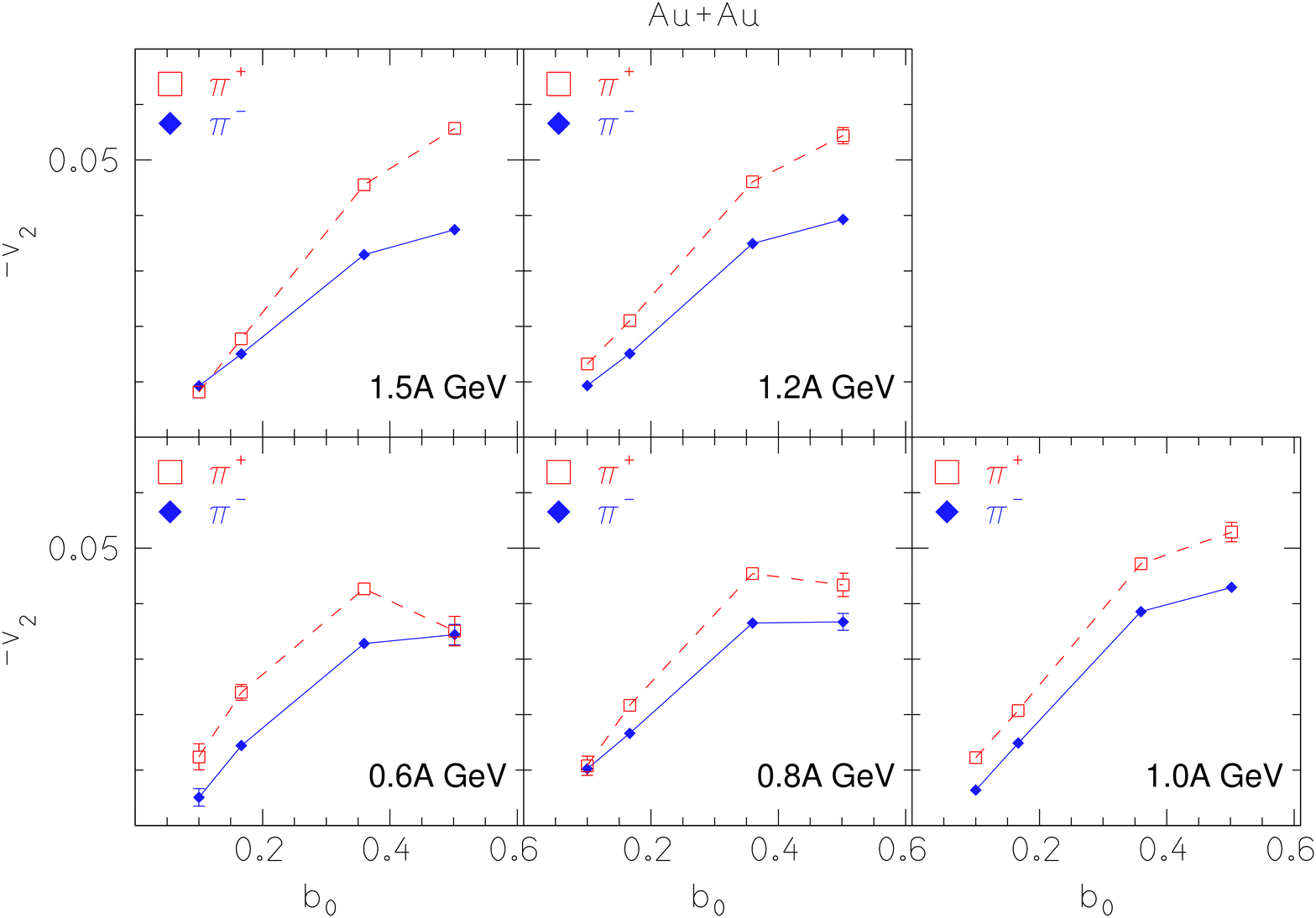,width=154mm}
\end{center}
\caption{
\small{
Centrality dependence of large acceptance elliptic flow $v_2$ 
($-1.8<y_0<0$ and $1<u_{t0}<4.2$) 
at various indicated
beam energies in the reaction Au+Au. 
Open squares joined by dashed lines: $\pi^+$, full diamonds: $\pi^-$.}
}
\label{v2-au}
\end{figure}

\begin{figure}
\begin{center}
\epsfig{file=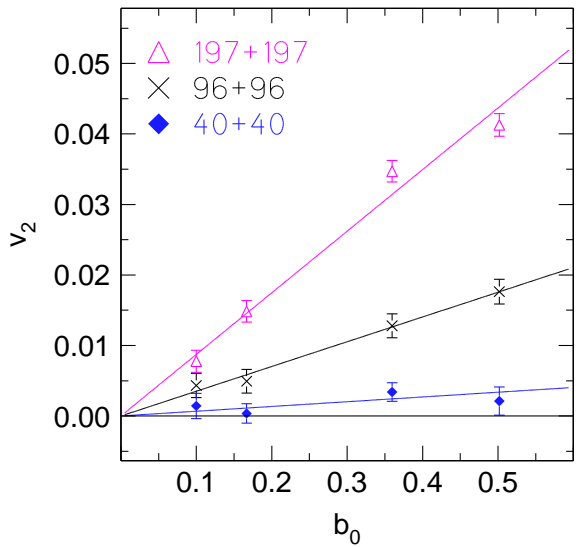,width=90mm}
\end{center}
\caption{
\small{
Centrality dependence of elliptic flow, $v_2$, 
($-1.8<y_0<0$ and $0.8<u_{t0}<4.2$) 
for various indicated system
masses.
The flow is averaged over $\pi^+$ and $\pi^-$.
The values for $A_p+A_t=96+96$ are an average between Zr+Zr and Ru+Ru.}
The straight lines are linear least square fits constrained to $v_2=0$ for
$b_0=0$.
}
\label{v2-asys}
\end{figure}

\begin{figure}
\begin{center}
\epsfig{file=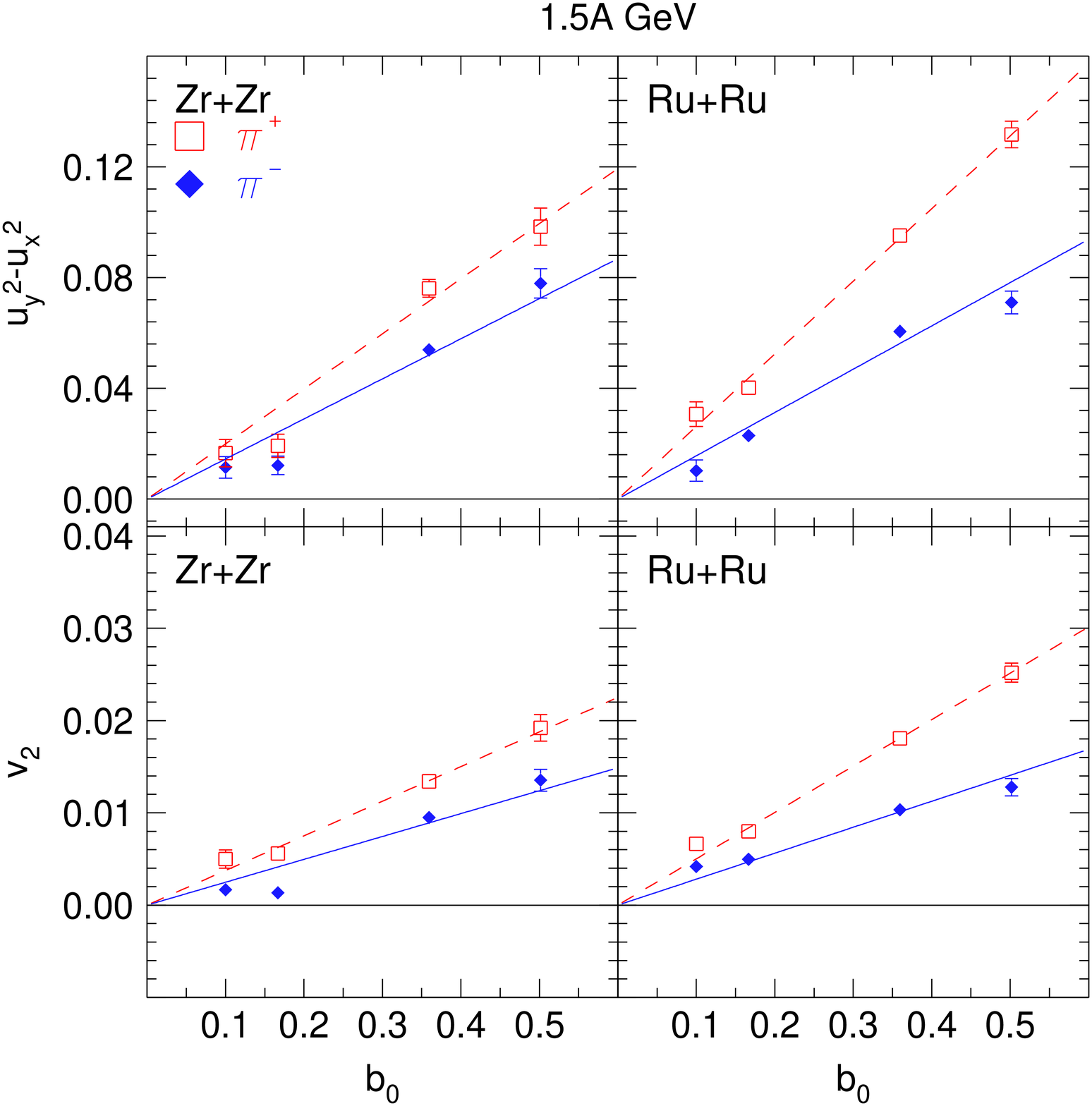,width=110mm}
\end{center}
\caption{
\small{
Comparison of the centrality dependence of elliptic flow ($v_2$ bottom and
$u_y^2-u_x^2$ top) in the systems $^{96}$Zr+$^{96}$Zr (left)
and $^{96}$Ru+$^{96}$Ru
($-1.8<y_0<0$ and $0.8<u_{t0}<4.2$).
Open squares joined by dashed lines: $\pi^+$, full diamonds: $\pi^-$.}
The straight lines are linear least square fits constrained to $v_2=0$ for
$b_0=0$.
}
\label{v2-iso}
\end{figure}

Beam energy dependences, system size and system isospin dependences are
established again in terms of large acceptance flow.
The outcome is summarized in Figs.~\ref{v2-au}, \ref{v2-asys}, \ref{v2-iso}.
Like $v_1$, $v_2$ evolves only slowly with energy (Fig.~\ref{v2-au}), $\pi^+$
elliptic flow always being somewhat larger then $\pi^-$ flow.
The size dependence of $v_2$ is rather pronounced (Fig.~\ref{v2-asys}) and
regular in contrast to $v_1$ (Fig.~\ref{v1-asys}).
Comparing the systems $^{96}$Ru+$^{96}$Ru and $^{96}$Zr+$^{96}$Zr,
Fig.~\ref{v2-iso}, one can make
the same comments as for $v_1$, Fig.~\ref{v1-iso}.
The $\pi^+ -\pi^-$ difference is larger for the system with the smaller $N/Z$.
If we parameterize the $b_0$ dependence by (one-parameter) straight lines
constrained to pass the origin, we see already by eye inspection (see
Fig.~\ref{v2-iso}) that the difference between the two systems in terms of
isospin differential  flow is larger than expected naively on account of
the $10\%$ Coulomb potential difference.
Using the slopes of these least squares fitted lines, we can put this in a
quantitative form.
For $v_2$ the slope difference ($\pi^+ - \pi^-$) is $0.0127\pm 0.0033$ 
(Zr+Zr) versus $0.0221 \pm 0.0029$ (Ru+Ru), and for $u_y^2 - u_x^2$ it is
$0.054 \pm 0.020$ (Zr+Zr) versus $0.106 \pm 0.010$ (Ru+Ru).
These observations suggest that a more extensive theoretical analysis of such
data could help establishing constraints on isospin dependences of high density
mean fields.

\begin{figure}
\begin{center}
\epsfig{file=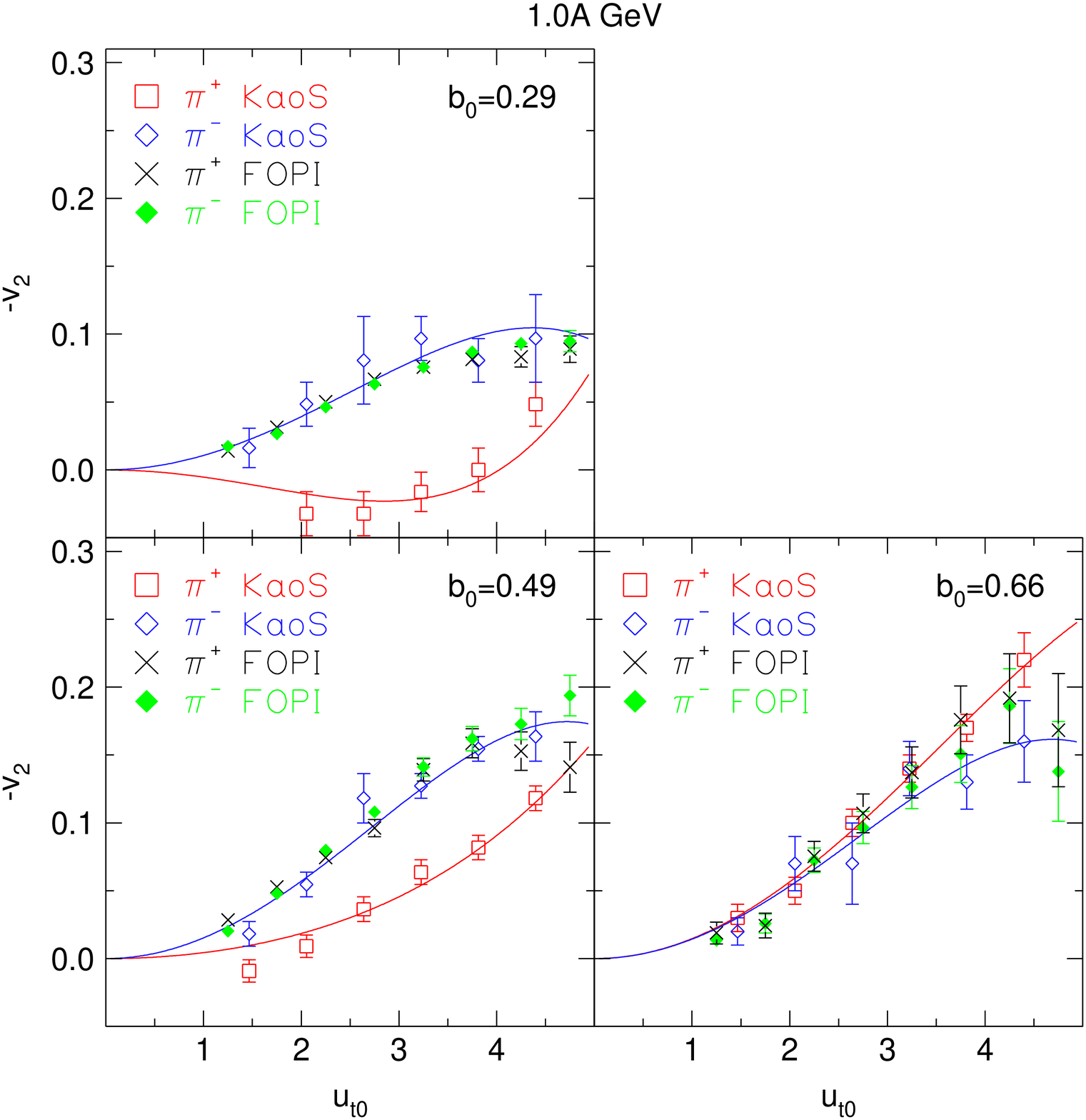,width=154mm}
\end{center}
\caption{
\small{
Transverse momentum dependence of elliptic pion flow, $v_2(u_{t0})$, for
incident beam energies of $1A$ GeV and various indicated centralities $b_0$.
The FOPI data for Au+Au are compared to the KaoS data~\cite{brill97} for Bi+Bi.
The smooth curves are least squares fits of the two-parameter polynomial
$v_{22} u_{t0}^2 + v_{24} u_{t0}^4$ to the KaoS data guiding the eye.
The data are taken in the rapidity interval $|y_0| < 0.5$.}
}
\label{brill}
\end{figure}

We close this subsection by comparing our data first to KaoS
data~\cite{brill97} and then to IQMD simulations.
KaoS published elliptic flow data for Bi+Bi at 0.4, 0.7 and $1A$ GeV.
As we do not have data at $0.7A$ GeV, and since the statistical significance of
our measurements at $0.4A$ GeV is rather modest, we limit ourselves to a
comparison at $1A$ GeV ignoring the difference between the systems Bi+Bi and
Au+Au.
The comparison is shown in Fig.~\ref{brill} for three centralities which were
dubbed MUL2 ($b_0=0.66$), MUL3 ($b_0=0.49$) and MUL4 ($b_0=0.29$) in
ref.~\cite{brill97}.
(The centrality selection of KaoS is based on charged particle multiplicities
registered in a 96-units hodoscope at polar angles between $12^{\circ}$ and
$48^{\circ}$.)
Again, we try to align the experimental conditions as much as possible,
choosing similar nominal centralities and rapidity cuts for our analysis.
The plotted KaoS data were inferred from Tables 1 and 2 of
ref.~\cite{brill97} which
tabulate $P_2 = 2v_2$ values uncorrected for resolution and then corrected
using correction factors from Table 4 of ref.~\cite{brill96}.
The average $b_0$ were inferred from Table 2~\cite{brill96} using $b_0=b/R$
with $R=13.468$ fm for Bi+Bi.
The shown KaoS data can be smoothened by a polynomial fit (see caption
Fig.~\ref{brill}).

We find an excellent agreement of the data for $\pi^-$ flow, but
not for $\pi^+$ flow at $b_0=0.29$ and $0.49$.
This is somewhat puzzling as the authors of ref.~\cite{brill97} stress that 
their data show no significant difference in the azimuthal emission pattern
of positively and negatively charged pions.
Under the conditions chosen to generate Fig.~\ref{brill} (i.e. limiting the
rapidity range to $|y_0|<0.5$) we indeed find that $\pi^+$ and $\pi^-$ flow are
very similar, although integration over a larger $y_0$ interval reveals a small
enhancement of $\pi^+$ flow as we saw above.
We also note that another statement in ref.~\cite{brill97} that {\em 'the pion
data indicate little dependence on the impact parameter'} is not 
properly characterizing our data for $b_0\leq 0.5$.

\begin{figure}
\begin{center}
\epsfig{file=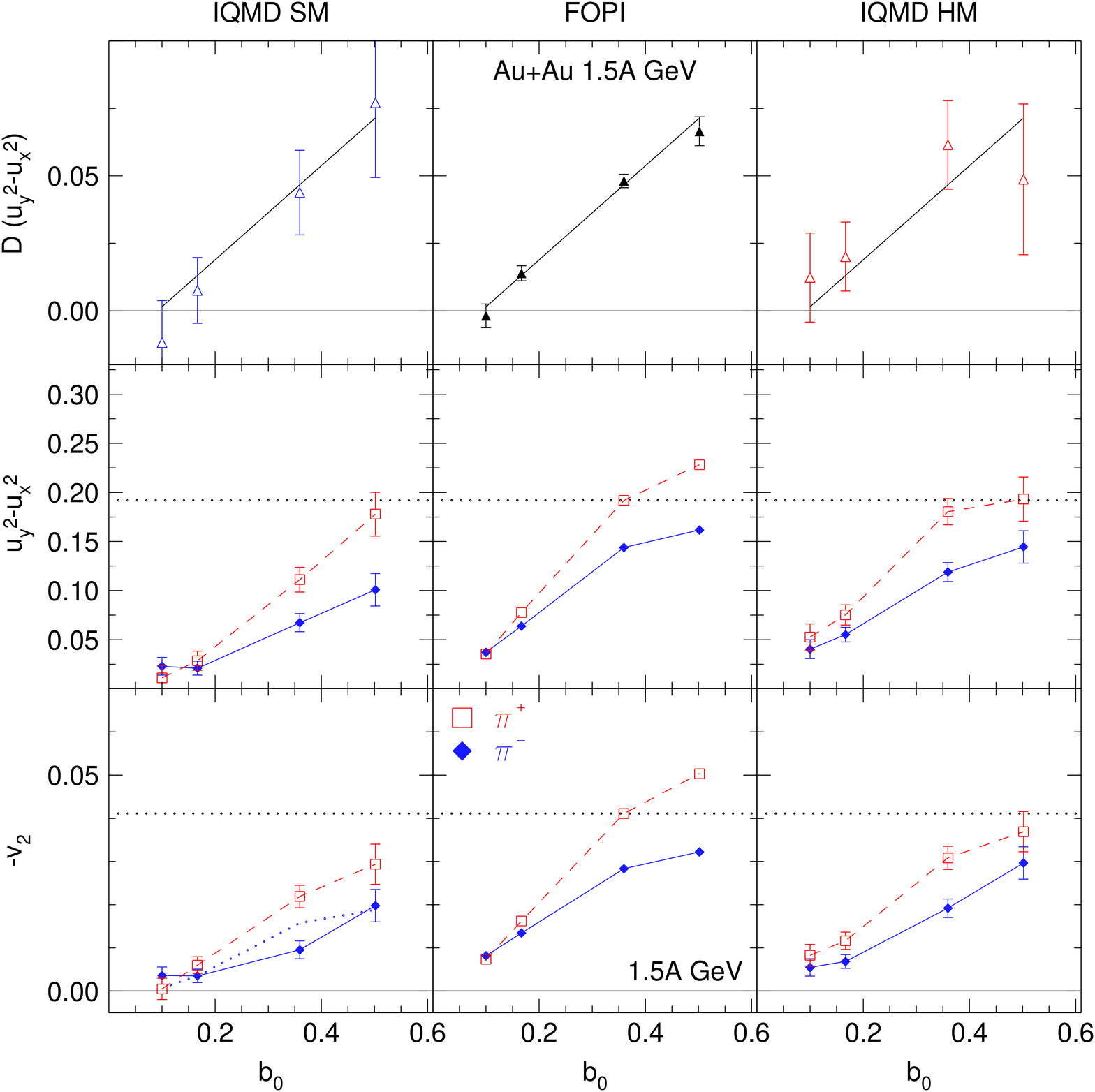,width=154mm}
\end{center}
\caption{
\small{
Centrality dependence of elliptic flow in the 
reaction Au+Au at $1.5A$ GeV (squares and dashed lines: $\pi^+$, full diamonds:
$\pi^-$).
Going from left to right, the middle panels ('FOPI') represent the measured
data for $v_2$, $u_y^2-u_x^2$ and the isospin differential flow 
$D (u_y^2-u_x^2)$.
The left (right) hand panels are the simulated data using a soft (stiff)
equation of state.
The dotted horizontal lines are  references to guide the eye.
The result of a linear least squares fit to the experimental
isospin-differential flow (top middle flow) is repeated in the adjacent upper
panels. 
The dotted curve in the lower left panel results from a calculation
(for $\pi^-$) with the phase shift prescription.}
}
\label{flow-v2}
\end{figure}

In Fig.~\ref{flow-v2} we present a comparison of measured elliptic flow with
results from simulations.
Following a similar line as for directed flow (Fig.~\ref{flow-v1})
we present large acceptance data for $(-v_2)$, $(u^2_{y0}-u^2_{x0})$ and 
isospin differential $(\pi^+-\pi^-)$ flow, $D (u^2_{y0}-u^2_{x0})$, in Au on
Au collisions at $1.5A$ GeV framed by simulations using a soft (left panels)
and stiff (right panels) EOS.
Qualitatively, the trends of the simulations are similar to those of the data,
the stiff EOS again being somewhat closer to the data, although still a little
on the short side.
The systematic $\pi^+ - \pi^-$ difference is well reproduced, see the upper row
of panels in Fig.~\ref{flow-v2}.
As already noticed for directed flow (Fig.~\ref{flow-v1}), the phase shift
prescription for the $\Delta$ baryon lifetime leads
to a moderate (and statistically
marginal) modification of the predicted elliptic flow (dotted curve in the
lower left panel for $\pi^-$).

\section{Summary} \label{summary}
This work presents for the SIS energy regime of heavy ion collisions the most
encompassing pion systematics available today.
While some of the presented data are just a (needed) confirmation of earlier
pioneering work, a significant number of observations are new.                

Among the data which urgently needed a confirmation, are pion multiplicities.
One important conclusion from our present study is that two pion detection
systems that were able to measure and publish pion multiplicities
in heavy ion reactions
in the $1A$ GeV regime with acceptances close to $4\pi$, the BEVALAC Streamer
Chamber and FOPI at SIS/Darmstadt, now offer highly consistent results
(Fig.~\ref{harris}).
In the larger framework of using heavy ion collision data to infer properties
of nuclear matter far off the ground state, this should encourage future
theoretical efforts to understand pion production in the $1A$ GeV regime on a
level of $10\%$, the experimental uncertainty that is, presently still, typical
for most absolute pion observables, and in particular the multiplicities.
Extending the truly pioneering Streamer Chamber data, our present data
give information on two pion charges, $\pi^+$ and $\pi^-$, and hence do not
require non-trivial assumptions on the isospin dependences in order to deduce 
the multiplicity of all pions.

Also, our systematics in terms of system sizes and system isospin is now
enlarged.
As a result of this improvement, we can now say that pion multiplicities per
participant are not strictly a constant for a given incident energy, but
show a measurable trend towards smaller values as the participant number is
increased, an effect that on the $10\%$ level does not depend on  how the
system size is varied, be it by varying the overlap zone
changing the centrality, or by modifying target and projectile
(Fig.~\ref{powerlaw}).

Most of the progress since the eighties, however, rests on the detailed, 
centrality selected truly three-dimensional momentum space populations that
have now become available.
In contrast with just integrated $4\pi$ multiplicities which require large
(and well understood) acceptances rather than a multitude of registered events, 
many subtle and relatively small effects, such as asymmetries in pion emission,
required electronic devices such as FOPI to be able to handle the wealth of
information connected with the variation of energy, centrality, system size and
system isospin.
Complemented with the studies of the TAPS and the KaoS Collaborations there
now exists a rather complete set of informations on pion emission.

In two (scaled) dimensions, longitudinal rapidity, $y_0$, and transverse
four-velocity, $u_{t0}$, the pions peak at mid-rapidity populating,
especially at low energy, a rather broad phase space extending over at least
twice the original rapidity gap.
In these scaled units the distributions become significantly more compact at the
higher end of the studied energy range (Fig.~\ref{uty}).

Rather than presenting the longitudinal (beam) direction in terms of a rapidity
distribution on a linear scale, and then switch to a logarithmic ordinate
scale to present the transverse direction in terms of a transverse mass or
momentum distribution, as is commonly done in the literature, we have compared
the two orthogonal directions more directly introducing the (one-dimensional)
transverse rapidity distributions.
The scaled variances of the transverse and the longitudinal rapidity
distributions decrease significantly with increasing beam energy
(Fig.~\ref{vary}),
the transverse variance always being the smaller of the two.
We have dubbed the ratio of the transverse to the longitudinal variance
'stopping' and find that this observable decreases steadily with increasing
incident energy, although more slowly than the individual variances, which in a
naive thermal model (equilibrium, no flow, no decay distortions) would relate
directly to (kinetic)
temperatures and would have to be independent of the direction.

This 'pion stopping' qualitatively follows the 'nucleon stopping'
\cite{reisdorf04a}, the latter decreasing however faster with energy,
probably a reflection, on average, of a less violent collision history, due in
part to a less perfect separation of spectator type influences.
However, we also found evidence for spectator influence on the pions, although
created particles originate in this energy regime from the overlap zone.
When studying the size dependence of pion stopping,
we found a remarkable phenomenon of two stopping branches:
while stopping in the most central collisions 
is almost independent of the system size, very much in contrast
to nucleon stopping~\cite{reisdorf04a},  and even seems to increase slightly as
the size decreases, an effect that is highly non-trivial, we find that stopping
decreases when significant spectator matter is present (i.e. in half-overlap,
$b_0=0.5$, collisions). 
We associate this observation (Fig.~\ref{vartlapart}) tentatively 
to a reacceleration of
pions penetrating and rescattering in the fast spectator (in the c.o.m.
system).
Since there was no significant indication of two corresponding branches in pion
{\em production}, one is tempted to conclude that pions are rescattered, rather
than absorbed by spectator matter.

Isospin dependence of pion stopping was not observed on a detectable level when
comparing $^{96}$Ru + $^{96}$Ru with $^{96}$Zr + $^{96}$Zr.

The partial transparency that is suggested by the observation that the
'stopping' observable is always less than one, is also evidenced by a marked
polar anisotropy of pion emission (Fig.~\ref{dndo}) which again shows the two
branches depending on the presence or absence of spectators
(Fig.~\ref{anisoapart}).
Anisotropies imply that extrapolations of midrapidity data to $4\pi$ that
assume isotropic emission underestimate pion yields typically by a factor 1.2
to 1.4.
When we apply this correction factor to TAPS or KaoS midrapidity data,
we find fair agreement with the data from large acceptance devices, however
some problem cases remain (Figs.~\ref{taps}, \ref{powerlaw}).

Our transverse momentum spectra, that agree in shape with KaoS
(Fig.~\ref{KaoS}),
and also with the IQMD simulation (Fig.~\ref{pt}), are characterized, besides
the variance (or apparent transverse temperature) by a marked difference
between $\pi^+$ and $\pi^-$ that is suggested by the simulation to originate
almost exclusively from the Coulomb fields (Fig.~\ref{pt}).
The latter also account for the main features of the systematics of ratios of
average momenta (Fig.~\ref{ptratio}).

Whether the symmetry part of the ground state EOS influences the yield ratios of
$\pi^-$ to $\pi^+$ in neutron rich nuclei is not clear.
The naive expectation from the first chance isobar model that this ratio should
grow quadratically with $N/Z$ is generally not fulfilled, as the ratio tends to
vary linearly with $N/Z$ due to a partial thermalization effect, except at the
lowest energy ($0.4A$ GeV). Simulations seem to miss this low energy trend,
however (Figs.~\ref{piratio}, \ref{bali}).

This work also provides a significant extension of data on pion
azimuthal correlations,
especially for the directed flow which is of special interest due to its
apparent sensitivity to the EOS (Fig.~\ref{flow-v1}) \cite{bass95a}.

The rapidity dependence $v_1(y_0)$ has the S-shape familiar from the early days
of flow measurements~\cite{reisdorf97} and is well described in terms of an odd
polynomial in $y_0$ including just a linear and a cubic term
(Fig.~\ref{v1-au1500}).
In contrast to the ultrarelativistic regime~\cite{NA49}
where the flow of pions appears to be mostly  opposite to proton flow
('antiflow'), we find in the $1A$ GeV regime  a more complex behaviour.
First, for high centralities, $b_0<0.4$, the flow of the pions follows the
nucleon flow although it  is considerably smaller in terms of $v_1$,
roughly by the ratio of pion to nucleon mass. 
For larger impact parameters there is a switch to antiflow
(Figs.~\ref{v1-au1500} and
\ref{kintner}).
Second, the flow of positively charged pions differs from that of negatively
charged pions, in particular the switch to antiflow occurs at smaller
centrality (Fig.~\ref{kintner}).
  
Although these features
confirm qualitatively what was found in an earlier study by
the EOS Collaboration~\cite{kintner97}, the quantitative agreement is not
satisfactory (Fig.~\ref{kintner}): whether the difference
is due to different detector acceptances is not clear.

The transverse velocity dependence $v_1(u_{t0})$ (integrated over a large
rapidity interval) is relatively flat with weak structures
(Fig.~\ref{v1ut-au1500}).

Despite theses detailed features which show that pion flow is a complex
phenomenon, it is useful for systematics purposes to try to characterize flow
by just one parameter.
One alternative is the midrapidity slope $dv_1/dy_0(y_0=0)$, for which we have
established a systematics varying the incident energy (Fig.~\ref{v1slop-au}):
this parameterization puts more weight on the midrapidity region.
Another alternative, stressing higher rapidities, consists of the 
large-acceptance
value of $v_1$, for which we have established size (Fig.~\ref{v1-asys}) and
isospin (Fig.~\ref{v1-iso}) dependences.
The isospin differential flow of pions is a systematic effect which is largest
for Au + Au at the lower incident energies (Fig.~\ref{v1slop-au}).
It seems to be dominated by Coulomb effects since it is quantitatively
reproduced by our simulation for Au+Au at $1.5A$ GeV (Fig.~\ref{flow-v1})
which did not include isospin dependent mean fields except for the Coulomb
field.
However, surprisingly, the isospin differential pion flow was found to be
significantly smaller for $^{96}$Zr + $^{96}$Zr ($N/Z=1.40$) than for
$^{96}$Ru + $^{96}$Ru ($N/Z = 1.18$), Fig.~\ref{v1-iso}.
Quantitatively, it seems necessary to introduce 
 a nuclear isospin field of
opposite sign to the Coulomb part of the field in order to explain this feature.
The simulation at $1.5A$ GeV (Fig.~\ref{flow-v1}) showed that the EOS
influences pion directed flow in a measurable way, favouring the stiffer EOS in
the IQMD version that we used.

In the $1A$ GeV energy range pion elliptic flow like nucleonic flow is
negative,
i.e. preferentially out-of plane .
The presence of spectator matter in the expansion phase of the fireball
is probably the most important 'geometrical' difference to the
ultrarelativistic regime~\cite{NA49}, our stopping studies indicating that
pion rescattering in spectator matter plays a role, as was already
suggested in ref.~\cite{bass95}.
We find that $v_2(y_0)$ 
is rather flat over a very large range of rapidities (Fig.~\ref{v2-au1500}).
In contrast to
directed flow,  $|v_2|$            increases nearly linearly 
with decreasing centrality  in the range covered by our data ($b_0 < 0.55$).
The transverse four-velocity dependence $v_2(u_{t0})$ (Fig.~\ref{v2ut-au1500})
is quadratic ($\pi^-$) or linear ($\pi^+$) for smaller momenta, 
flattening out at higher momenta.
We confirm the tabulated KaoS data~\cite{brill97}
 for $\pi^-$ (Fig.~\ref{brill}), but not for $\pi^+$.
While the dependence on incident energy is relatively weak, Fig.~\ref{v2-au},
the system-size dependence is rather strong for all centralities,
Fig.~\ref{v2-asys}.
As for directed flow, the system $^{96}$Ru + $^{96}$Ru shows a larger
difference between $\pi^+$ and $\pi^-$ elliptic flow than the system
$^{96}$Zr + $^{96}$Zr, so the same comments apply.
In comparing with the IQMD simulations we find experimentally a somewhat larger
elliptic flow, even when using the stiff EOS (Fig.~\ref{flow-v2}).
The isospin differential
flow in the heavy system Au+Au is again well reproduced (Fig.~\ref{v2-iso}).

At this time, however we do not wish to draw firm conclusions on the issue
of the
stiffness of the EOS from the pion flow, as we expect other features of the
simulation, namely the treatment of the $\Delta$ baryon propagation
in the medium, to
influence the predicted flow.
Shorter effective $\Delta$ baryon lifetimes in the model 
are expected to decrease the
efficiency of the assumed dominant pion absorption mechanism via $N\Delta
\rightarrow NN$ and hence the pion yields at freeze out, as well as the
influence of the mean field in the time between elementary collisions, and
hence the observed flow.
This is an interesting aspect of in-medium physics in its own right and deserves
theoretical investigations with the aim of reaching a consensus on how this
problem should be treated properly in transport codes.

Much work still needs to be
done, to coordinate the transport theoretical efforts with the aim of achieving
code independent conclusions.
Such efforts were started in ref.~\cite{hartnack98} and continued on a larger
scale in ref.~\cite{kolomeitsev05} which also contains further literature
citations (and transport codes) for the interested reader

The request for reproduction of finer details, such as the full 3D features of
the momentum space population under exclusive conditions, subtle system size and
isospin effects, is premature at this time as long as global features, such as
production, stopping and integrated flow are not under sufficient control.
This is a task for future theoretical work beyond the scope of the present
experimental work.
From the richness of our observations it follows that this represents  a
challenge and hopefully
will eventually contribute to a better understanding of nuclear
medium properties, which cannot simply be inferred from 
incoherently superimposing
experimental information on so-called 'elementary' hadron-hadron reactions.

\begin{ack}
This work has been supported by the German BMBF,
contract 06HD154 and within the framework of the WTZ
program (Project RU8 02/021), by the DFG (Project 446-KOR-113/76),
the DAAD (PPP D /03/44611) and the IN2P3/GSI agreement 97/29.
This work was also supported by a Korea Research Foundation grant 
(KRF-2005-041-C00120).
\end{ack}
\newpage



\begin{thebibliography}{30}

\bibitem{lee02} T.-S. Lee, R. P. Redwine,
Annu. Rev. Nucl. Part. Sci. 52 (2002) 23.                 

\bibitem{mayer93} R. S. Mayer, et al.,
Phys. Rev. Lett. 78 (1997) 4165).

\bibitem{bertsch88}G. F. Bertsch, S. Das Gupta,
Phys. Rep.  160 (1988) 189.

\bibitem{harris87}J. W. Harris, et al.,
Phys. Rev. Lett. 58 (1987) 463.

\bibitem{stock86}R. Stock, 
Phys. Rep.  135 (1986) 259.

\bibitem{bertsch84}G. F. Bertsch, H. Kruse, S. Das Gupta,
Phys. Rev. C 29 (1984) 673.           

\bibitem{kruse85} H. Kruse, B. V. Jacak, H. St\"{o}cker,
Phys. Rev. Lett. 54 (1985) 289.  

\bibitem{kitazoe86} Y. Kitazoe, M. Sano, H. Toki, S. Nagamiya,
Phys. Lett. B 166 (1986) 35.         

\bibitem{bass95a} S. A. Bass, C. Hartnack, H. St\"{o}cker, W.Greiner,
Phys. Rev. C 51 (1995) R12.

\bibitem{fung78}S. Y. Fung, et al.,
Phys. Rev. Lett. 52 (1978) 292.  

\bibitem{sandoval80}A. Sandoval, et al., 
Phys. Rev. Lett.  45 (1980) 874.

\bibitem{harris85}J. W. Harris, et al.,
Phys. Lett. B 153 (1985) 377.

\bibitem{alard87} J. P. Alard, et al.,
Nucl. Instr. Meth. A 261 (1987) 379.

\bibitem{gobbi93} A. Gobbi, et al.,
Nucl. Instr. Meth. A 324 (1993) 156.

\bibitem{ritman95}J. Ritman,
Nucl. Phys.  B 44 (1995) 708.

\bibitem{wieman91} H. Wieman, et al.,
Nucl. Phys. A 525 (1991) 617c.

\bibitem{senger93}P. Senger, et al.,
Nucl. Instr. Meth. A 327 (1993) 393.

\bibitem{novotny91} R. Novotny,           
IEEE Trans. Nucl. Sci. 38 (1991) 379.

\bibitem{senger99}P. Senger, H. Str\"{o}bele,
J. Phys. G  25 (1999) R59.

\bibitem{pelte97au} D.Pelte, et al.,
Z. Phys. A  357 (1997) 215.

\bibitem{pelte97ni} D.Pelte, et al.,
Z. Phys. A  359 (1997) 55.  

\bibitem{hong97} B. Hong, et al.,
Phys. Lett. B 407 (1997) 115.

\bibitem{hong98} B. Hong, et al.,
Phys. Rev. C  57 (1998) 244.   

\bibitem{hong05} B. Hong, et al.,
Phys. Rev. C  71 (2005) 034902.

\bibitem{stockmeier02}Marc R. Stockmeier,
Ph.D thesis, University of Heidelberg (2002), Germany.

\bibitem{reisdorf97a}W. Reisdorf, et al.,
Nucl. Phys. A 612 (1997) 493.

\bibitem{hartnack98}C. Hartnack, et al.,
Eur. Phys. J. A  1 (1998) 151.

\bibitem{aichelin91} J. Aichelin,
Phys. Rep.  202 (1991) 233.

\bibitem{bass93} S. A. Bass, C. Hartnack, R. Mattiello, H. St\"{o}cker,
 W.Greiner,
Phys. Lett. B 302 (1993) 381.

\bibitem{bass93a} S. A. Bass, C. Hartnack, H. St\"{o}cker, W.Greiner,
Phys. Rev. Lett. 71 (1993) 1144.

\bibitem{bass94} S. A. Bass, C. Hartnack, H. St\"{o}cker, W.Greiner,
Phys. Rev. C 50 (1994) 2167.

\bibitem{bass94a} S. A. Bass, M. Hofmann, C. Hartnack, H. St\"{o}cker, 
W.Greiner,
Phys. Lett. B 335 (1994) 289.

\bibitem{bass95} S. A. Bass, C. Hartnack, H. St\"{o}cker, W.Greiner,
Phys. Rev. C 51 (1995) 3343.

\bibitem{bass95b} S. A. Bass, C. Hartnack, H. St\"{o}cker, W.Greiner,
Z. Phys. A 351 (1995) 359.

\bibitem{hjort97} E. L. Hjort, et al.,
Phys. Rev. Lett. 79 (1997) 4345.

\bibitem{eskef98} M. Eskef, et al.,
Eu. Phys. J A 3 (1998) 335. 

\bibitem{matulewicz00} T. Matulewicz, et al.,
Eur. Phys. J. 9 (2000) 69.

\bibitem{bali04} Bao-An Li, C. B. Das, S. Das Gupta and C. Gale,
Phys. Rev. C 69 (2004) 011603(R); Nucl. Phys. A 735 (2004) 563.

\bibitem{baran05} V. Baran, M. Colonna, V. Greco, and M. Di Toro,
Phys. Rep. 410 (2005) 335.

\bibitem{gaitanos} T. Gaitanos, et al.,
Nucl. Phys. A 650 (1999) 97;
C. Fuchs, T. Gaitanos,
Nucl. Phys. A 714 (2003) 643.

\bibitem{cassing00} W. Cassing, S. Juchem,
Nucl. Phys. A 665 (2000) 377;
Nucl. Phys. A 672 (2000) 417;
Nucl. Phys. A 677 (2000) 445.

\bibitem{larionov03} A. B. Larionov, U. Mosel,                    
Nucl. Phys. A 728 (2003) 135.                  

\bibitem{danielewicz95} P. Danielewicz, 
Phys. Rev. C 51 (1995) 716.

\bibitem{qfli05} Qingfeng Li, Zhuxia Li, Sven Soff, Marcus Bleicher, Horst
St\"{o}cker,
J. Phys. G. 32 (2006)151.                      

\bibitem{bass98} S. A. Bass, et al.,
Prog. Part. Nucl. Phys. 41 (1998) 255.

\bibitem{reisdorf04b}W. Reisdorf, et al.,
Phys. Lett. B 595 (2004) 118. 

\bibitem{geant}GEANT-Detector description and Simulation Tool,
CERN program library Long Writeup W5013 (CN Division, CERN, 1993).

\bibitem{reisdorf04a}W. Reisdorf, et al.,
Phys. Rev. Lett.  92 (2004) 232301.

\bibitem{averbeck03}R. Averbeck, R. Holzmann, V. Metag, R. S. Simon,
Phys. Rev. C 67 (2003) 024903.

\bibitem{wolf79} K. L. Wolf, et al.,
Phys. Rev. Lett. 42 (1979) 1448.

\bibitem{nagamiya81} S. Nagamiya, et al.,
Phys. Rev. C 24 (1981) 971; 
J. Chiba, et al.,
Phys. Rev. C 20 (1979) 1332; 
K. Nakai, et al.,
Phys. Rev. C 20 (1979) 2210. 

\bibitem{holzmann96} R. Holzmann, et al. (TAPS
Collaboration),
Phys. Lett. B 366 (1996) 63.  

\bibitem{tyminska06} K. Tymi\'{n}ska, T. Matulewicz, K. Piasecki (TAPS
Collaboration),
Acta Phys. Polonica B 37 (2006) 161.  

\bibitem{brockmann84} R. Brockmann, et al.,
Phys. Rev. Lett. 53 (1984) 2012.

\bibitem{gosset77} J. Gosset, et al.,                     
Phys. Rev. C 16 (1977) 629.

\bibitem{stock82}R. Stock, et al., 
Phys. Rev. Lett.  49 (1982) 1236.

\bibitem{wagner96} A. Wagner,              
Phd thesis, University of Darmstadt (1996) (in German).

\bibitem{martinez99} G. Martinez, et al.,
Phys. Rev. Lett. 83 (1999) 1538.

\bibitem{averbeck97}R. Averbeck, et al., 
Z. Phys. A 359 (1997) 65.

\bibitem{gazdzicki95}M. Gazdzicki, D. R\"{o}hrich,
Z. Phys. C 65 (1995) 215.   

\bibitem{schwalb94} O. Schwalb, et al.,
Phys. Lett. B 321 (1994) 20. 

\bibitem{ashery81} D. Ashery, I. Navon, G. Azuelos, H. K. Walter, H. J.
Pfeiffer, F. W. Schlep\"{u}tz,
Phys. Rev. C 23 (1981) 2173.

\bibitem{ahle00} L. Ahle, et al.,
Phys. Lett. B 476 (2000) 1.

\bibitem{kolomeitsev05} E. E. Kolomeitsev, et al.,
J. Phys. G  31 (2005) S741.

\bibitem{wigner55} E. P. Wigner,         
Phys. Rev. 98 (1955) 145. 

\bibitem{danielewicz96} P. Danielewicz, S. Pratt,
Phys. Rev. C 53 (1996) 249.

\bibitem{vandalen05} E. N. E. van Dalen, C. Fuchs, A. Faessler,
Phys. Rev. Lett.  95 (2005) 022302.

\bibitem{kaos} F. Uhlig (KaoS Collaboration),
private communication.

\bibitem{wagner98} A. Wagner, et al.,
Phys. Lett. B 420 (1998) 20.

\bibitem{bali03} Bao-An Li,                    
Phys. Rev. C 67 (2003) 017601.                 

\bibitem{gaitanos04} T. Gaitanos, M. Di Toro, S. Typel, V. Baran, C. Fuchs, 
V. Greco, H. H. Wolter,
Nucl. Phys. A 732 (2004) 24.

\bibitem{kapusta77} J. J. Kapusta,
Phys. Rev. C 16 (1977) 1493.

\bibitem{gustafsson84}H. A. Gustafsson, et al.,
Phys. Rev. Lett.  52 (1984) 1590.

\bibitem{renfordt84}R. E. Renfordt, et al.,
Phys. Rev. Lett.  53 (1984) 763.   

\bibitem{danielewicz85} P. Danielewicz, G. Odyniec,
Phys. Lett. B 157 (1985) 168.

\bibitem{ollitrault98} J.-Y. Ollitrault, 
Nucl. Phys. A 638 (1998) 195c.

\bibitem{bhalerao03}R. S. Bhalerao, N. Borghini, J.Y. Ollitrault,
Nucl. Phys. A 727 (2003) 373; Phys. Lett. B 580 (2004) 157.

\bibitem{bastid05} N. Bastid et al.,
Phys. Rev. C 72 (2005) 011901 and to be published.

\bibitem{gosset89} J. Gosset, et al.,
Phys. Rev. Lett. 62 (1989) 1251.

\bibitem{kintner97} J. C. Kintner {\it  et al.},
Phys. Rev. Lett. 78 (1997) 4165.

\bibitem{bali94} Bao-An Li,                    
Nucl. Phys. A 570 (1994) 797.                  

\bibitem{brill93}D. Brill, et al.,
Phys. Rev. Lett. 71 (1993) 336. 

\bibitem{venema93} L. B. Venema, et al.,
Phys. Rev. Lett.  71 (1993) 835.

\bibitem{brill97}D. Brill, et al.,
Z. Phys. A 357 (1997) 207.

\bibitem{wagner00} A. Wagner, et al.,
Phys. Rev. Lett.  85 (2000) 18.

\bibitem{E877} J. Barrette, et al. (E877 Collaboration),
Phys. Rev. C 55 (1997) 1420.

\bibitem{NA49} C. Alt, et al. (NA49 Collaboration),
Phys. Rev. C 68 (2003) 034903.

\bibitem{brill96}D. Brill, et al.,
Z. Phys. A 355 (1996) 61. 

\bibitem{reisdorf97}W. Reisdorf, H. G. Ritter,
Annu. Rev. Nucl. Part. Sci.  47 (1997) 663;
N. Herrmann, J. P. Wessels, T. Wienold,
{\it ib.}  49 (1999) 581.


\end{thebibliography}
\end{document}